\documentclass[11pt]{article}

\usepackage{amsmath,amssymb,amsfonts,graphicx,pifont}
\usepackage{epstopdf}
\usepackage{color}
\usepackage{cancel}
\usepackage[normalem]{ulem}
\usepackage{comment}
\RequirePackage{slashed}
\usepackage{tikz}

\def\dpdf#1{F_{#1}}

\newcommand{\bit}{\begin{itemize}}
\newcommand{\eit}{\end{itemize}}

\def\be{\begin{equation}}
\def\ee{\end{equation}}
\def\bea{\begin{eqnarray}}
\def\eea{\end{eqnarray}}

\newlength\savedwidth

\begin{document}




\title{Correlations in Double Parton Distributions:  \\
Perturbative and Non-Perturbative effects}

\author{Matteo Rinaldi\footnote{Corresponding author.
email: matteo.rinaldi@pg.infn.it
}~\footnote{From 
October $1^{st}$, Departament de Fisica Te\`orica, Universitat de Val\`encia
and Institut de Fisica Corpuscular, Consejo Superior de Investigaciones
Cient\'{\i}ficas, 46100 Burjassot (Val\`encia), Spain}, Sergio Scopetta\\
Dipartimento di Fisica e Geologia,
Universit\`a degli Studi di Perugia, 
\\ 
and Istituto Nazionale di Fisica Nucleare,
Sezione di Perugia, 
\\
via A. Pascoli, I - 06123 Perugia, Italy
\\
Marco Traini\\
Institut de Physique Th\'eorique CEA-Saclay, F-91191 Gif-sur-Yvette, France,\\
INFN - TIFPA, Dipartimento di Fisica, Universit\`a degli Studi di Trento,\\ 
Via Sommarive 14, I-38123 Povo (Trento), Italy
\\Vicente Vento\\
Departament de Fisica Te\`orica, Universitat de Val\`encia\\
and Institut de Fisica Corpuscular, Consejo Superior de Investigaciones
Cient\'{\i}ficas,\\ 46100 Burjassot (Val\`encia), Spain
}

\date{\today}

\maketitle

\begin{abstract}

The correct description of
Double Parton Scattering (DPS), which represents a background in 
several channels
for the search of new Physics at the LHC, requires the knowledge
of double parton distribution functions
(dPDFs).
These quantities represent also a
novel tool for the study of the three-dimensional nucleon structure,
complementary to the possibilities offered by electromagnetic 
{{probes}}.
In this paper we analyze dPDFs using Poincar\'e covariant
predictions obtained by using 
a Light-Front {{constituent quark}} model proposed in a
recent paper, and QCD evolution.
We study to what extent factorized expressions for dPDFs,
which neglect, at least in part, two-parton correlations,
can be used.
We show that they fail in reproducing the calculated
dPDFs, {{in particular}} in the valence region. Actually
measurable processes at existing facilities
occur at low longitudinal momenta of the interacting partons;
to have contact with these processes we have analyzed correlations between
pairs of partons of different kind, finding that,
in some cases, they are strongly suppressed at low 
longitudinal momenta, while for other distributions they can
be sizeable.
For example, the effect of gluon-gluon correlations can be
as large as 20 $\%$.
We have shown that these behaviors can be understood
in terms of a delicate interference of non-perturbative correlations, 
generated by the dynamics of the model, and perturbative ones,
generated by the model independent evolution procedure.
Our analysis shows that at LHC kinematics two-parton correlations can
be relevant in DPS, and therefore we
address the possibility to study them experimentally.

\end{abstract}

\section{\label{sec:intro} Introduction}

Multi Parton Interactions (MPI) occur when more than one 
parton scattering
takes place in one hadron-hadron collision.
They have been defined
long time ago \cite{paver}, have been recently 
rediscovered 
and are presently attracting remarkable
attention, thanks to 
the activity of Large Hadron Collider (LHC), where specific
signatures are expected to
be observed (see Refs. 
\cite{gaunt,diehl_1,manohar_1,bansal,ciurek} for recent reports).

In particular, the cross section for hard double parton scattering (DPS), 
the simplest MPI process,
depends on 
non-perturbative objetcs, the double parton distribution functions (dPDFs),
describing the number density of two partons 
located at a given transverse separation in coordinate space and
with given longitudinal momentum fractions. 
dPDFs encode, for example, the novel information on
the probability that partons which are
close to each other are faster, or slower,
than those which are far from each other.
They are therefore naturally related to parton correlations, as noticed several 
years ago \cite{calucci}, 
and represent a novel tool to access   the 
three-dimensional (3D) nucleon 
structure, presently studied using electromagnetic probes \cite{gui, dupre}.
The  correlations in DPS are presently deeply investigated (see, e.g., 
\cite{diehl_1,cattaruzza,muld}). 
 
In addition to this non perturbative information, the knowledge
of dPDFs, DPS and MPI in general could be very useful to
constrain the background to the search of new Physics at the LHC,
making their study very timely.
No data are presently
available for dPDFs and their calculation using non perturbative 
methods is cumbersome.
A few model calculations,
able in principle to grasp the most relevant 
features of dPDFs, 
have been therefore performed 
\cite{manohar_2,noi1,noi2,ruiz,ruiz1}. 
In particular, in Ref. \cite{noi2}, a Light-Front (LF) 
Poincar\'e covariant approach,
reproducing the essential sum rules of dPDFs
without ad hoc assumptions and containing
natural two-parton correlations, has been described.
We note in passing that,
although it has not yet been possible to extract dPDFs from data,
the so called  ``effective cross section'', $\sigma_{eff}$,
{{the ratio of the product of two single parton scattering cross sections
to the DPS cross section with the same final states,}}
has been extracted, in a model dependent way, in several experiments 
\cite{afs,data0,data2,data3,data4,data5}.
Despite of large error bars, the present experimental scenario 
is consistent with the idea 
that $\sigma_{eff}$ is constant w.r.t. the center-of-mass energy 
of the collision.
In Ref. \cite{plb}
we have presented a predictive study of  $\sigma_{eff}$,
making use of the LF quark
model  approach to dPDFs developed in Ref. \cite{noi2}.
It was found that the order of magnitude of the measured
$\sigma_{eff}$ is correctly reproduced by the model
and, more interestingly, in the valence region, a clear
dependence is predicted on the longitudinal momentum fractions
of the proton carried by the two partons. If measured, this feature
could represent a first access to the observation
of 2-partons correlations in the proton. 

Beyond these intriguing results, already found
in the valence region, one should check if similar
possibilities survive at LHC kinematics, 
dominated by low-$x$ partons, at very high energy scales.
In this paper, using our model predictions, we plan therefore:
\begin{itemize}
\item[i)]
to test the validity of factorization assumptions,
which basically neglect at least part of the correlations
between the partons, often
used in dPDFs studies, at the scale of the model and after
evolution to experimental energy scales;
\item[ii)]
to test if correlations in longitudinal and transverse
momenta survive the evolution procedure;
\item[iii)]
to develop an extension of our approach to include, at the low energy
scale of the model, sea quarks and gluon degrees of freedom;
\item[iv)]
to study the importance of 2-body correlations between different kinds
of partons (valence quarks, sea quarks and gluons) at values
of longitudinal momenta and energy scales close to the
experimental ones, to establish the possibility to
observe them at the LHC.
\end{itemize}
The paper is structured as follows.
The first section is dedicated to present
a short summary of the formalism and the results obtained
in Ref. \cite{noi2}. The second section is dedicated to
compare our results,
where correlations
are naturally produced by the dynamics of the model, with
a few factorized forms of dPDFs. 
In the third section, we study how QCD evolution
to high momentum scales affects the results of the model.
In the following section we describe a strategy to introduce
sea quarks and gluons at the low momentum scale
of the model.
In section five we quantify, within our scheme,
how large are the correlation effects between
different kind of partons at very low
values of longitudinal momentum fractions and
very high energy scales. This is very important
to address measurable signatures of two-partons
correlations. We end by drawing some conclusions of our study.

\section{\label{sec:dPDFs-LF}Calculating Double Parton Distribution Functions}

Recently dPDFs have been explicitly calculated by us within a Light-Front (LF) 
approach \cite{noi2}. 
The method is fully covariant and 
is based on a fixed-number Light-Front $SU(6)$-symmetric Hamiltonian making 
use of an Hypercentral potential introduced in 
Ref. \cite{LF2} as a generalization of a 
non-relativistic constituent quark model proposed 
in Ref.
\cite{santop}. 
The approach is particularly suitable for the description of 
Deep Inelastic Scattering (DIS) processes which find their natural 
environment in a LF - description. The numerous applications to a 
large varieties of DIS observables like polarized 
\cite{LF2} and unpolarized 
\cite{PasquiniTrainiBoffi2002,Traini2012-2014,marco2} structure functions, 
spin and angular momentum distributions 
\cite{CanoFaccioliTraini2000,CanoFaccioliScopettaTraini2002}, 
helicity-independent and dependent GPDs 
\cite{BoffiPasquiniTraini2003-2004-2005}, demonstrate the reliability and 
flexibility of the approach.

\subsection{The Light-Front formulation}

Let us briefly summarize the main steps for the LF-evaluation of 
the dPDFs. In terms of the Light-Cone (LC) quantized fields 
$q_i$ for a quark of flavor $i$, helicity $\lambda$ in an unpolarized proton, 
the dPDFs in  momentum space, often called
"$_2GPDs$" in the literature \cite{blok_1,blok_2}, read 
(see, e.g., \cite{manohar_2,noi1})
\begin{eqnarray}
\label{eq:LF-dPDF} 
&&\dpdf{ij}^{\lambda_1,\lambda_2}(x_1,x_2,{\vec k}_\perp) = \nonumber \\
&& (-8 \pi P^{+}) \dfrac{1}{2} \underset{\lambda}\sum \int\! d 
\vec z_\perp \, e^{i \vec z_\perp
\cdot \vec k_\perp} \hat {\mathcal T}_i^1 \hat {\mathcal T}_j^2 \times 
\nonumber \\ 
& \times &
\int \left [ \underset{l}{\overset{3}\prod}  \dfrac{
d z_l^-}{4 \pi} \right]\,
e^{i x_1 P^+ z_1^-/2}\,e^{i x_2 P^+ z_2^-/2}\,e^{-i x_1 P^+ z_3^-/2} \times
\nonumber \\
& \times &
\langle \lambda, \vec P = \vec 0  \big|
\hat {\mathcal T}_i^1 \hat {\mathcal T}_j^2 \big|  
\vec P = \vec 0,\lambda \rangle~,
\end{eqnarray}
where
\bea
\hat {\mathcal T}_i^k & = & \hat{\mathcal{O}}_i^k 
\left( z_1^-\dfrac{\bar n}{2},z_3^- \dfrac{\bar n}{2}
+ \vec z_\perp \right) \equiv \hat {\mathcal T}_i^k(z,z')= \nonumber \\
& = &  \hat{\mathcal O}_i^k(z,z') = \bar q_i(z) \hat{O}(\lambda_k) q_i(z')\,,
\eea
and
\begin{eqnarray}
\hat O(\lambda_k) =
\dfrac{\bar n   \hskip-0.2cm /}{2}    
\dfrac{1 + \lambda_k \gamma_5}{2}~.
\end{eqnarray}
In the above equations, both the light-like four vector  $\bar n  = (1,0,0,-1)$ and 
the rest frame state of the nucleon with helicity
$\lambda$, $\big| \vec P = \vec 0,\lambda \rangle$,
have been introduced. 
The ``$\pm$'' components
of a four-vector $b$ are defined according to $b^\pm = b_0 \pm b_z$ and 
$x_i = {k_i^+}/{P^+}$ is the
fraction of the system momentum carried by the parton ``i'',
while the notation $\tilde b = (b^+, \vec b_\perp)$ is used
for light-cone vectors.
The LC free quark fields are defined as
\bea
q_i(\xi) &=& \underset{r}\sum \int \dfrac{d \tilde k}{2 (2 \pi)^3 
\sqrt{k^+}} \theta(k^+) e^{-i \xi^- k^+} a_{\tilde k,r}^i~ 
u_{LF}(\tilde k,r)~, \nonumber \\
\eea
where the operator
$a_{\tilde k,r}^i$ destroys a quark of flavor $i$, helicity $r$
and LC momentum  $\tilde k$. The spinors are indicated by 
$u_{LF}(\tilde k,r)$ 
(we adhere to the definitions and notations of Ref. \cite{LF0}).
The proton state $\big|  \vec P = \vec 0,\lambda \rangle$ 
can be expanded in its Fock components 
retaining only the first (valence) contribution 
(the short-hand notation $(\{\alpha_i\})$
is adopted, here and in the following, 
for $(\alpha_1, \alpha_2, \alpha_3)$,
where $\alpha_i = x_i, \vec k_{i\perp} ,\lambda_i^f, \tau_i $):
\begin{eqnarray}
\label{f-nstate}
&& |\vec 0, \lambda \rangle \simeq 
|\vec 0, \lambda^f, val \rangle = \nonumber \\
&=& \underset{\lambda_i^f \tau_i}\sum
\int\left[ \underset{i=1}{\overset{3}\prod}  \dfrac{d x_i}{\sqrt{x_i}}   
\right]
\delta \left( 1- \underset{i=1}{\overset{3}\sum}x_i \right) \nonumber \\
&\times & \left[
\underset{i=1}{\overset{3}\prod}  \dfrac{d \vec
k_{i\perp}}{2(2\pi)^3}  \right]2(2\pi)^3
\delta \left(  \underset{i=1}{\overset{3}\sum}\vec k_{i\perp} \right)
\nonumber \\
&\times&\Psi^{[f]}_{\lambda}
(\lbrace x_i, \vec k_{i\perp} ,\lambda_i^f, \tau_i \rbrace) 
\underset{i=1}{\overset{3}\prod} |\tilde k_i, \lambda_i^f,
\tau_i \rangle~,
\end{eqnarray}
in terms of the LF one-quark states of isospin $\tau_i$,
$|\tilde k_i, \lambda_i^f,
\tau_i \rangle $.

At variance the same proton state can be described in terms of canonical, Instant-Form (IF),
one-quark states $|\vec k_i, \lambda_i^c, \tau_i \rangle$,
\begin{eqnarray}
&& |\vec 0, \lambda \rangle \simeq
|\vec 0, \lambda^c, val \rangle = \nonumber \\ 
&=& \underset{\lambda_i^c \tau_i}\sum
\int\left[ \underset{i=1}{\overset{3}\prod}  d \vec k_i 
\right]\delta \left(  \underset{i=1}{\overset{3}\sum}\vec k_{i} \right) \nonumber \\
&\times& \Psi^{[c]}_{\lambda}(\lbrace  \vec k_{i} ,\lambda_i^c,
\tau_i \rbrace) \underset{i=1}{\overset{3}\prod} |\vec k_i, \lambda_i^c,
\tau_i \rangle~.
\label{I-nstate}
\end{eqnarray}
The two descriptions are related by Melosh rotations \cite{Melosh-rotation}. 

Following our previous developments 
(e.g. Refs. \cite{LF2,BoffiPasquiniTraini2003-2004-2005})
the considerations made for {\it free} canonical states can be 
generalized to interacting quarks 
in a proton, by means of a suitable representation of the 
Poincar\'e operators, namely the Bakamjian-Thomas construction 
\cite{BakamjinThomas1953}. 
 The extension to interacting systems requires, in fact, a dynamical 
representation of the
Poincar\'e group. One way to achieve this result 
is to add an interaction V to the 
free mass operator $M_0$ to obtain the mass operator 
$M = M_0 + V$.  
Since
the LF boosts we use are interaction independent,
all the other definitions 
remain unaffected.
All required commutation relations 
are satisfied if the
mass operator commutes with the total spin  and with the kinematic generators. 
In practice,
the conditions are realized if:
\begin{itemize}
\item[i)] $V$ is independent on the total momentum $\tilde {\bf P}$;
\item[ii)] $V$ is invariant under ordinary rotations. 
\end{itemize}

{\it Summarizing}: in the LF formulation of the quark dynamics, the intrinsic 
momenta of the quarks (${k}_i$) can be obtained from the corresponding momenta 
(${p}_i$) in a generic frame through a LF boost
($K_i = u(P) \cdot p_i$, $P \equiv \sum_{i_1}^3 p_i$) such that the 
Wigner rotations reduce to the identity. 
The spin and spatial degrees of freedom are described by the wave function
\be
\Psi = {1 \over \sqrt{P^+}} \delta 
\left( \tilde P - \tilde p\right) \chi(\{{\bf k}_i,\mu_i\})\,,
\ee
where $\mu_i$ refers to the eigenvalue of the LF spin, 
so that the spin part of the wave function is transformed by 
the tensor product of three independent Melosh rotations, 
namely ${\cal R}^\dag = \prod_{\otimes i=1}^3 R^\dag({\bf k}_i,m_i)$. 
The internal wave function is an eigenstate of the baryon mass operator 
$M = M_0 + V$, with $M_0 = \sum_{i=1}^3 \sqrt{{\vec k_i^2} + m_i^2}$ and 
where the interaction term $V$ must be independent on the total momentum 
$\tilde P$ and invariant under rotations. The nucleon state is 
then characterized by isospin (and its third component), parity, 
Light-Front (non-interacting) angular momentum operators with well 
defined projection along the quantization axis.

The relativistic mass equation chosen is built according to 
such a dynamical construction \cite{LF2}. 
Thanks to the correct kinematical conditions on the longitudinal 
momentum fraction carried by the quark as described by the 
LF-approach, dPDFs vanish in the forbidden kinematical region, $x_1+x_2 > 1$.
(see Ref. \cite{noi2} for further details).

\subsection{Light-Front results at the low scale of the model}

Reducing Eq. (\ref{eq:LF-dPDF}) to the first (valence) Fock components 
and specializing the result to the $u$ quarks as an example, 
one has ($\lambda_1, \lambda_2 \equiv \uparrow(\downarrow) $)
\begin{eqnarray}
&&  u_V^{\uparrow(\downarrow)}u_V^{\uparrow(\downarrow)} (x_1,x_2, k_\perp) 
= \nonumber \\
&&  2(\sqrt{3})^3  \int  
 d \vec k_{1\perp} d \vec k_{2\perp}  ~ {1 \over j}
\dfrac{E_1 E_2 E_3}{k_1^+ x_1 x_2 
(1-x_1-x_2) } \nonumber \\
 &\times&  
\langle \tilde P_1^{\uparrow(\downarrow)}  \rangle 
 \langle \tilde P_2^{\uparrow(\downarrow)}  \rangle \,
 \psi^* 
\left(\vec k_1 +
\dfrac{\vec k_\perp}{2}, \vec k_2 -
\dfrac{\vec k_\perp}{2}, -\vec k_1 -\vec k_2 \right)  \nonumber \\
 &\times&  
\psi \left(\vec k_1 -
\dfrac{\vec k_\perp}{2}, \vec k_2 +
\dfrac{\vec k_\perp}{2}, -\vec k_1 -\vec k_2 \right)~,
\label{eq:uVuVpol}
\end{eqnarray}
with
\begin{eqnarray}
k_1^+ & = & \left\{x_1 \left[ m^2 \left(1+\dfrac{x_1}{x_2}+
\dfrac{x_1}{1-x_1-x_2} \right) + \right. \right.\nonumber \\
&+& \left. \left. k_{1 \perp}^2 + \dfrac{x_1}{x_2}
k_{2\perp}^2 +\dfrac{x_1}{1-x_1-x_2}  k_{3\perp}^2\right] \right\}^{1/2}~, 
\nonumber 
\nonumber \\
k_2^+ & = & \dfrac{x_2}{x_1} k_1^+, ~~ k_3^+ = \dfrac{1-x_1-x_2}{x_1} k_1^+, 
\nonumber \\
k_{iz} & =&  -\dfrac{m^2+k_{i\perp}^2}{2 k_i^+} +\dfrac{k_i^+}{2}~, 
\end{eqnarray}
\begin{eqnarray}
E_i & = & \sqrt{m^2+k_{iz}^2+\vec k_{i\perp}^2}~, \nonumber \\
j & = & \left|
\dfrac{m^2+ k_{1 \perp}^2}{2 k_1^{+2}} +\dfrac{m^2+ k_{2 \perp}^2}{2
\dfrac{x_2}{x_1} k_1^{+2}}+\dfrac{m^2+ k_{3 \perp}^2}{2
\dfrac{1-x_1-x_2}{x_1} k_1^{+2}} + \dfrac{1}{2 x_1} \right|~. \nonumber 
\end{eqnarray}
The spin projector values are determined by the  Melosh rotations $\hat D_i$:

\begin{eqnarray}
\langle \tilde{P}^{\uparrow(\downarrow)}_i \rangle & = & \langle \hat D_i 
\hat P^{\uparrow(\downarrow)}(i)
\hat D_i^\dagger\rangle = \nonumber \\
& = & \langle \hat D_i 
\left(  \dfrac{1 \pm 
\sigma_z(i)}{2}  \right) 
\hat D_i^\dagger\rangle \,,
\end{eqnarray}
to be calculated using the canonical spin-isospin states corresponding 
to the SU(6) symmetric matrix elements.

In particular the combinations
\begin{eqnarray}
u_Vu_V(x_1,x_2, k_\perp,\mu_0^2) 
& = &  
u^{\uparrow}_Vu^{\uparrow}_V(x_1,x_2, 
k_\perp,\mu_0^2) 
\nonumber \\ 
& + &  
u^{\downarrow}_V u^{\downarrow}_V(x_1,x_2, k_\perp,\mu_0^2) \, 
\nonumber \\ 
& + &  
{{u^{\uparrow}_V u^{\downarrow}_V(x_1,x_2, k_\perp,\mu_0^2)}} 
\nonumber \\ 
& + &  
{{u^{\downarrow}_V u^{\uparrow}_V(x_1,x_2, k_\perp,\mu_0^2)}} \,, 
\label{eq:uVuVunpol}
\end{eqnarray} 
will describe two unpolarized $u$-valence quarks, and
\begin{eqnarray}
 \label{eq:pol}
&& \Delta u_V \Delta u_V(x_1,x_2, k_\perp,\mu_0^2)  =  \nonumber \\ 
&& u^{\uparrow}_V u^{\uparrow}_V(x_1,x_2, k_\perp,\mu_0^2)
+u^{\downarrow}_V u^{\downarrow}_V(x_1,x_2, k_\perp,\mu_0^2)+ \nonumber \\
&-&  
u^{\uparrow}_V u^{\downarrow}_V(x_1,x_2, k_\perp,\mu_0^2) -
u^{\downarrow}_V u^{\uparrow}_V(x_1,x_2, k_\perp,\mu_0^2)\,, 
\end{eqnarray}
two (longitudinally) polarized $u$-valence quarks. These two distributions 
only contribute to the total cross section of events involving unpolarized 
proton targets.

In Fig. \ref{fig:xxuVuV-kperp} the numerical results of the 
Eqs. 
(\ref{eq:uVuVpol})-(\ref{eq:uVuVunpol})
for two unpolarized 
$u$-valence quarks. The dPDFs vanish in the region $x_1+x_2 > 1$ and the 
correlations in $x_1,x_2$ are dictated by the LF-quark dynamics, 
which governs also the
{{dependence}} in $k_\perp$, 
clearly seen in the right panel of the same figure.
Since the Fock expansion of the proton state has been restricted to the 
three valence quarks (cfr. Eq. (\ref{f-nstate})), it is natural that the 
full momentum is carried by those quarks and the resulting appropriate 
energy scale remains quite low (the so-called
hadronic scale, $\mu_0^2 \approx 0.1\, {\rm GeV}^2$  \cite{trvlc}) 
as indicated by analogous $Leading-Order$ calculations 
(see, e.g., Refs. \cite{LF2,Traini2012-2014,marco2}). 
That scale is  clearly indicated in the resulting expressions 
(\ref{eq:uVuVunpol}) and 
(\ref{eq:pol}) and in both panels of Fig. \ref{fig:xxuVuV-kperp}.

\subsection{\label{sec:factorization} 
Factorization and approximations at the low scale of the model}

In the present Section we will compare our approach to
a number of strategies used in the 
literature to calculate dPDFs, strategies  that we call, in general,
"factorization schemes". 
Differences and analogies will help 
in understanding the role of correlations and their dependence on the 
evolution scale.

\subsubsection{\label{sec:pheno-factorization}Phenomenological factorizations}

As a first illustrative example we can restrict the discussion to the  
approach proposed 
by Diehl, Kasements and Keane 
in Ref. 
\cite{dkk}, which has motivated {{in part 
the discussion presented in this section}}.

\begin{figure}[tbp]
\centering\includegraphics[width=8.2cm,clip=true,angle=0]
{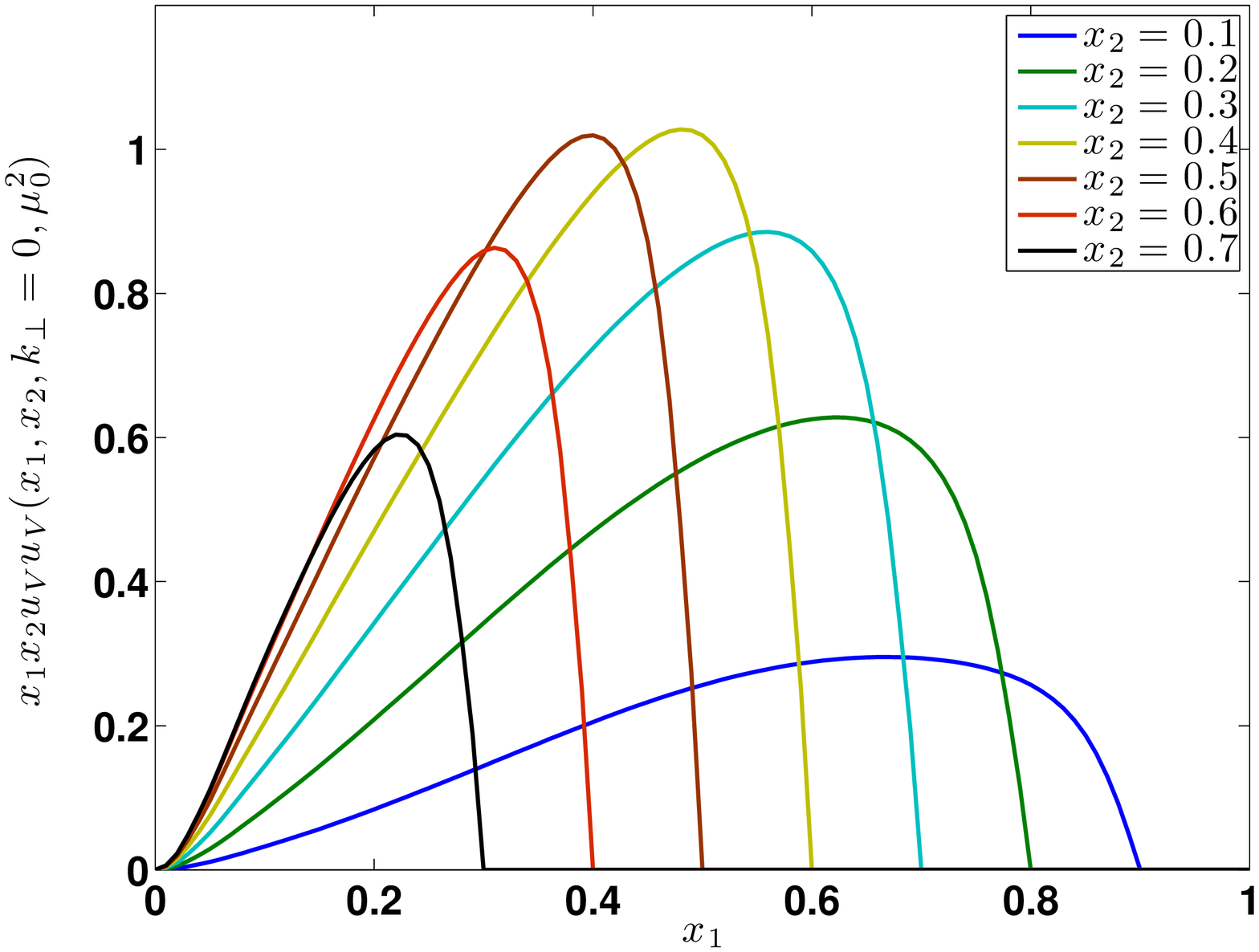}
\centering\includegraphics[width=8.2cm,clip=true,angle=0]
{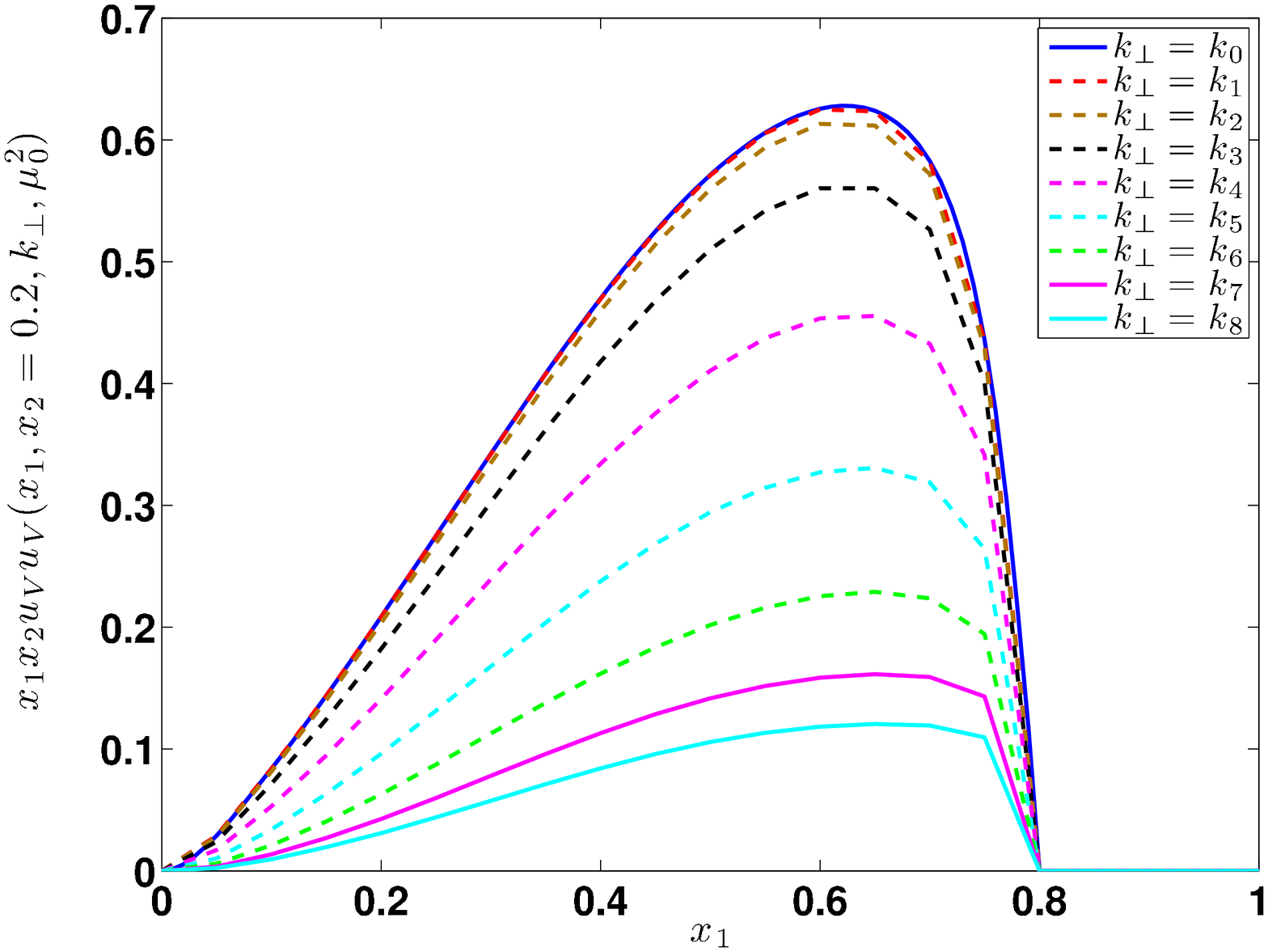}
\caption{
\small  {Left panel:}  
$x_1x_2u_Vu_V(x_1,x_2,k_\perp=0,\mu_0^2)$ as function of $x_1$ at fixed values 
of $x_2$.
{Right panel:} 
$x_1x_2u_Vu_V(x_1,x_2=0.2,k_\perp,\mu_0^2)$ as function of 
$x_1$ at various values of $k_\perp$ (
$(k_0,k_1,...,k_8) 
\simeq 
(0, 0.03, 0.14, 0.32, 0.57 ,0.85 ,1.15, 1.43, 1.68)
\,\,{\rm GeV}$, which are Gaussian points between 0 and 2 GeV).}

\label{fig:xxuVuV-kperp}
\vspace{-1.0em}
\end{figure}

In fact the interest of those authors is on the influence of the evolution 
scale on correlation effects, precisely one of the goals of the present work. 
The model they propose refers to the description of the dPDFs at the starting 
scale, where they assume independent partons. In that case, in fact, the 
dPDFs in coordinate space
can be simply written
as a convolution 
of $f_{a,b}(x, \vec b)$ functions, which are 
impact parameter dependent generalized parton 
distributions
(see, e.g, Ref. \cite{diehl_1}):
\be
F_{ab}(x_1,x_2, \vec y) = \int d^2 \vec b\,f_a(x_1, \vec  b + \vec y) 
f_b(x_2, \vec b)~, 
\label{13}
\ee
with $a,b$ denoting parton species. This idea has been firstly presented in 
Refs. \cite{blok_1,blok_2}.
The authors of Ref. \cite{dkk} assume a Gaussian $\vec b$ dependence with an 
$x$-dependent width, 
namely
\be
f_a(x,\vec b) = f_a(x) {1 \over 4 \pi h_a(x)} 
\exp{\left[-{\vec b^2 \over 4 h_a(x)}\right]}~, 
\label{eq:ansatz}
\ee
where
$f_a(x)$ denotes the usual parton densities (taken from the LO set of the MSTW
2008 analysis \cite{MSTW2008}), while Eq. (\ref{eq:ansatz}) is assumed to be 
valid at the starting scale $Q^2_0 = 2$ GeV$^2$.  Diehl {\it et al} stress 
that the approach is tailored 
for the region  $x_1,x_2 < 0.1$  and its
parameters are specified for gluons and for the sum, $q^+ = q + \bar q$, and 
difference, $q^- = q -  \bar q$, of quark and antiquark distributions.
{{
The expressions for $h_a(x)$ are found in Ref.
\cite{dkk} and not reported here;
the  parameters which are necessary to define $h_a(x)$
are fixed so that the resulting parton densities 
are in tentative agreement with phenomenology.
}}

The final expression for the unpolarized dPDFs Eq. (\ref{13}) reads:
\bea
F_{ab}(x_1, x_2, \vec y,Q_0^2) & = &f_a(x_1,Q_0^2) f_b(x_2,Q_0^2) 
{ 1 \over  4 \pi h_{ab}(x_1, x_2)} \times \nonumber \\
& \times & \exp\left[{-{{\vec y}^2 \over  4 h_{ab}(x_1, x_2)}}\right] 
\;\;\;\;\;\;\;\; \;\;\;\; \;\;\label{eq:Fab-y}
\eea
and, as a consequence, one has, for the Fourier transform,
\bea
F_{ab}(x_1,x_2,k_\perp,Q_0^2) & = & f_a(x_1,Q_0^2) f_b(x_2,Q_0^2) 
{h_{ab}(x_1,x_2) \over \pi} \times \nonumber \\
& \times & \exp\left[-h_{ab}(x_1, x_2)  k_\perp^2\right] .
\label{eq:Fab-k}
\eea
The term
\bea
h_{ab}(x_1, x_2) & = & h_a(x_1) + h_b(x_2) = \nonumber \\
& = &  \alpha '_a \ln {1 \over x_1} + \alpha '_b \ln {1 \over x_2}
+ B_a + B_b\,  \;\;\; \label{eq:hab-form}
\eea
is assumed, at the same scale, to introduce correlations between 
$x_1$ and $x_2$, in fact Eq. (\ref{eq:hab-form}) does not 
factorize into separate contributions from each of the two partons, 
$a$ and $b$ (the values of the parameters in Eq. (\ref{eq:hab-form}) 
can be found in Ref. \cite{dkk}).
The combinations $u^-$ and $u^+$ are taken as representatives of the quark 
sector. 

\subsubsection{Factorization by means of Generalized Parton Distributions}

In Ref. \cite{diehl_1}, a systematic study of relations 
between single parton and double parton
distributions has been performed. To reduce $F_{ab}$ to single-particle 
distributions the authors find it more convenient 
to work in the transverse-momentum $\vec k_\perp$ space, rather 
than transverse distance $\vec y$ representation and the result reads:
\bea
&& F_{q q}(x_1,x_2,\vec k_\perp,Q^2)  \approx \nonumber \\
&& \approx H^q(x_1,\xi=0,-k_\perp^2,Q^2)  H^q(x_2,\xi=0,-k_\perp^2,Q^2) 
+ \nonumber \\
&& + \,{k_\perp^2 \over 4 M_p^2}\,E^q(x_1,\xi=0,-k_\perp^2,Q^2)  
E^q(x_2,\xi=0,-k_\perp^2,Q^2)~,\nonumber \\ 
\label{eq:HH}
\eea
where $M_p$ is the proton mass and $H^q(x,\xi,t)$ and $E^q(x,\xi,t)$ 
are   Generalized Parton Distributions (GPDs) 
(see, e.g., 
\cite{diehlgpd} 
and references therein). $H^q$ generalize the unpolarized quark 
densities $q(x)$ while  $E^q$ is related to unpolarized quarks 
in a transversely polarized proton.  The first term in Eq. (\ref{eq:HH}) 
depends on $H^q$ only and it corresponds to the simplest approximation 
of the two-parton distribution as a product of single-parton distributions 
(cfr. Eq. (\ref{eq:Fab-k})).

In Ref. \cite{diehl_1} one can read: 
{\it "Although the relation between multiparton distributions and GPDs 
is an approximation whose accuracy is not easy to
estimate (and although our current knowledge of GPDs is far less advanced 
than that of ordinary parton densities) this
relation provides opportunities to obtain information about multiple 
interactions that is hard to get by other means"}.

The question on the accuracy is particularly relevant in view of possible 
experimental studies of multi-parton effects, and the LF-approach we are 
presenting can shed some light on the approximation  (\ref{eq:HH}), 
including 
the role played by the $E$ correction term. Since GPDs have been studied, 
precisely within the same LF-approach, by 
Pasquini, Boffi and Traini \cite{BoffiPasquiniTraini2003-2004-2005} some 
years ago, one can check directly the 
accuracy of Eq. (\ref{eq:HH})
\footnote{We are indebted to Markus Diehl who brought our attention to 
Eq. (4.48) of Ref. \cite{diehl_1}, corresponding to 
Eq. (\ref{eq:HH}) and for his useful suggestions.}. 
Because of the natural normalization of the expression 
Eq. (\ref{eq:uVuVunpol}):
\begin{equation}
\int d x_1 d x_2 \, u_Vu_V(x_1,x_2,k_\perp=0,\mu_0^2) = 2\,,
\end{equation}
and the normalization of the $H^{u_V}$ GPDs
\begin{equation}
\int d x \, H^{u_V}(x,\xi=0,-k_\perp^2) = 2\,,
\end{equation}

\begin{figure}[tbp]
\centering\includegraphics[width=8.2cm,clip=true,angle=0]
{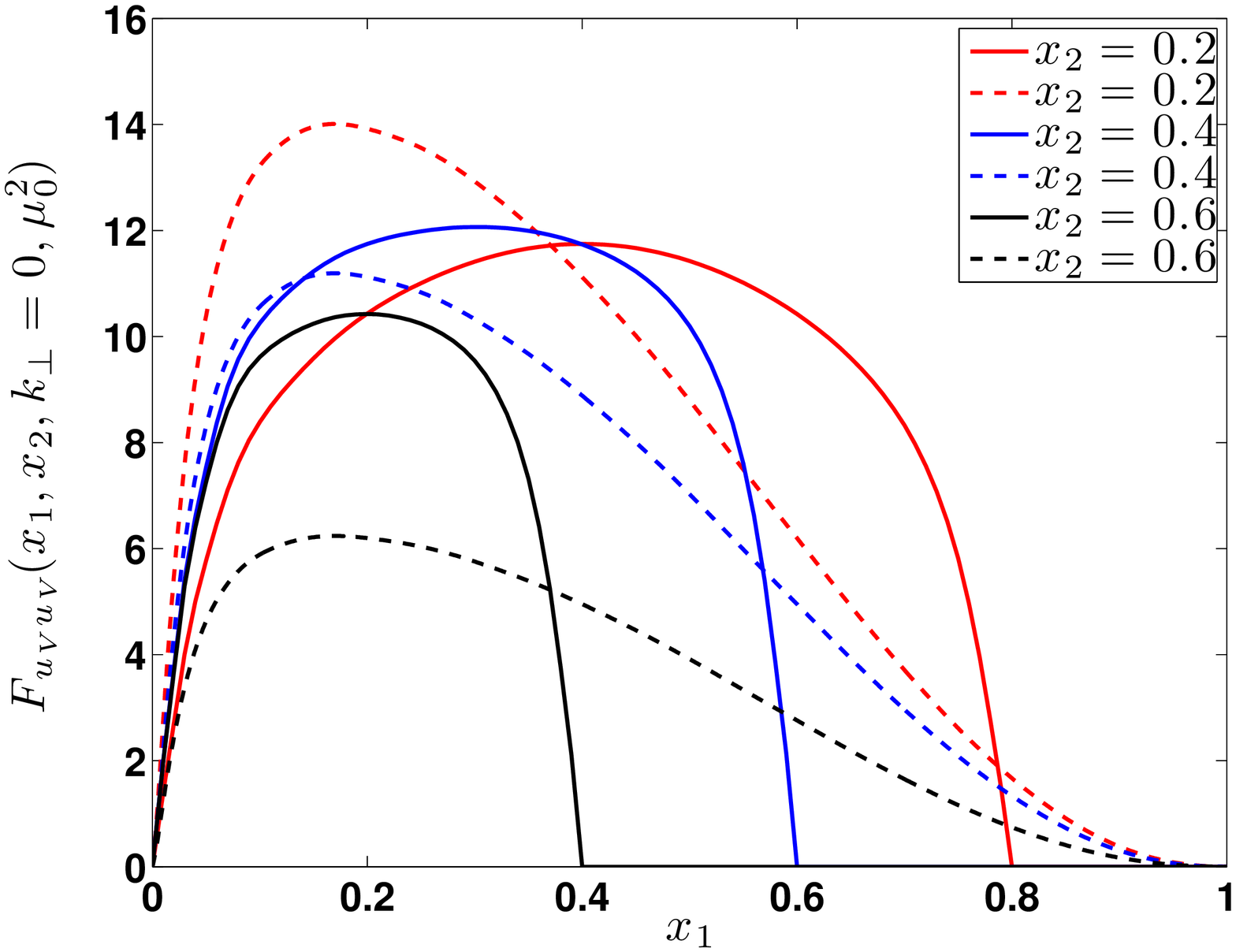}
\centering\includegraphics[width=8.2cm,clip=true,angle=0]
{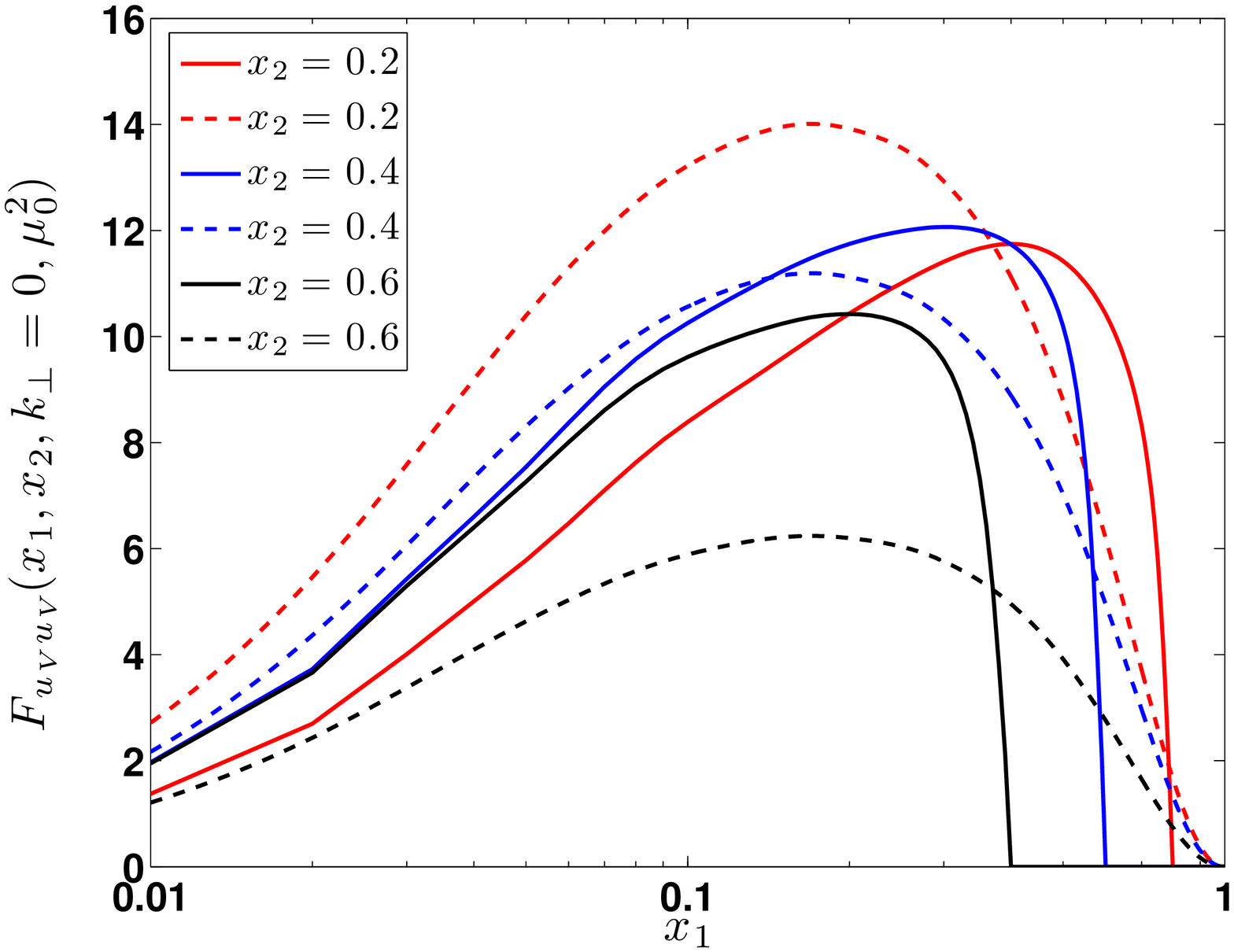}
\caption{\small  {Left panel:}  
$F_{u_Vu_V}(x_1,x_2,k_\perp=0,\mu_0^2)$ as function of $x_1$ at fixed values of 
$x_2=0.2,0.4,0.6$. The continuous lines represent the results obtained within 
the LF-approach ($F_{u_Vu_V} = 2 \times u_Vu_V$ of Eqs. (\ref{eq:uVuVpol}), 
(\ref{eq:uVuVunpol})), the dot-dashed lines the results of the approximation  
(\ref{eq:HH}). See text for discussion.
{Right panel:} As in the left panel, in logarithmic $x$-scale to emphasized 
the 
low-$x$ behavior. }
\label{fig:uVuV-k0_mu0_Hyp}
\vspace{-1.0em}
\end{figure}
%
%
\begin{figure}[tbp]
\centering\includegraphics[width=8.2cm,clip=true,angle=0]
{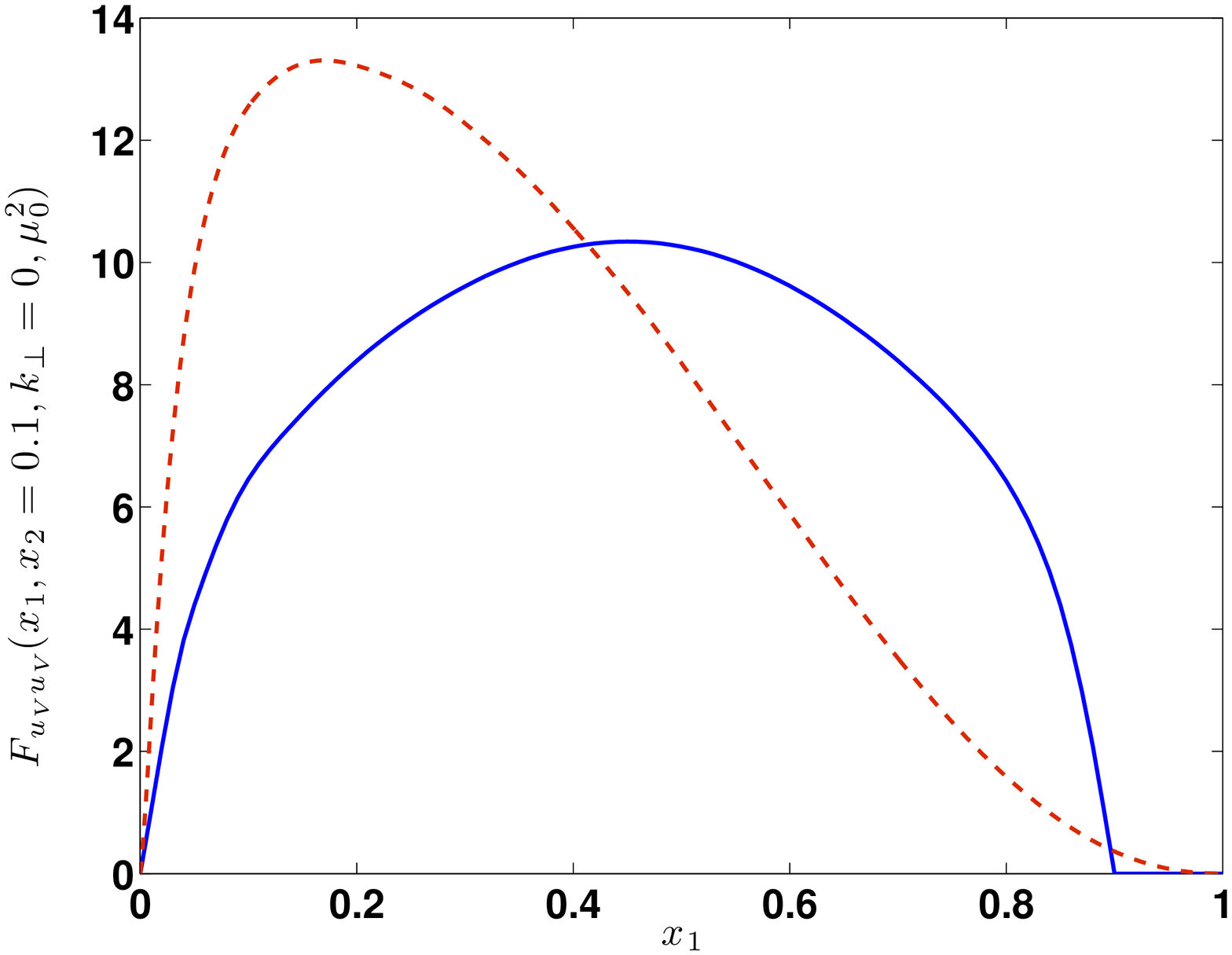}
\centering\includegraphics[width=8.2cm,clip=true,angle=0]
{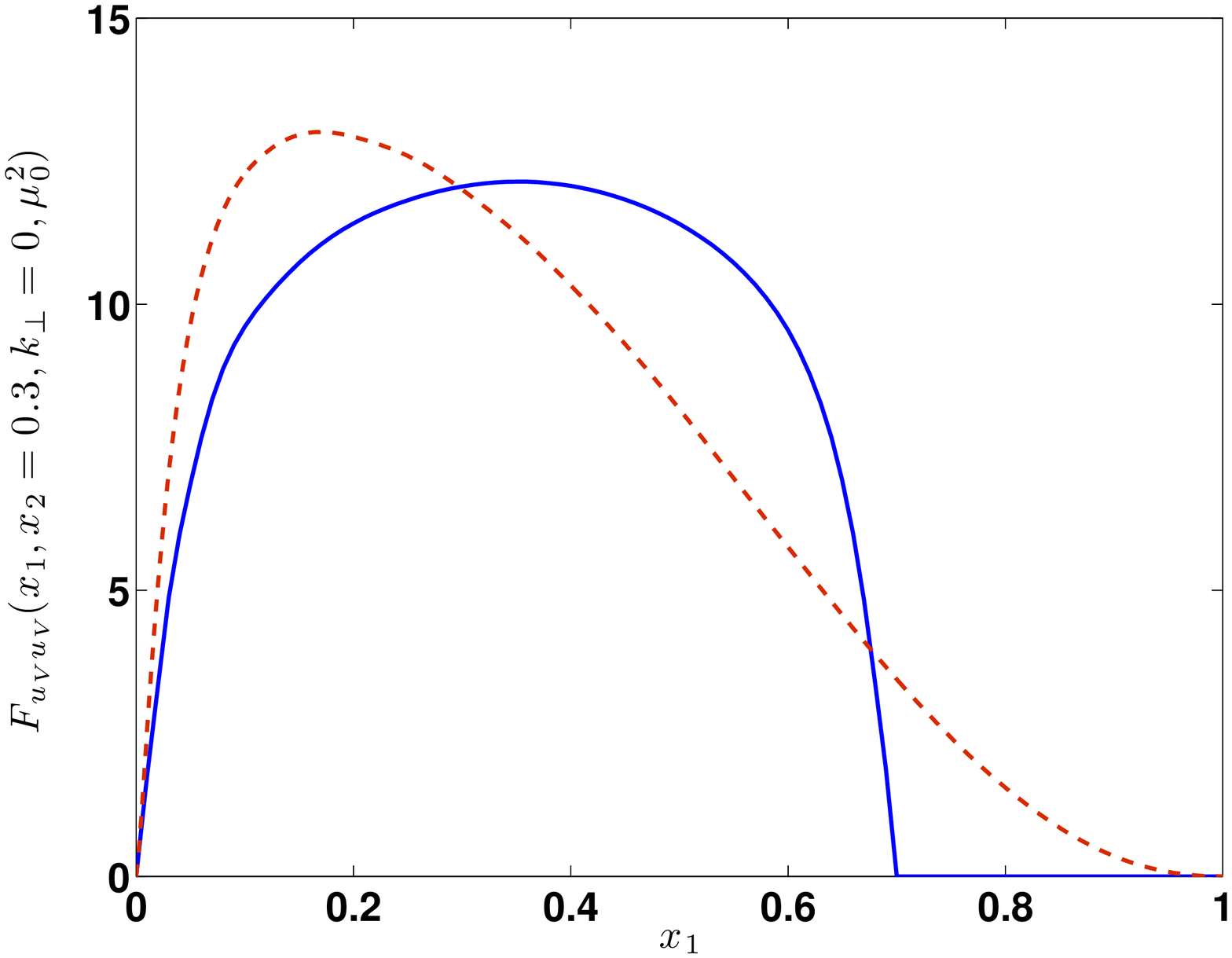}
\caption{\small  {Left panel:} 
$F_{u_Vu_V}(x_1,x_2,k_\perp=0,\mu_0^2)$ as function $x_1$ at fixed $x_2=0.1$ and 
$k_\perp=0$. 
{Right panel:} As in the left panel, at fixed $x_2=0.3$. The continuous line 
($F_{u_Vu_V} = 2 \times u_Vu_V$ of Eqs. (\ref{eq:uVuVpol}), 
(\ref{eq:uVuVunpol})), crosses the dashed line (approximation  (\ref{eq:HH})) 
for $x_1 \approx 0.3$. See text for discussion.}
\label{fig:Fuu2_HEu_x10103_k0}
\vspace{-1.0em}
\end{figure}


\begin{figure}[tbp]
\centering\includegraphics[width=8.2cm,clip=true,angle=0]
{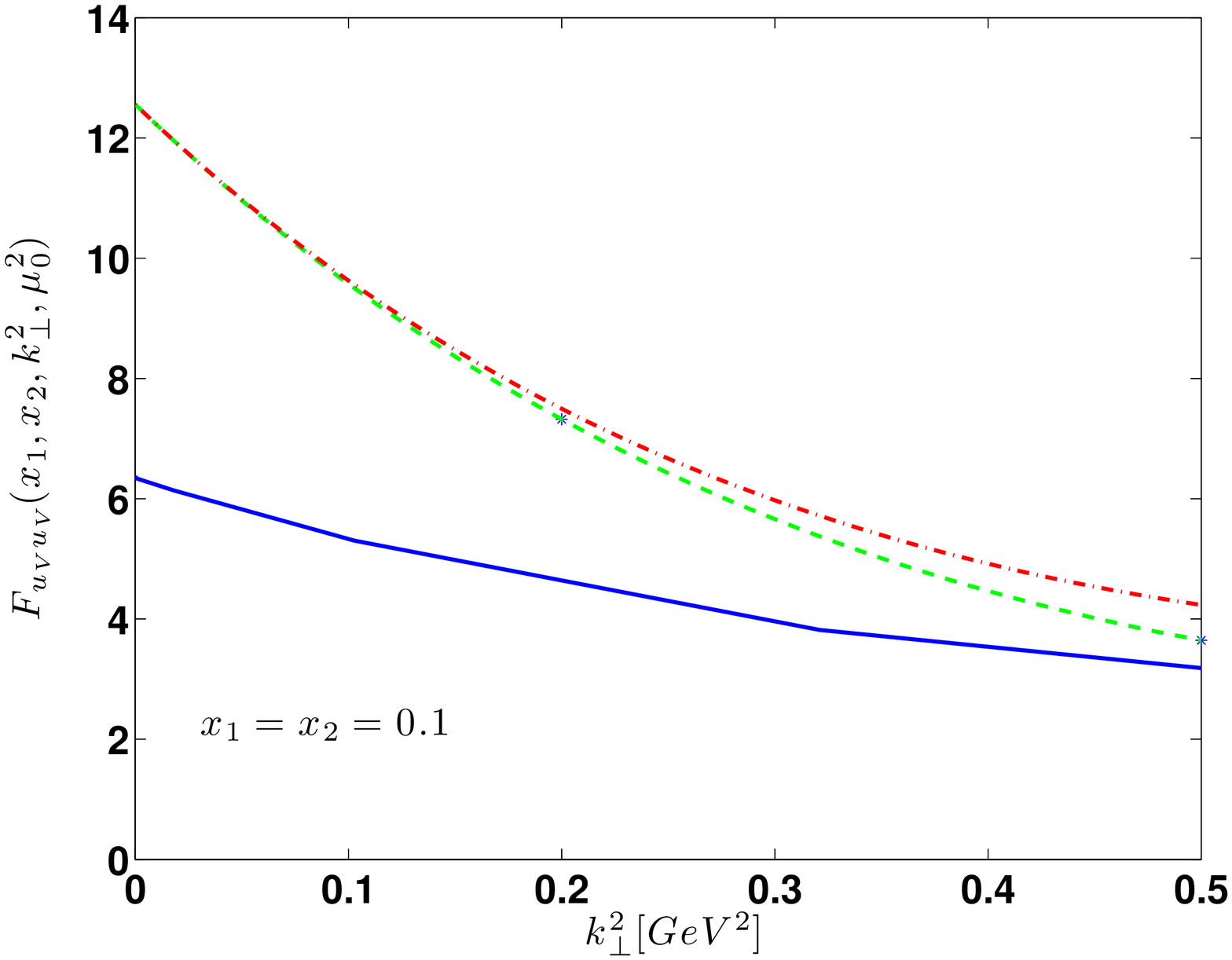}
\vspace{-1.0em}
\centering\includegraphics[width=8.2cm,clip=true,angle=0]
{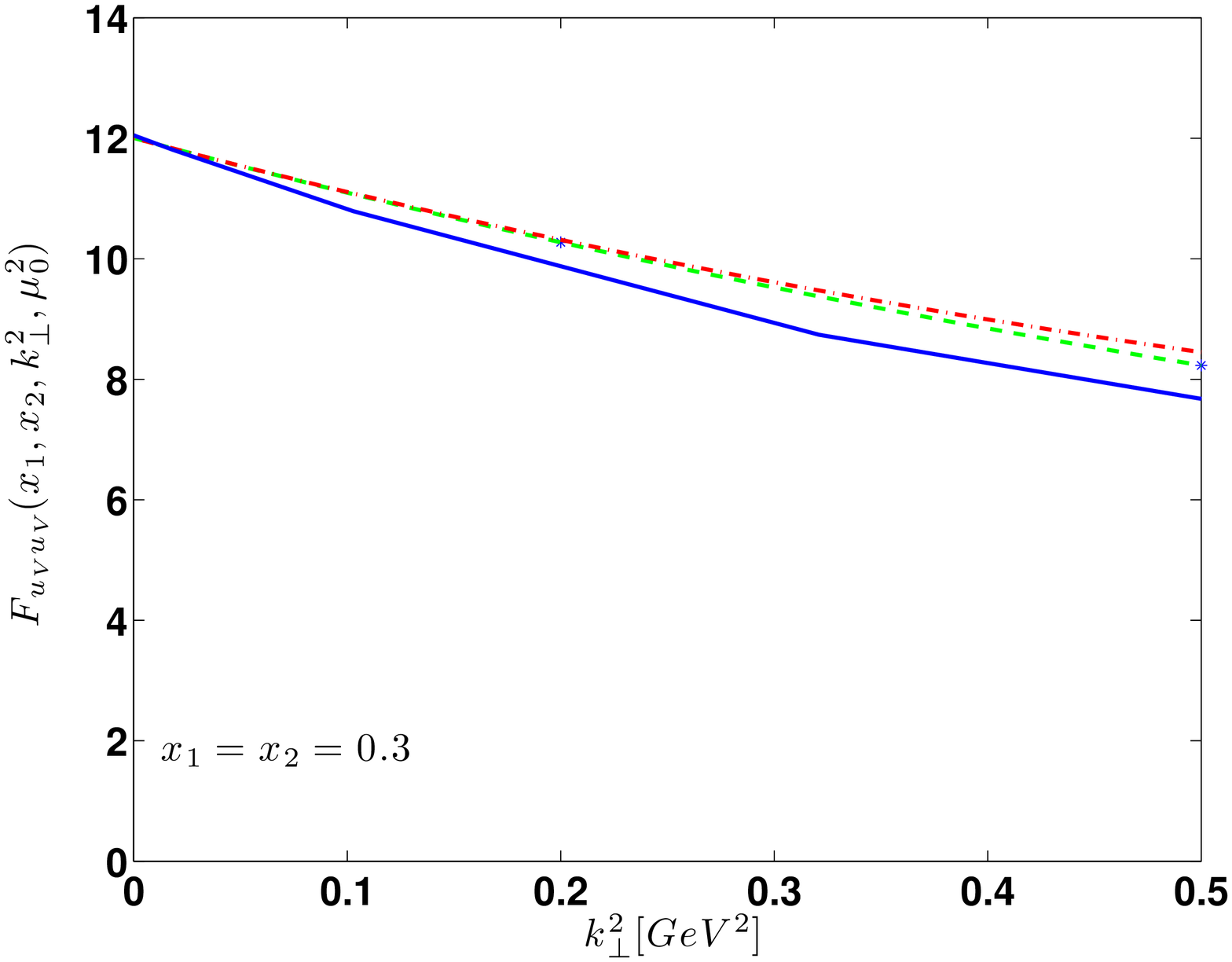}
\caption{\small  {Left panel:}  
$F_{u_Vu_V}(x_1,x_2,k_\perp,\mu_0^2)$ as function of $k_\perp$ at 
fixed $x_1=x_2=0.1$. The continuous line represents the results obtained 
within the LF-approach ($F_{u_Vu_V} = 2 \times u_Vu_V$ of 
Eqs. (\ref{eq:uVuVpol}), (\ref{eq:uVuVunpol})), the dashed lines the results 
of the approximation  (\ref{eq:HH}), the dot-dashed lines neglect the 
corrections due to the $k_\perp^2$-term in Eq. (\ref{eq:HH}). 
See text for discussion.
{Right panel:} As in the left panel, at fixed $x_1=x_2=0.3$.}
\label{fig:Fuu2_HEu0103}
\vspace{-1.0em}
\end{figure}
\noindent the comparison holds for 
\bea
&& F_{u_Vu_V}(x_1,x_2,k_\perp,\mu_0^2)  =  
2 \times u_Vu_V(x_1,x_2,k_\perp,\mu_0^2) \approx 
\nonumber \\ && 
\approx H^{u_V}(x_1,\xi=0,-k_\perp^2,\mu_0^2) 
 H^{u_V}(x_2,\xi=0,-k_\perp^2,\mu_0^2) + \nonumber \\
&& + \,{k_\perp^2 \over 4 M_p^2}\,E^{u_V}(x_1,\xi=0,-k_\perp^2,\mu_0^2)  
\nonumber \\ && 
E^{u_V}(x_2,\xi=0,-k_\perp^2,\mu_0^2)\,.\nonumber \\
\label{eq:HHuVuV}
\eea
{ {Fig. \ref{fig:uVuV-k0_mu0_Hyp} shows a first comparison at $k_\perp =0$, 
where the correction due to the presence of $E^q$ contributions vanishes.
A clear conclusion emerges:}} the 
approximations  (\ref{eq:HH}) or  (\ref{eq:HHuVuV}) 
can have some validity in the restricted regions $x_1+x_2 < 1$, 
the range where the dPDFs do not vanish. For $x_1+x_2 > 1$ the dPDFs must 
vanish while the single parton responses $H$ and $E$ do not.  
\\


{{ A detailed comparison is shown in 
Fig. \ref{fig:Fuu2_HEu_x10103_k0} and} }  
Fig. \ref{fig:Fuu2_HEu0103} for two specific values of $x_2$, 
namely $x_2 = 0.1$ and $x_2=0.3$. In these two cases the comparison is not 
restricted  to $k_\perp=0$ only (see Fig. \ref{fig:Fuu2_HEu_x10103_k0}), 
but it extends to the kinematical region up to $k_\perp^2 = 0.5$ GeV$^2$ 
(see Fig. \ref{fig:Fuu2_HEu0103}). 
Once again no systematic agreement is 
found. 
The only weak improvement, 
for $k_\perp > 0$, is due to the 
$E^{u_V}$ dependent 
correction term. 

\section{\label{sec:correlations}Scale impact on correlations}


\vskip3cm
\begin{figure}
 \includegraphics{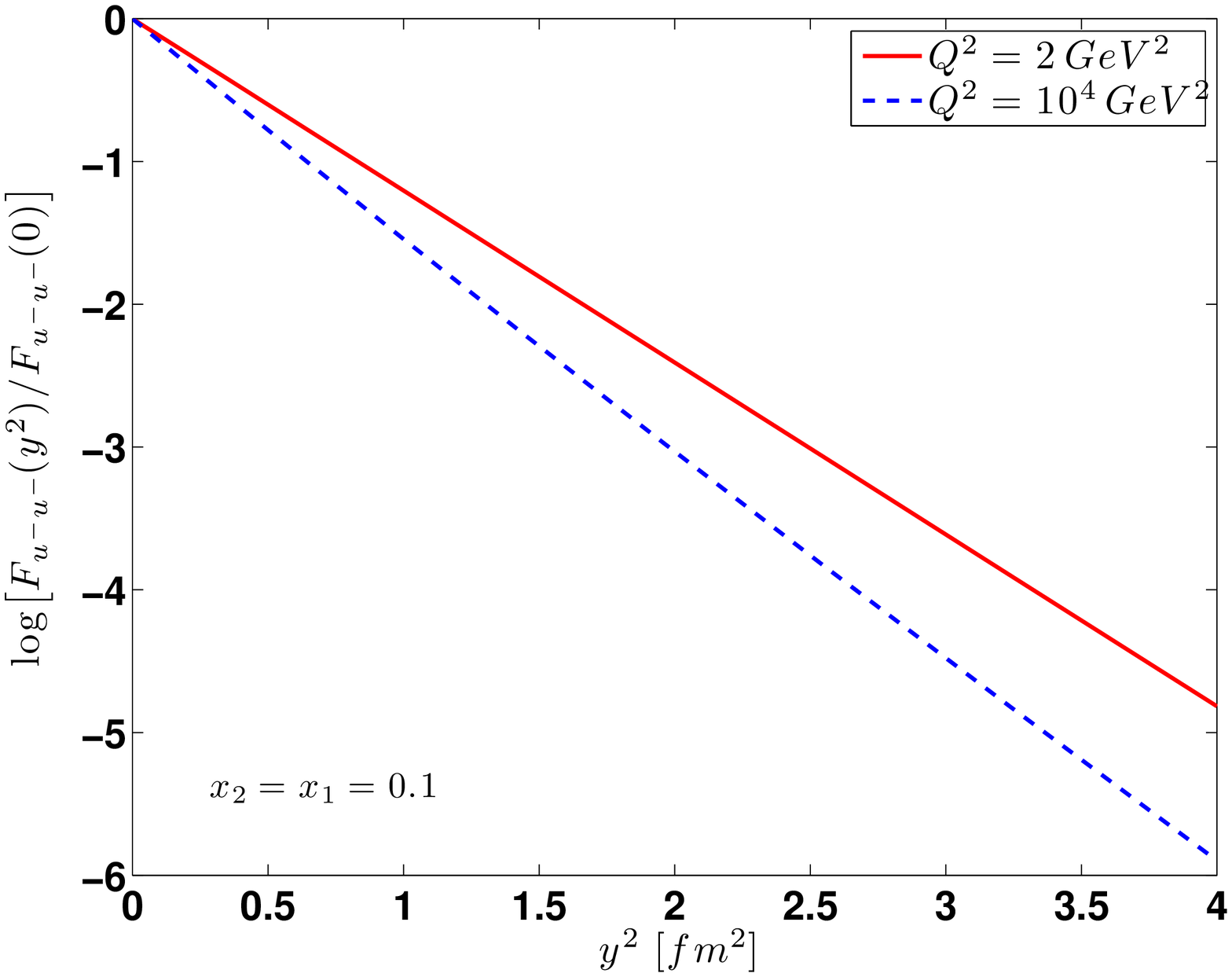}
\vskip 2cm
\includegraphics{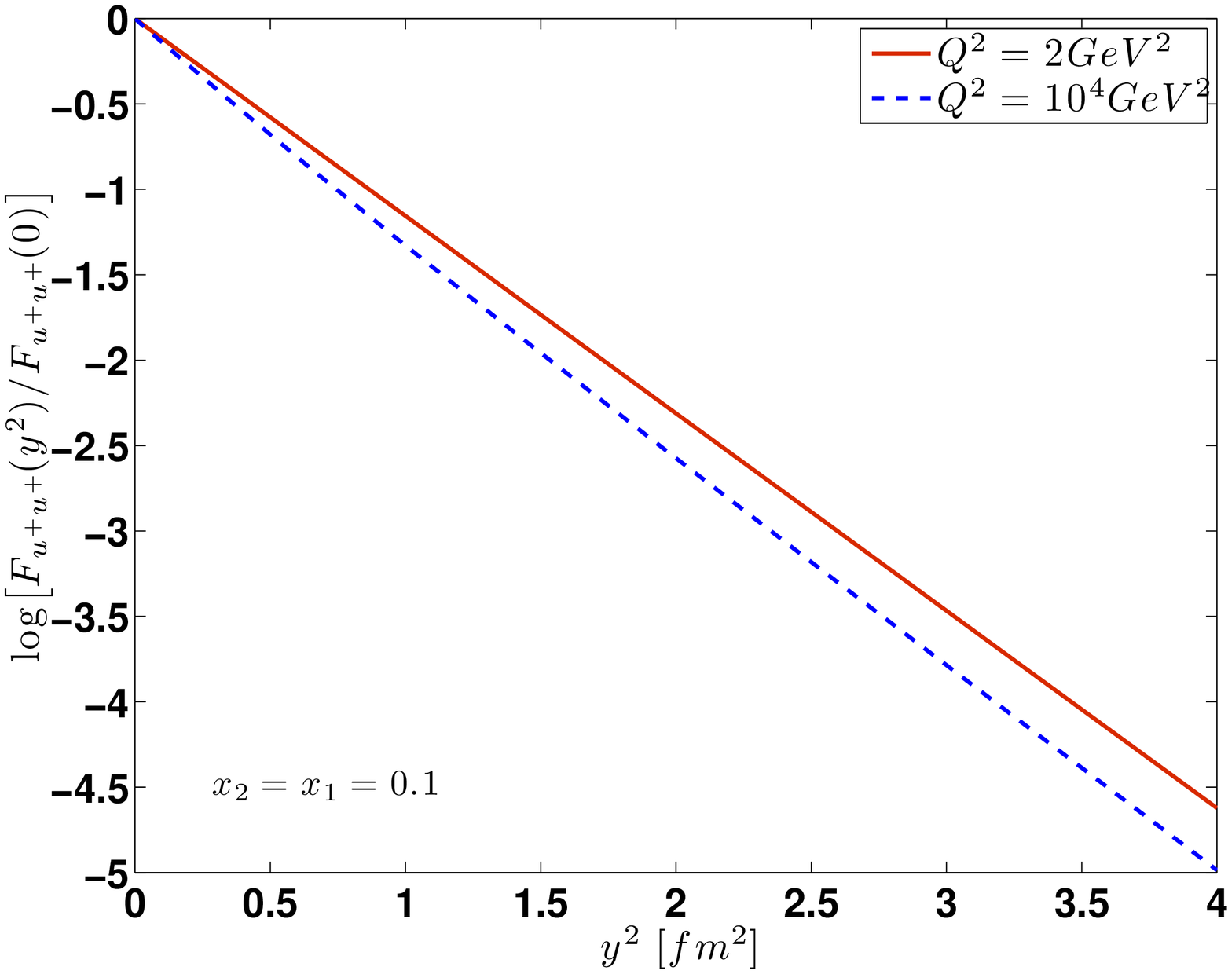} 
\vskip 2cm
 \includegraphics{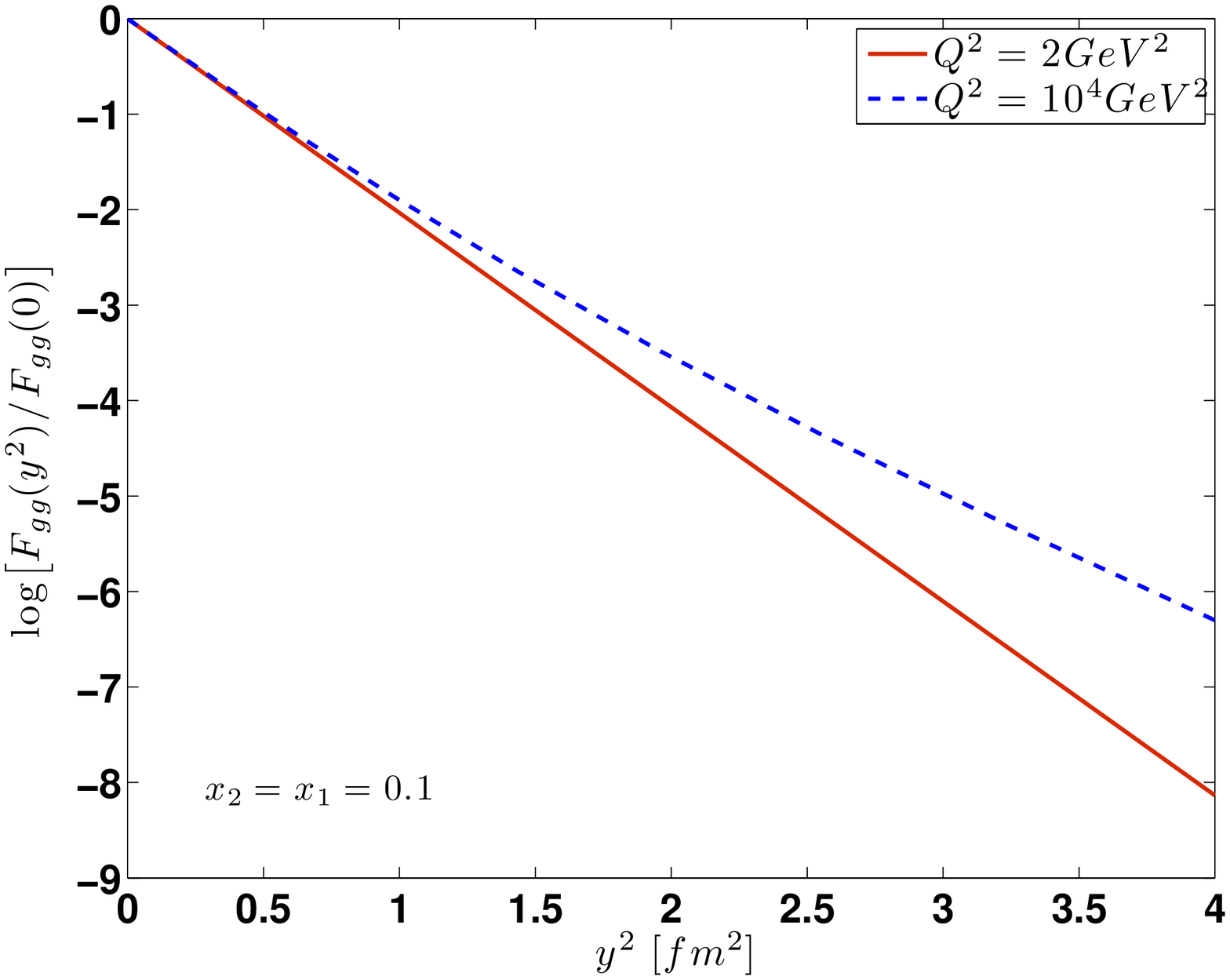}
\vskip 14cm
\caption{\small  Effects of evolution on correlations 
according to the scheme of Ref. \cite{dkk}.
{Upper panel:} $\ln [F_{u^-u^-}(\vec y^2)/F_{u^-u^-}(0)]$
at $x_2=x_1=0.1$ as function of $\vec y^2$ [fm$^2$] at fixed values of $Q^2$ 
and following the assumptions of Ref. \cite{dkk}.
{Middle panel:} As in the upper panel for 
$\ln [F_{u^+u^+}(\vec y^2)/F_{u^+u^+}(0)]$.{ Lower panel:} As previous panels, 
for $\ln [F_{gg}(\vec y^2)/ F_{gg}(0)]$.}
\label{fig:FabDKK01}
\vspace{-1.0em}

\end{figure}

\vskip -2.8cm

\subsection{\label{sec:Diehletal}Analysis of the approach 
of Ref. \cite{dkk}
}

Let us first analyze the scale 
dependence of the correlations introduced at $Q_0^2$ within the assumptions  
Eqs. (\ref{eq:Fab-y}) and (\ref{eq:Fab-k}), 
as proposed by Diehl {\it et al}. in Ref. \cite{dkk}. 
To this end, we study the QCD-evolution of the dPDFs.
In particular one could ask oneself to which 
extent the Gaussian $y$-dependence (or $k_\perp$-dependence) of the starting 
scale is preserved under evolution. Quantities particularly suitable to this 
end are the ratios
\be
\ln \left[ F_{ab}(x_1=x_2,\vec y^2,Q^2) \over F_{ab}(x_1=x_2,\vec y^2=0,Q^2)\right]_{Q_0^2} = -{{\vec y}^2 \over  4 h_{ab}(x_1, x_2)}\,, \label{eq:ratio-ab-y} \\
\ee
\be
\ln \left[ F_{ab}(x_1=x_2,\vec k_\perp^2,Q^2) \over F_{ab}(x_1=x_2,\vec k_\perp^2=0,Q^2)\right]_{Q_0^2} = -{h_{ab}(x_1, x_2)\,{\vec k_\perp}^2}\,, \label{eq:ratio-ab-k}  \\ 
\ee
which, at $Q_0^2$ and $x_1=x_2= constant$, are just 
straight lines as functions of $\vec y^2$ or $k_\perp^2$.

Perturbative evolution of the dPDFs is summarized and discussed in 
Appendix \ref{App:pQCD}; however let us anticipate the results in this 
example, proposed in Ref. \cite{dkk}.
As it is done also in Ref. \cite{dkk}, only the homogeneus part of dPDFs 
evolution
is implemented, for the moment being, in our scheme.
According to some studies, the inhomogeneus part
could play some role in this phenomenology \cite{blok_2,1x2};
its analysis is beyond the scope of the present paper.

In Fig. \ref{fig:FabDKK01} we show the ratios Eqs. (\ref{eq:ratio-ab-y}) 
for different quark and gluon combinations:
$F_{u^+u^+}$,  $F_{u^-u^-}$ and $F_{gg}$ at $x_1=x_2=0.1$ at different scales, 
namely the starting scale 
$Q_0^2= 2$ GeV$^2$ and $Q^2 = 10^4$ GeV$^2$.
One can check that, for quarks,  the shape remains approximately Gaussian 
(a straight line) up to scales as high as $Q^2 = 10^4$ GeV$^2$ even if the 
slope changes rather strongly. For gluons, also the Gaussian property is not 
preserved.

\vskip 6cm
\begin{figure}[h]
 \includegraphics{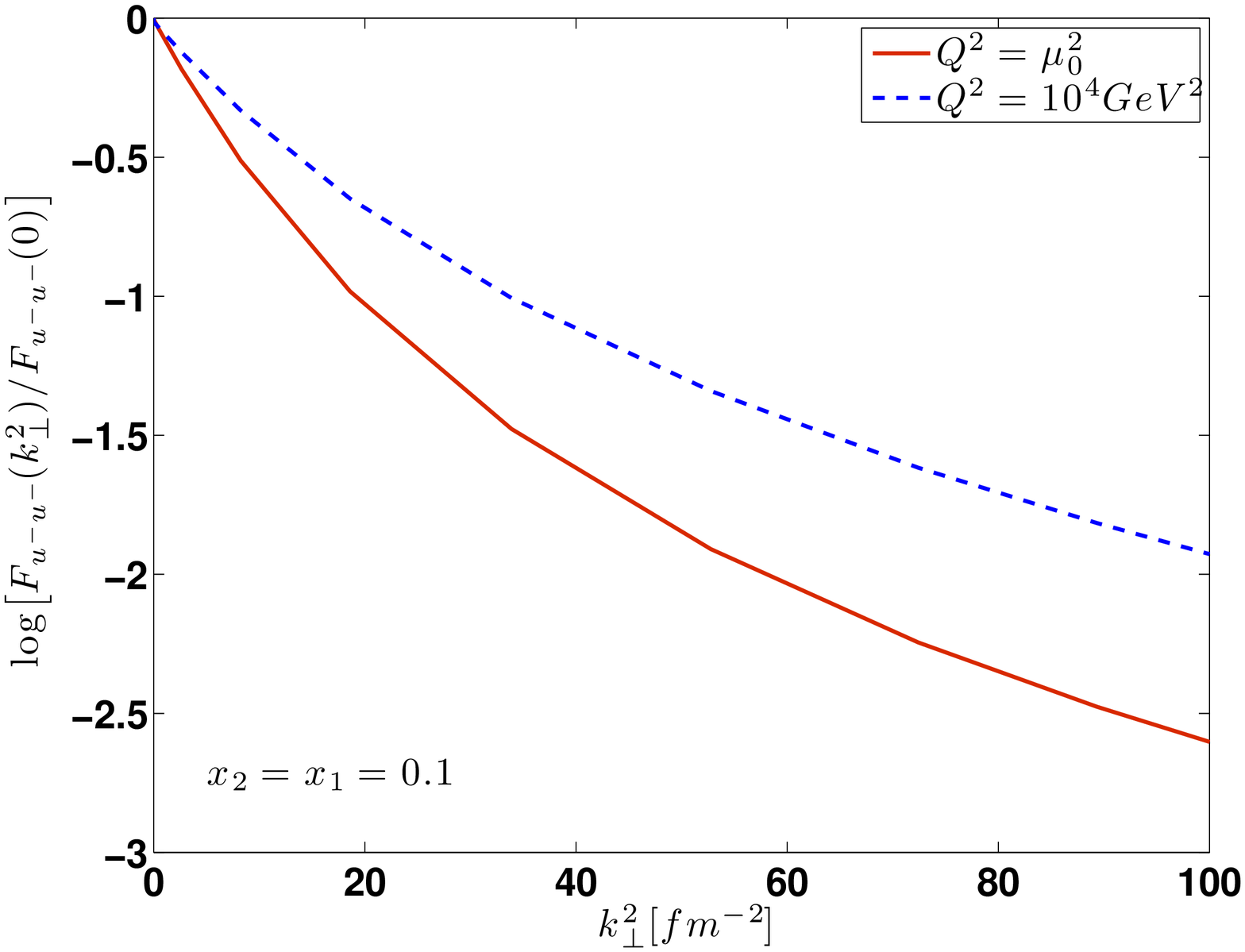}
\vskip 2cm
\includegraphics{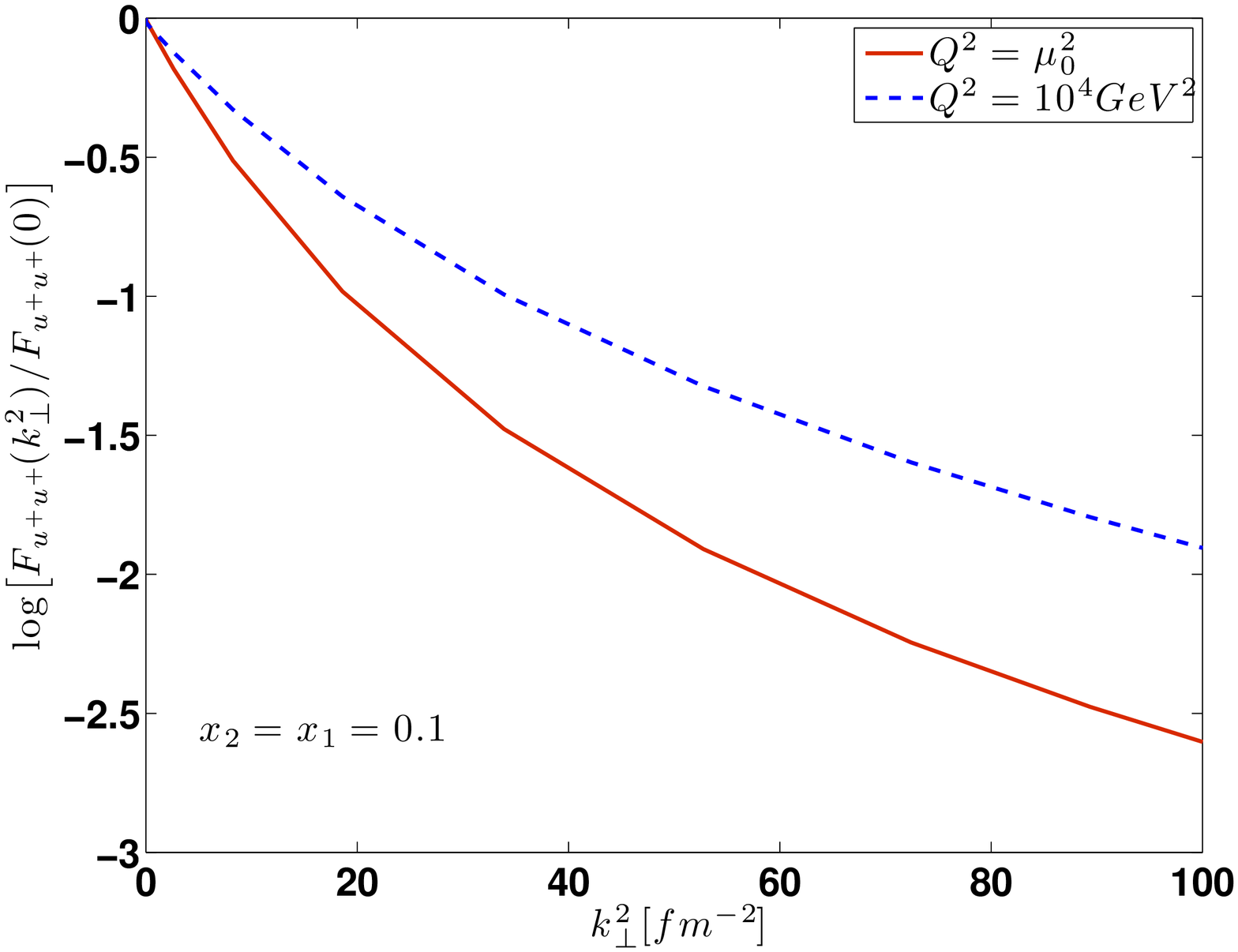} 
\vskip 2cm
 \includegraphics{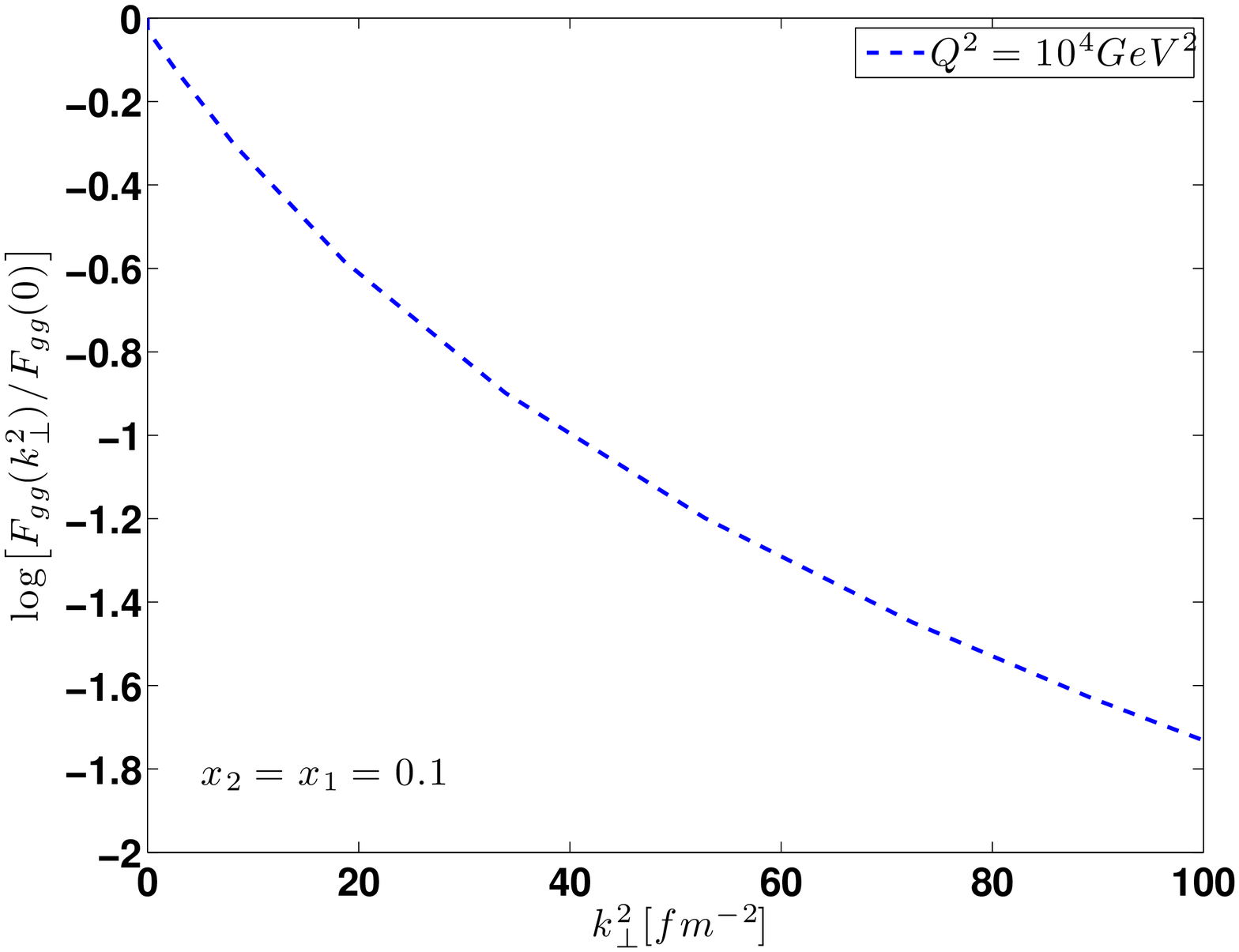}
\vskip 10cm
\caption{\small  Effects of evolution on correlations for the 
LF-Hypercentral approach. {Upper panel} $\ln 
[F_{u^-u^-}(k_\perp^2)/F_{u^-u^-}(0)]$
at $x_2=x_1=0.1$ as function of $\vec k_\perp^2$ [fm$^{-2}$ ] and fixed values 
of $Q^2$.
{Middle panel:} $\ln [F_{u^+u^+}(k_\perp^2)/F_{u^+u^+}(0)]$ with analogous 
notations.
{Lower panel:} $\ln [F_{gg}(k_\perp^2)/F_{gg}(0)]$ with the same notations. The 
gluon distribution $F_{gg}$ vanishes identically at $Q^2 = \mu_0^2$ 
(cfr. Eqs. (\ref{eq:Fab-mu0})).}
\label{fig:FabHyp01}
\vspace{-1.0em}

\end{figure}

\vskip4cm

In particular the upper panel of Fig. \ref{fig:FabDKK01} shows the valence 
components of $F_{ab}$ ($F_{u^-u^-} = F_{(u - \bar u)(u - \bar u)} 
\equiv F_{u_Vu_V}$). For those distributions, only the non-singlet evolution  
is 
relevant. In the case of $F_{u^+u^+} = F_{(u + \bar u)(u + \bar u)} = 
F_{(u_V + 2 \bar u)(u_V + 2 \bar u)} = F_{u_Vu_V} + 2 [F_{u_V \bar u}+
F_{\bar u u_V}] + 4 F_{\bar u \bar u}$ (middle panel), the singlet components 
are contributing in a substantial way; $F_{gg}$ (lower panel) is purely 
singlet.
The distributions are defined as functions of the distance $|\vec y|$, 
the Fourier transform would give the distributions as functions of 
$\vec k_\perp$ without adding more information. The choice to show the
$\vec y\,^2$- dependence makes easier the comparison of the results shown 
in Fig. \ref{fig:FabDKK01} with the calculation of Diehl {\it at al.} as 
illustrated in their 
Figs. 1(a), 1(c), and 1(e).

\vskip3cm
\subsection{\label{sec:LFapproach} Scale dependence within the LF-approach}

In the LF-approach the Fock decomposition of the proton state
at the lowest scale $\mu_0^2$ includes valence quarks only and one 
remains with the following reductions
\bea
F_{u^-u^-} & \equiv&  F_{u_Vu_V} \nonumber \\
& \to & 2 u_Vu_V(x_1,x_2, k_\perp, \mu_0^2)\,,\nonumber \\ 
F_{u^+u^+} &\equiv& F_{u_Vu_V} + 2 [F_{u_V \bar u}+F_{\bar u u_V}] + 4 F_{\bar u \bar u}
\nonumber \\
& \to & 2 u_Vu_V(x_1,x_2, k_\perp, \mu_0^2)\,, \nonumber \\
F_{gg} &\to & 0\,,\label{eq:Fab-mu0}
\eea
with $u_Vu_V(x_1,x_2, k_\perp, \mu_0^2)$ given by 
Eqs. (\ref{eq:uVuVpol}) and (\ref{eq:uVuVunpol}).

The distributions are now function of $\vec k_\perp$ since in the  
LF-approach they are defined in momentum space. 
The relation with $\vec y$ is a simple Fourier transform (cfr. 
Eqs. (\ref{eq:Fab-y}) and (\ref{eq:Fab-k}));
however their functional forms,
entirely determined by the dynamical structure of the 
LF-wavefunctions, are far from being Gaussian.

The distributions at the starting point are strongly simplified as indicated 
by Eqs. (\ref{eq:Fab-mu0}), but they evolve in a complicated way as combination 
of non-singlet ($Valence = V_i = q_i - \bar q_i$, $T_3= u^+-d^+$, 
$T_8=u^++d^+-2 s^+$) as well as singlet components ($\Sigma = u^+ 
+d^+ +s^+ = \sum_i q_i^+$, $gluons$). 

In Fig. \ref{fig:FabHyp01}, the results are shown at fixed $x_2=x_1=0.1$, 
$Q^2 = 10^4$ GeV$^2$, as function of $k_\perp^2$. 

The form is clearly non-Gaussian, since non-Gaussian is its functional form 
at $\mu_0^2$, fact which is related to the dynamics of the LF, 
not on the value of $\mu_0^2$. The complete results are shown in 
Fig. \ref{fig:FabHyp01}, following the same notations and criteria of  
Fig. \ref{fig:FabDKK01}.

Comparing the results of the two set of Figures (\ref{fig:FabDKK01} 
and \ref{fig:FabHyp01}), it is evident that the evolution effects are 
similar in the two different cases, but it is also evident that the Gaussian 
ansatz is rather arbitrary and not supported by  LF dynamics.

\section{\label{sec:sea-LF} 
Adding sea quarks and gluons at a low energy scale}

In the previous sections, our Light-Front approach has been focused on the 
study of valence degrees of freedom at 
low momentum scale. Other partons, and their correlation 
effects, emerge from radiated gluons 
in the perturbative QCD-evolution of the dPDFs. 
In the present Section we enlarge the 
perspective studying how sea quarks and gluons can be included at a 
low-momentum scale and within the same LF framework. An example 
(e.g. Refs. \cite{Traini2012-2014, marco2}) is given by 
inclusive DIS, where the 
(non-perturbative) meson degrees of freedom 
can be introduced by means of a description of the meson cloud and the 
scattering of the virtual photon off the 
constituents of the mesons (Sullivan process). Analogous  approach can be 
applied to the explicit evaluation of meson cloud effects on GPDs 
(e.g. Refs. \cite{BoffiPasquiniTraini2003-2004-2005,PasquiniBoffi2006}).

Hereafter we will propose a simplified approach in 
which the effects of the valence degrees of freedom 
(producing the largest part 
of the dPDFs at low-momentum scale) are calculated using 
Eqs. (\ref{eq:uVuVpol}) and
(\ref{eq:uVuVunpol}), while the non-perturbative sea and gluons components are 
evaluated 
by means of a factorized approximation of the kind discussed in Sect.
\ref{sec:factorization}. In order to minimize the hypothesis on 
factorization let us start discussing the limiting case $k_\perp=0$. Let us 
first 
illustrate, as an example, the $u u$ dPDFs:
\bea
F_{u u}(x_1,x_2,k_\perp=0, Q_0^2)     
&=& F_{(u_V+\bar u)(u_V+\bar u)}(x_1,x_2,k_\perp=0, Q_0^2)  \nonumber \\
& = & F_{u_Vu_V}(x_1,x_2,k_\perp\!=\!0,Q_0^2) + \label{eq:uVuVQ0} \\
& + & \left[F_{u_V \bar u}(x_1,\!x_2,k_\perp\!= \!0,Q_0^2)  + \right.\nonumber \\
  \phantom{} &+& \left. F_{\bar u  u_V}(x_1,\!x_2,k_\perp\!= \!0,Q_0^2)
\right] +\label{eq:uVbaruQ0} \\
& + & F_{\bar u \bar u}(x_1,x_2,k_\perp=0,Q_0^2) \label{eq:barubaruQ0}\,.
\eea
The pure valence (and dominant) term, the expression (\ref{eq:uVuVQ0}) in 
the above equation,
\be
 F_{u_Vu_V}(x_1,x_2,k_\perp,Q_0^2)  =  2 \times 
u_Vu_V(x_1,x_2,k_\perp,Q_0^2)\,,
\ee
can be evaluated in a direct way within the LF-approach described in the 
previous sections.
In order to calculate  the residual terms, Eqs. (\ref{eq:uVbaruQ0}) and 
(\ref{eq:barubaruQ0}), one can assume  factorized forms 
(see e.g. Ref. \cite{diehl_1}).

The complete (approximate) expression for $F_{uu}$ becomes: 
\bea
F_{uu}(x_1,x_2,k_\perp \!= \!0,Q_0^2) &\approx& \label{eq:approxuu} \\
& = & F_{u_V u_V}(x_1, x_2,k_\perp\!=\!0,Q_0^2) + \label{eq:uVuV}  
\label{eq:uuVV} \\
& + & \left\{\left[u_V(x_1,Q_0^2) \bar u(x_2,Q_0^2) + 
\bar u(x_1,Q_0^2) u_V(x_2,Q_0^2) \right] + \right. \nonumber \\
& + & \left. \bar u(x_1,Q_0^2) \bar u(x_2,Q_0^2)\right\} (1-x_1-x_2)^n 
\theta(1-x_1-x_2)\,. \nonumber \\
\label{eq:correlationQ0}
\eea
Few comments are in order:  

\begin{itemize}

\item[i)] the contribution Eq. (\ref{eq:uuVV}) is the term  due to 
valence quarks, 
{\it it is not approximated by a factorized procedure} and  it is based on the 
calculated expressions 
Eqs. (\ref{eq:uVuVpol}) and (\ref{eq:uVuVunpol});

\item[ii)]  the residual contributions imply the knowledge of the singlet 
component $\bar u(x,Q_0^2)$ and fulfill the correct kinematical conditions 
for $x_1+x_2>1$, owing to the constraints  introduced by the phenomenological 
function $(1-x_1-x_2)^n \theta(1-x_1-x_2)$. The exponent $n$ has to be fixed 
phenomenologically, as seen in Sect. \ref{sec:pheno-factorization} 
in the case of the model of Ref. \cite{dkk} and will 
be discussed in the next Sections for the LF-approach;

\item[iii)]$\bar u(x,Q_0^2) = u_{sea} (x,Q_0^2)$ has, at the low 
momentum scale 
$Q_0^2$, a non-perturbative origin,
basically due to the meson cloud surrounding the nucleon; 

\item[iv)] the scale $Q_0^2$ is not to be identified with 
$\mu_0^2$, i.e. the scale of the {\it bare}
nucleon, where only the three valence quarks contribute.

\end{itemize}
In the following sections we will discuss a straight-forward 
(phenomenological) 
way of introducing meson and gluon degrees of freedom at 
the low-momentum non-perturbative scale. QCD evolution will be used to
reach the high energy scale of the LHC experiments.

\subsection{\label{sec:LF-factorization} Factorization procedures within 
the LF-approach at $k_\perp=0$}
 
The advantage of the approach we are discussing is based on a complete 
calculation of correlation effects within the LF-dynamics, in the restricted 
space of valence degrees of freedom. At the same time it allows to discuss 
the role of the factorization procedure and its validity, 
comparing our approach with phenomenological factorized models.
This comparison aims to identify the coherence and
self-consistency of the factorization schemes.
In the following we will give 
three examples: i) the identification of the exponent $n$ to fix  the 
correlating function Eq. (\ref{eq:correlationQ0}) [see Sect. \ref{sec:n_exp}]; 
ii) the introduction of a larger number of degrees of freedom at $k_\perp=0$ 
[Sect. \ref{sec:MSTW}]; iii) the extension to 
$k_\perp >0$ of the sea and gluon contributions to dPDFs 
[Sect. \ref{sec:LF-factorization_kperp}]. 

\subsubsection{\label{sec:n_exp} Fixing the factorization form}

The optimization of factorization 
procedures for dPDFs is not a simple issue. 
The most relevant constraints are related to momentum and quark number sum 
rules \cite{gaunt}. 
Our LF-approach, on the contrary, fulfills 
such sum rules by construction and therefore one does not need 
to implement phenomenological assumptions required
to build factorized dPDFs.

As an example the resulting valence dPDF $u_Vu_V(x_1,x_2,k_\perp=0, \mu_0^2)$, 
as well as the single PDFs (sPDFs) calculated within the same 
LF dynamical approach, fulfill the momentum and quark number sum rules. 
One can take advantage from such fundamental  properties to fix the order 
of magnitude of the phenomenological exponent  $n$ in Eq. 
(\ref{eq:correlationQ0}), trying to combine the knowledge of sPDFs and dPDFs 
in the following (factorized) relation:
\bea
& & F_{u_V u_V}(x_1,x_2,k_\perp=0,\mu_0^2) = \nonumber \\
&=& \left.2 \cdot  u_Vu_V(x_1,x_2,k_\perp=0,\mu_0^2) \right|_{\rm LF} = \nonumber \\
&\approx& \left. \left. u_V(x_1,\mu_0^2)\right|_{\rm LF}u_V(x_1,\mu_0^2)\right|_{\rm LF} \times \nonumber \\
&\times& (1-x_1-x_2)^n \theta(1-x_1-x_2)\,. \label{eq:sP-dP-comparison}
\label{32}
\eea
The restricted validity of the factorization approach has been already 
discussed in Sect. \ref{sec:factorization}, therefore one cannot expect 
Eq. (\ref{eq:sP-dP-comparison}) to be satisfied with a high degree of accuracy. 
We expect, however, indications for the value of the exponent $n$ to be used 
for building the additional sea and gluon contributions to the dPDFs.
The value $n=2$ has been discussed in the past as a good choice 
(see, e.g., Ref. \cite{snig1} and references therein). 
More recent arguments (see, e.g., Refs.
\cite{dkk},
and
\cite{gaunt})
are in favor of more sophisticated parametrizations. 
Given the restricted use we are going to make of the factorization assumption, 
we prefer to remain within the straight-forward 
formulation Eq. (\ref{eq:sP-dP-comparison}).
{{Our numerical analysis confirms a
limited validity of the factorization 
and, at the same time, suggests $n\approx 0.2$
(more precisely, values within the range $0.1 <  n < 0.5$; $n=0.2$ is our 
optimal choice).}}

\subsubsection{\label{sec:MSTW} Sea and gluon contribution 
according to Ref. \cite{MSTW2008}}

The advantages of the 
approximation Eq. (\ref{eq:approxuu}) are now clear: 
the largest contributions are
due to the valence components and the LF approach has the merit of 
preserving at $\mu_0^2$ quark number and momentum sum rules. The perturbative 
evolution needed to reach the new low-momentum scale $Q_0^2$ to integrate 
new degrees of freedom preserves those constraints. At the same time the 
residual terms can be approximated within a clear and self constrained 
factorized approach able to select the form of the factorization as discussed 
in the previous subsection. 

In the following we discuss the introduction of sea and gluon degrees of 
freedom by means of one of the most used phenomenological parametrization of 
sPDF, the $LO$ MSTW2008  parametrization (see Table 4 of Ref. 
\cite{MSTW2008}).
The parametrization is valid at $Q_0^2 = 1.0$ GeV$^2$. The fact that we are 
proposing $LO$ parametrization is specifically due to the evolution 
properties of the dPDFs, known at $LO$ only.

\begin{figure}[tbp]
\centering\includegraphics[width=10cm,clip=true,angle=0]
{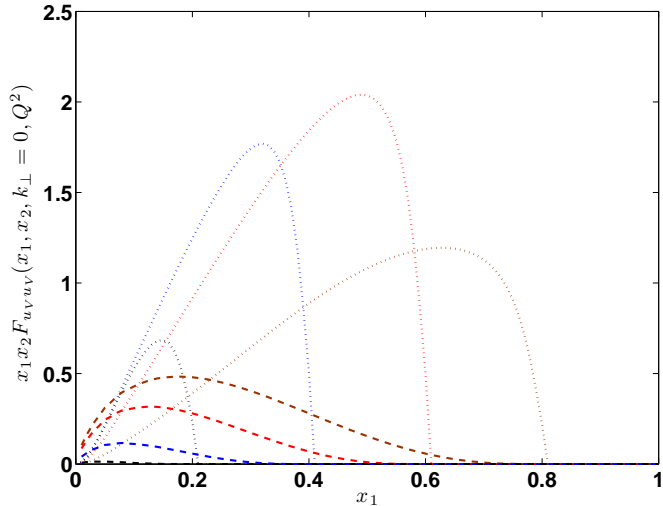}
\caption{\small  $x_1 x_2 F_{u_Vu_V}(x_1,x_2,k_\perp=0,Q^2)$ 
as function of $x_1$ and $x_2=0,2,0.4,0.6,0.8$
and for two values of $Q^2$, namely the extreme low momentum scale 
$\mu_0^2$ (dotted lines) and the scale (dashed lines) of the 
MSTW parametrization, $Q^2 = Q_0^2 = 1.0$ GeV$^2$.}
\label{fig:xxFuVuVQ0_MSTW}
\end{figure}

At the scale $Q_0^2$ , the sPDF MSTW2008 parametrization is characterized 
by the presence of partons like $u_V$, $d_V$, $\bar u$, $\bar d$, $s$,
$\bar s$ and $gluons$. The total momentum is shared among such degrees of 
freedom and one has:
\bea
\int dx\,x\, \left[u_V(x,Q_0^2) +d_V(x,Q_0^2)\right] = 0.452 \,; 
\eea
\bea
\int dx \, x \, Sea(x,Q_0^2) & = & 
\int dx\, x \, \left[(2 \bar u (x,Q_0^2) +2\bar d (x,Q_0^2)) \right.
\nonumber \\ & + & \left. 
s(x,Q_0^2) + \bar s(x,Q_0^2) \right] = 0.108 \,
\eea 
\bea
\int dx\,x\, g(x,Q_0^2)= 0.431\,.  
\eea
Since at the scale $\mu_0^2$ the system is determined by the valence degrees 
of freedom only, one has:
$\int dx x \left[u_V(x,\mu_0^2) +d_V(x,\mu_0^2)\right] \!\!= \!\!1$;
$Sea(x,\mu_0^2) \!= \! 0$;  $g(x,\mu_0^2)  =  0$.
As a consequence, to use Eq. (\ref{32}) at the scale $Q_0^2$, the dPDF  
$u_Vu_V(x_1,x_2,k_\perp=0,\mu_0^2)$ 
 at the scale $\mu_0^2$ has to be evolved  to $Q_0^2$. 
The evolution is performed by means of the Non-Singlet reduction of the QCD 
evolution as described in Ref. \cite{noi2} 
and summarized in the appendix \ref{App:pQCD}, where the complete Mellin 
procedure we are proposing for both Singlet and Non-Singlet sectors 
is illustrated in some detail.
The result is shown in Fig. \ref{fig:xxFuVuVQ0_MSTW} for four 
selected values of $x_2$. 
\begin{figure}[tbp]
\centering\includegraphics[width=10cm,clip=true,angle=0]
{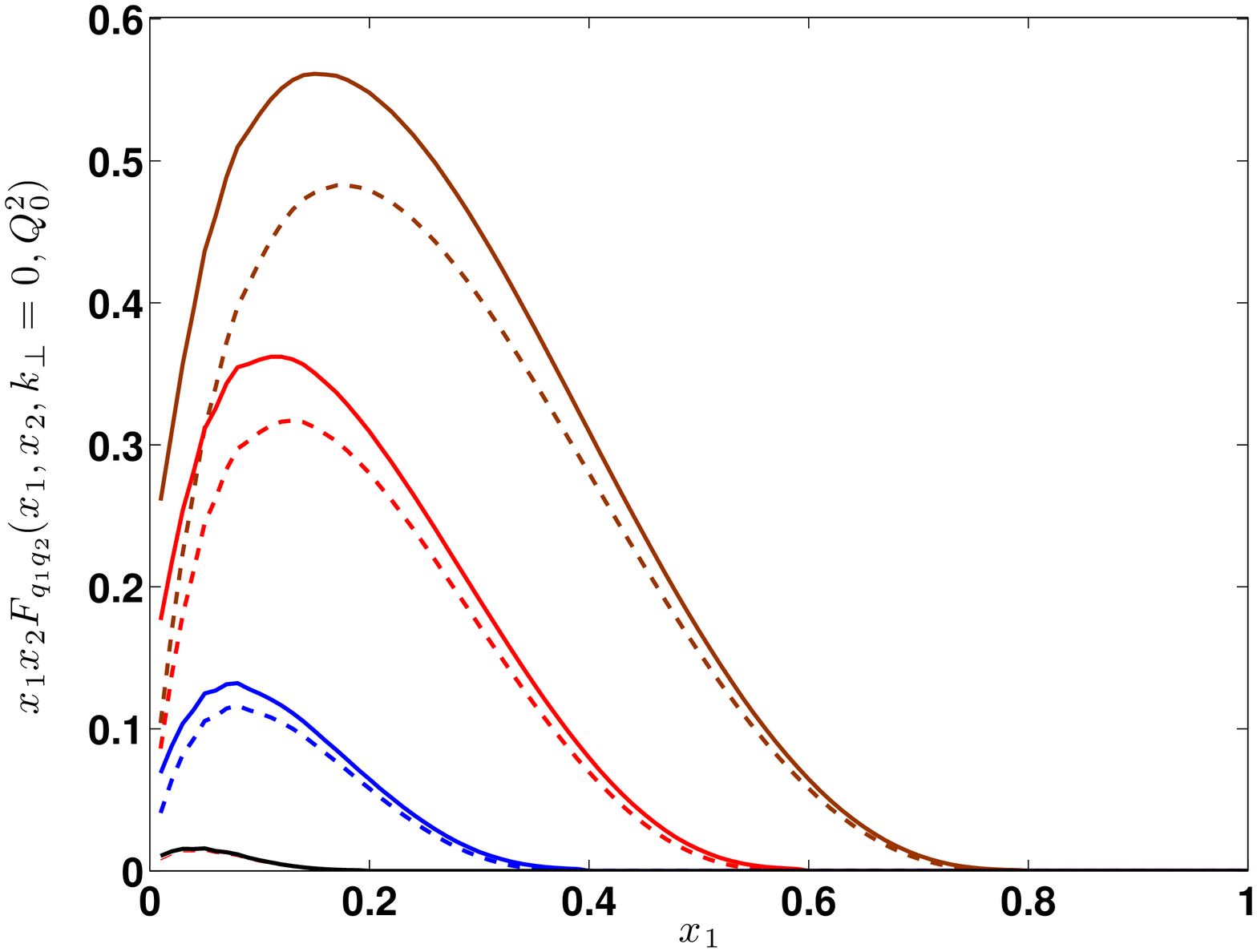}
\caption{\small  $x_1 x_2 F_{u_Vu_V}(x_1,x_2,k_\perp=0,Q_0^2)$ 
(dashed lines) of Fig. \ref{fig:xxFuVuVQ0_MSTW} is compared with 
$x_1 x_2 F_{u u}(x_1,x_2,k_\perp=0,Q_0^2)$ which contains the additional 
sea contributions from $\bar u$ ($u = u_V + \bar u$, see text) 
(continuous lines).}
\label{fig:xxFuVuV-uuQ0_MSTW}
\vspace{-1.0em}
\end{figure}


\begin{figure}[bp]
\centering\includegraphics[width=10cm,clip=true,angle=0]
{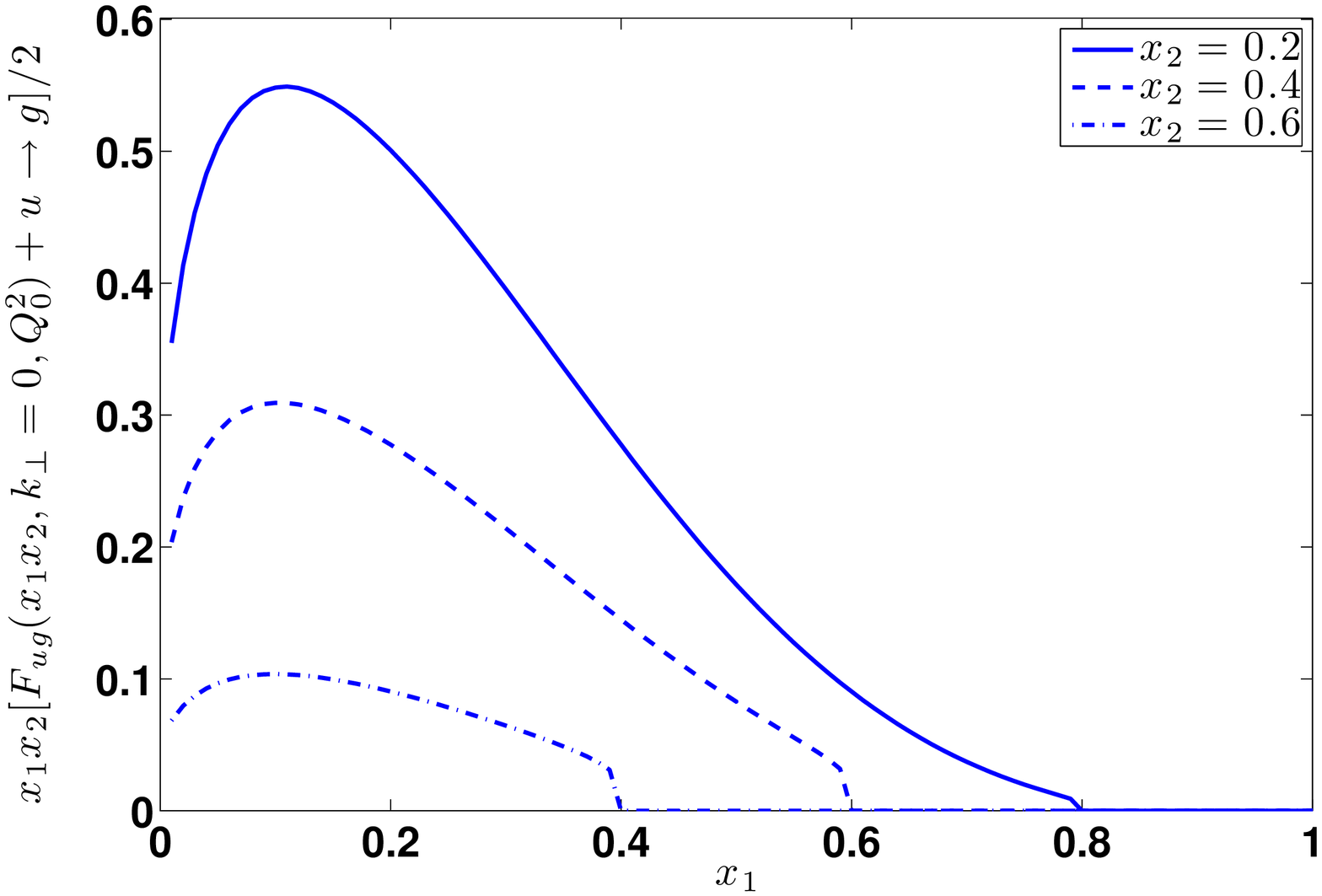}
\caption{\small The dPDFs (at $Q_0^2$ and $k_\perp = 0$) due to the 
interference of $u$ quarks and gluons as function of $x_1$ and selected values 
of $x_2$.}
\label{fig:xxFugQ0_MSTW}
\vspace{-1.0em}
\end{figure}
In Fig. \ref{fig:xxFuVuV-uuQ0_MSTW} and Fig. \ref{fig:xxFugQ0_MSTW}
we show the complete set of dPDFs involving the
$u$-quark at the starting 
scale $Q_0^2$. In particular, the combination  $F_{u u} = F_{(u_V+\bar u) 
(u_V+\bar u)} = F_{u_Vu_V} + [F_{u_V \bar u} + F_{\bar u u_V}] + 
F_{\bar u \bar u}$ is shown in Fig. \ref{fig:xxFuVuV-uuQ0_MSTW} giving 
explicit 
evidence to the contribution due to $\bar u$ quarks. The largest effects of 
the sea component are clearly evident for smallest values of the momentum 
fraction. In Fig. \ref{fig:xxFugQ0_MSTW} we show
the dPDFs $(F_{u g} + F_{g u})/2$ containing 
valence, sea and gluon contributions. The order of magnitude of those 
components is comparable with the valence part $F_{u_Vu_V}$ at the scale 
$Q_0^2$ (cfr. Fig. \ref{fig:xxFuVuVQ0_MSTW}).

\subsection{\label{sec:LF-factorization_kperp} 
Factorization procedures within the LF-approach at $k_\perp > 0$}

The dPDFs in a pure valence scenario have been discussed in previous 
sections and the dependence on  $k_\perp$ has been 
explicitly investigated (cfr. for 
example, Figs. \ref{fig:xxuVuV-kperp}, \ref{fig:Fuu2_HEu0103}, and 
\ref{fig:FabHyp01}). As a result they do not admit simple factorized forms.
However, as a first attempt to go beyond the valence scenario 
at $k_\perp \ne 0$,
we could add the other degrees of freedom 
using factorized expressions. To this aim,
the knowledge of the exact LF valence component at 
$k_\perp=0$ helps to define the additional, 
factorized contributions. In this section we find 
a reasonable factorized approximation
to the exact valence LF dPDFs. In this way
we fix the parameters which will be
 used for the non-valence degrees of freedom

\begin{figure}[tpb]
\centering\includegraphics[width=8.2cm,clip=true,angle=0]
{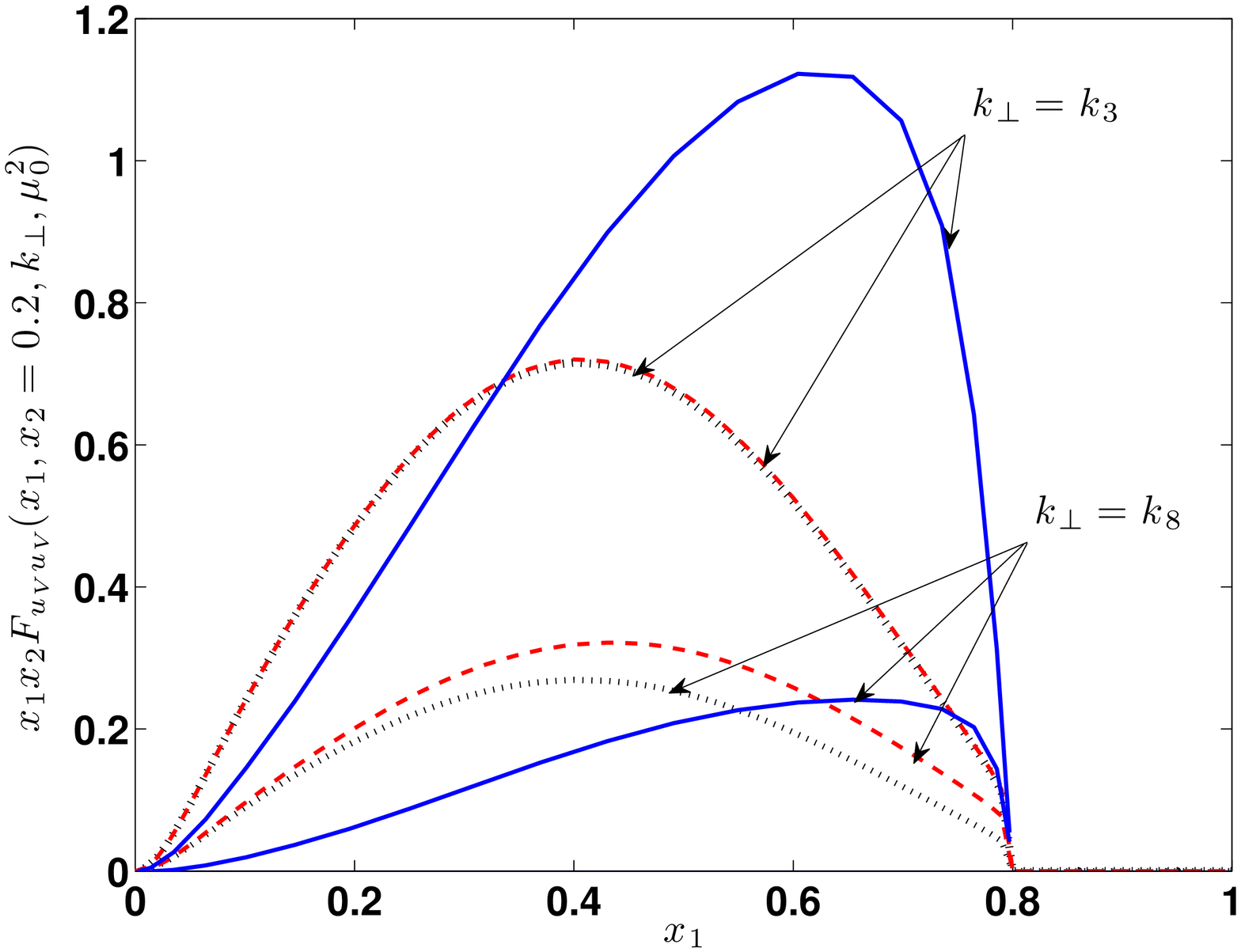}
\centering\includegraphics[width=8.2cm,clip=true,angle=0]
{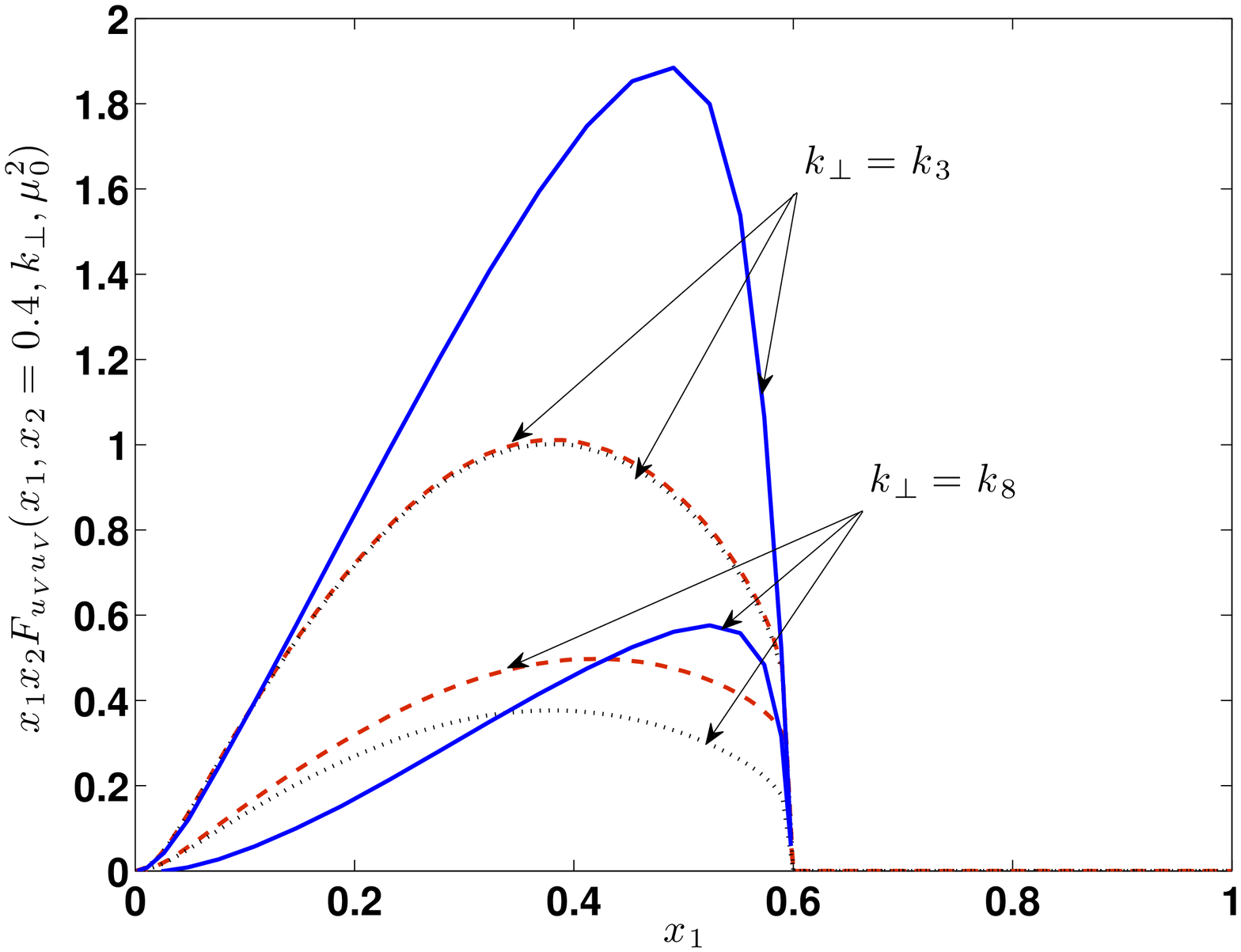}
\caption{\small The exact LF dPDFs $x_1x_2 F_{u_Vu_V}(x_1,x_2,k_\perp,\mu_0^2)$
(continous lines) and its factorized approximation Eq. 
(\ref{eq:sP-dP-comparison_kperp}) (scenario $B$ Eq. (\ref{eq:B}) , dashed 
lines, scenario $A$ Eq. (\ref{eq:A}), 
dotted, almost  indistinguishable for $k_\perp = k_3$), as function of 
$x_1$ at $x_2=0.2$ ({left panel}) and $x_2=0.4$ ({right panel}); 
$k_3 =0.32$ GeV and $k_8 = 1.68 $ GeV (cfr. the caption of 
Fig. \ref{fig:xxuVuV-kperp}). We remind that $n=0.2$ and $b_A = b_B = 0.6$ 
GeV$^{-1}$ (see text).}
\label{fig:bfit}
\end{figure}

In practice, we want to generalize to $k_\perp > 0$
Eqs. (\ref{eq:correlationQ0}) and (\ref{eq:sP-dP-comparison}), 
valid at $k_\perp=0$.
For instance, Eq. (\ref{eq:sP-dP-comparison}) becomes
\bea
& & F_{u_V u_V}(x_1,x_2,k_\perp,\mu_0^2) = \nonumber \\
&= & \left. 2 \cdot u_Vu_V(x_1,x_2,k_\perp,\mu_0^2) \right|_{\rm LF} 
\approx \nonumber \\
&\approx& \left. \left. u_V(x_1,\mu_0^2)\right|_{\rm LF}u_V(x_1,\mu_0^2)
\right|_{\rm LF} \times \nonumber \\
&\times& \left(1-x_1-x_2\right)^n \, \phi(x_1,x_2,k_\perp) \, 
\theta(1-x_1-x_2) \,, \label{eq:sP-dP-comparison_kperp}
\eea
and simple choices are (cfr. Eq. (\ref{eq:Fab-k})),
\bea
A) \;\;\; && \phi_A(k_\perp)  =  \exp\left[-b_A^2 \, k_\perp^2 \right]\,, 
\label{eq:A}\\
\nonumber \\
B) \;\;\;&& \phi_B(x_1,x_2,k_\perp)  =   \nonumber \\
&=& \exp\left[-b_B^2 \,(1-x_1-x_2)^n k_\perp^2 \right]\,. \label{eq:B}
\eea
Within scenario $A$ of Eq. (\ref{eq:A}), no correlations between 
$x_1,x_2$ and $k_\perp$ have been introduced:
$\phi_A$ depends on $k_\perp$ and it does not depend on $x_1,x_2$;
in scenario $B$ of Eq. (\ref{eq:B}) the exponent depends 
on $x_1,x_2$, similarly to Eq. (\ref{eq:Fab-k}).
The knowledge of  $\left. u_Vu_V(x_1,x_2,k_\perp,\mu_0^2) \right|_{\rm LF}$ 
from Eq. (\ref{eq:uVuVunpol}) and of the sPDF $\left. u_V(x_1,\mu_0^2)
\right|_{\rm LF}$ with the additional information $n = 0.2$ from the analysis 
of Sect. \ref{sec:n_exp}, can be used to optimize the fit Eq. 
(\ref{eq:sP-dP-comparison_kperp}). 
The results of the optimization procedure are shown in Fig. \ref{fig:bfit} 
for selected and extreme examples. Our recommended values are 
$b_A = b_B = 0.6$ GeV$^{-1}$ and the quality of the fit is, once again, 
quite poor with a slight preference for the full correlated approximation of 
scenario $B$ (Eq.(\ref{eq:B})).  The approximation is crude for the 
valence-valence correlations, but it is sound and the next subsection will be 
devoted to the implementation 
of additional degrees of freedom on the basis offered by the 
factorization Eq. (\ref{eq:sP-dP-comparison_kperp}) and scenario $B$.

\subsubsection{\label{sec:MSTW_kperp} Sea and gluon contribution at $k_\perp>0$}

The valence-valence dPDFs do not allow for a 
simple factorization. The approximation proposed in 
Eq. (\ref{eq:sP-dP-comparison_kperp}) is therefore quite poor in that case, 
as Fig. \ref{fig:bfit} explicitly shows.
However, we do not need to approximate valence-valence correlations; we can 
resort to the exact calculation also in the case of $k_\perp > 0$ 
(cfr. Fig. \ref{fig:xxuVuV-kperp} and 
Eqs. (\ref{eq:uVuVpol}),
(\ref{eq:uVuVunpol}))
and, using the best factorization scheme 
($n = 0.2$ and $b_A = b_B = 0.6$ GeV$^{-1}$ in 
Eqs. (\ref{eq:A}), (\ref{eq:B})) to introduce the additional degrees 
of freedom at the scale $Q_0^2$ and $k_\perp>0$. We have to follow once again 
the steps i)-iv) of Sect.~\ref{sec:sea-LF} and the procedure described in 
Sect.~\ref{sec:MSTW}. Numerically they are more challenging and, 
in a sense, incomplete since the evolution in $k_\perp$ is still 
an open problem \cite{diehl_1}. We simply evolve at fixed $k_\perp$ 
applying the scale evolution of Appendix~\ref{App:pQCD}.
\begin{figure}[bpt]
\centering\includegraphics[width=8.2cm,clip=true,angle=0]
{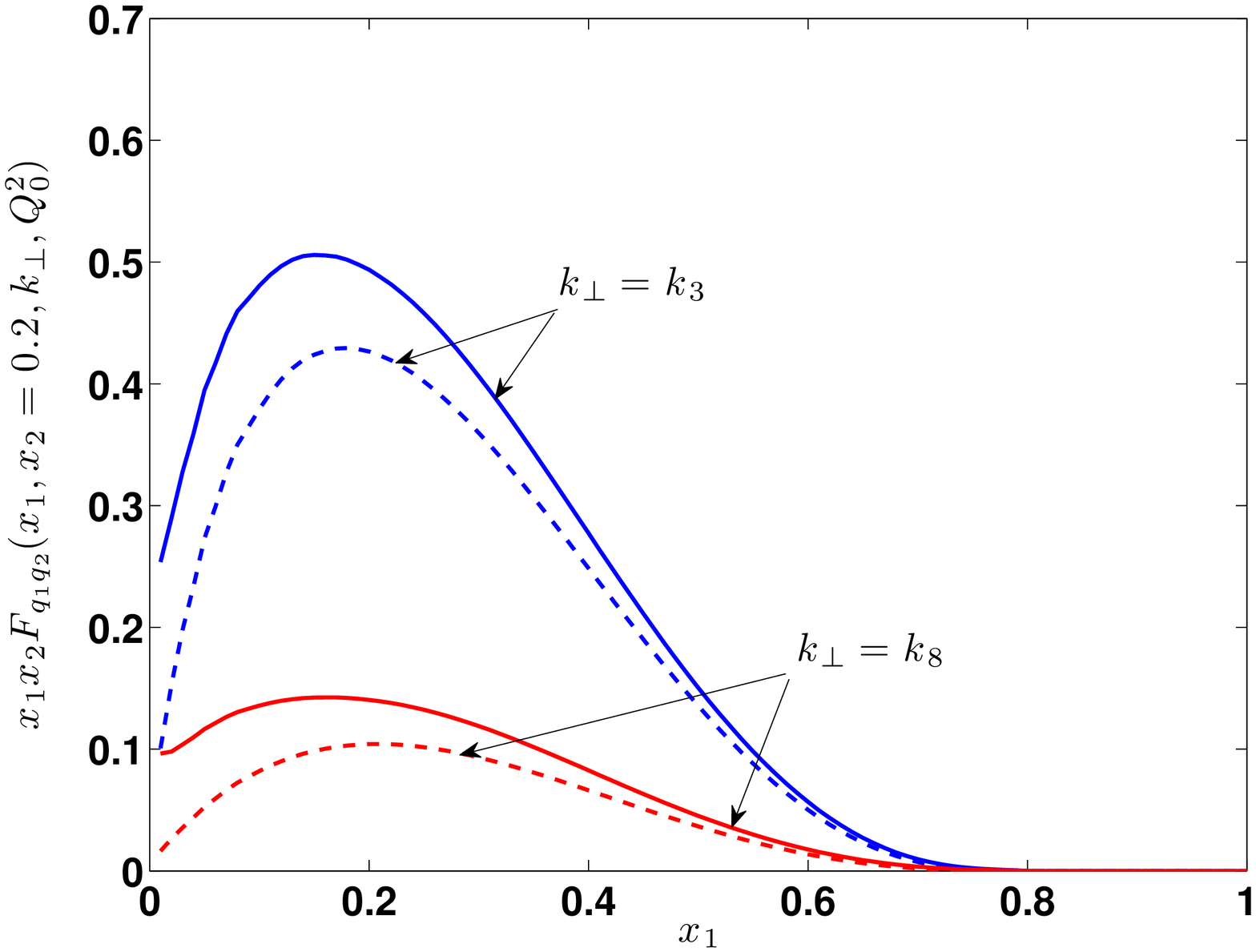}
\centering\includegraphics[width=8.2cm,clip=true,angle=0]
{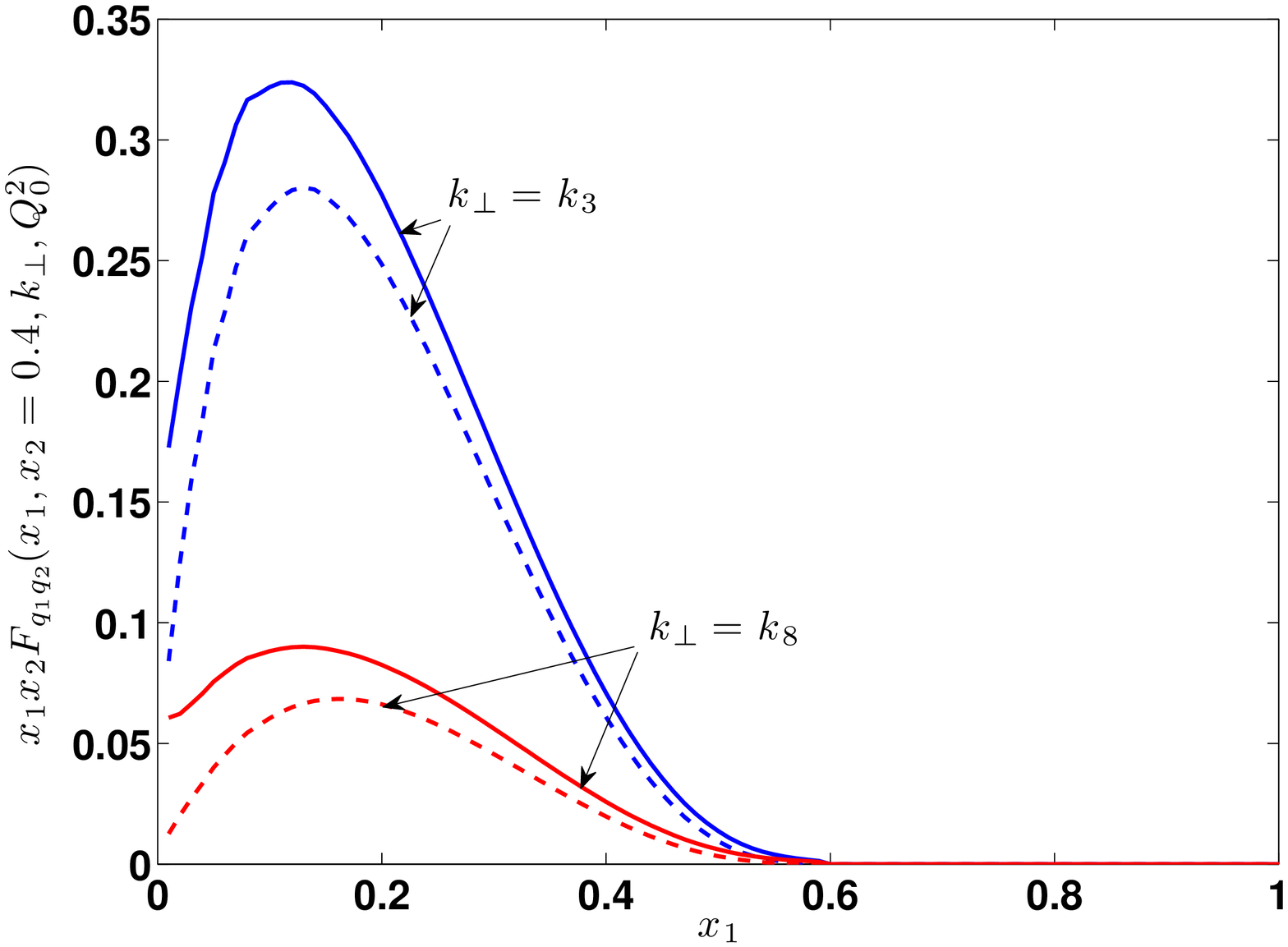}
\caption{\small The exact LF dPDFs $x_1x_2 F_{u_Vu_V}(x_1,x_2,k_\perp,Q_0^2)$ 
(Eq.(\ref{eq:uuVVk}), dashed lines lines) as function of $x_1$ and at fixed
$x_2 = 0.2$ ({left panel}) and $x_2 = 0.4$ ({right panel}).
$x_1x_2F_{uu}(x_1,x_2,k_\perp,Q_0^2)$ (continuos lines) are obtained from the expression 
(\ref{eq:approxuuk}) and scenario $B$ Eq. (\ref{eq:B});  $k_3 =0.32$ GeV and 
$k_8 = 1.68 $ GeV (cfr. the caption of Fig. \ref{fig:xxuVuV-kperp}).}
\label{fig:xxFuuQ0k_3_8_x20204}
\end{figure}
As an  example we show in Fig. \ref{fig:xxFuuQ0k_3_8_x20204} 
the distribution $F_{uu}(x_1,x_2,k_\perp, Q_0^2)$ at the scale $Q_0^2$ 
which generalizes Eq. (\ref{eq:correlationQ0}) and 
Fig. \ref{fig:xxFuVuV-uuQ0_MSTW}, and whose resulting expression 
reads (from now on, we will discuss the scenario $B$ of Eq. (\ref{eq:B}) 
only):
\bea
F_{uu}(x_1,x_2,k_\perp ,Q_0^2) &\approx& \label{eq:approxuuk} \\
& = & 2 \cdot u_Vu_V(x_1, x_2,k_\perp,Q_0^2) + \label{eq:uVuV}  \label{eq:uuVVk} \\
& + & \left\{\left[u_V(x_1,Q_0^2) \bar u(x_2,Q_0^2) + u_V(x_2,Q_0^2) \bar u(x_1,Q_0^2) \right] + \right. \nonumber \\
& + & \left. \bar u(x_1,Q_0^2) \bar u(x_2,Q_0^2)\right\} (1-x_1-x_2)^n \times \nonumber \\
&\times & \phi_B(x_1,x_2,k_\perp)\,\, \theta(1-x_1-x_2)\,;
\label{eq:correlationQ0k}
\eea
\begin{figure}[tbp]
\centering\includegraphics[width=8.2cm,clip=true,angle=0]
{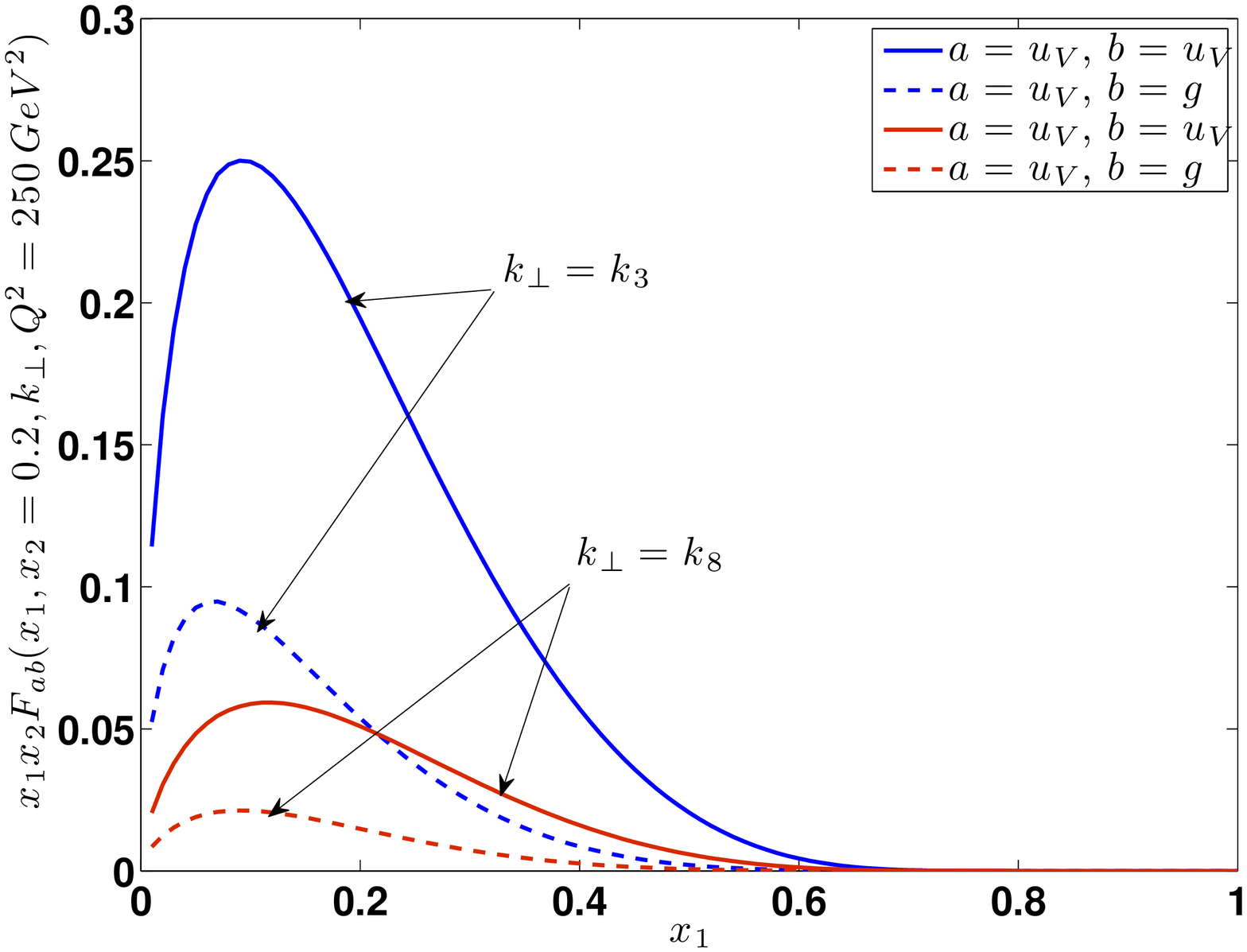}
\centering\includegraphics[width=8.2cm,clip=true,angle=0]
{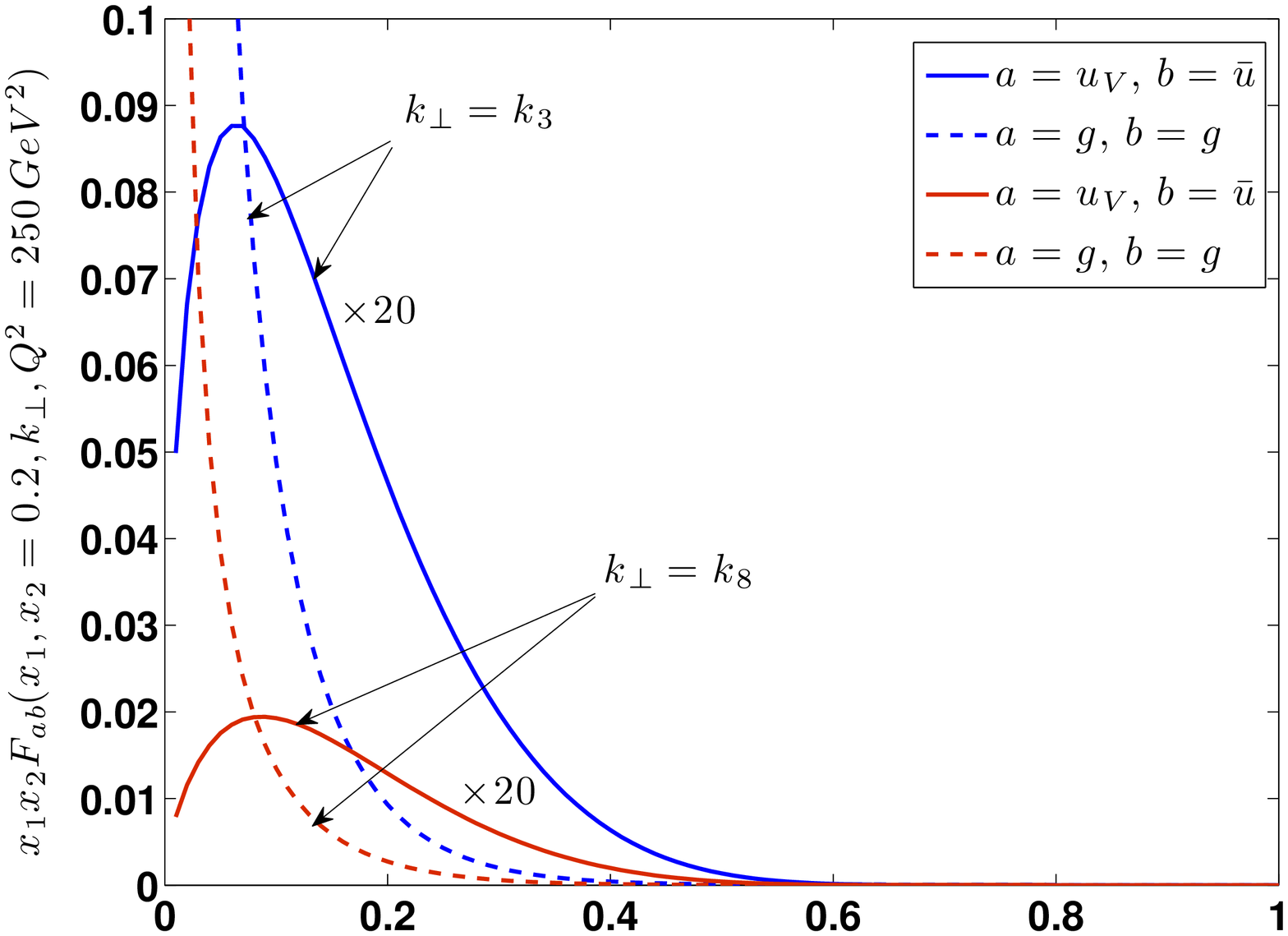}
\caption{\small  {Left panel:} 
The $x_1x_2 F_{u_V u_V}$  dPDFs obtained from perturbative evolution from 
the lowest scale $\mu_0^2$ to $Q^2 = 250$ GeV$^2$ (continuous lines) as 
function of $x_1$ and at fixed $x_2 = 0.2$ (and for two values of $k_\perp$), 
are compared with  $x_1 x_2 F_{u_V g}$ dPDFs (dashed lines) at the same 
high scale. {Right panel:} The $x_1x_2 F_{u_V \bar u}$ dPDFs (amplified 20 
times) at 
the scale $Q^2 = 250$ GeV$^2$ are compared with the $x_1x_2F_{gg}$ 
dPDFs at the same scale, and kinematical conditions. ($k_3 \simeq 0.32$ GeV 
and $k_8 \simeq 1.68 $ GeV).}
\label{fig:xxFabmu0Q2k_3_8_x202}
\end{figure}
\begin{figure}[tbp]
\centering\includegraphics[width=8.2cm,clip=true,angle=0]
{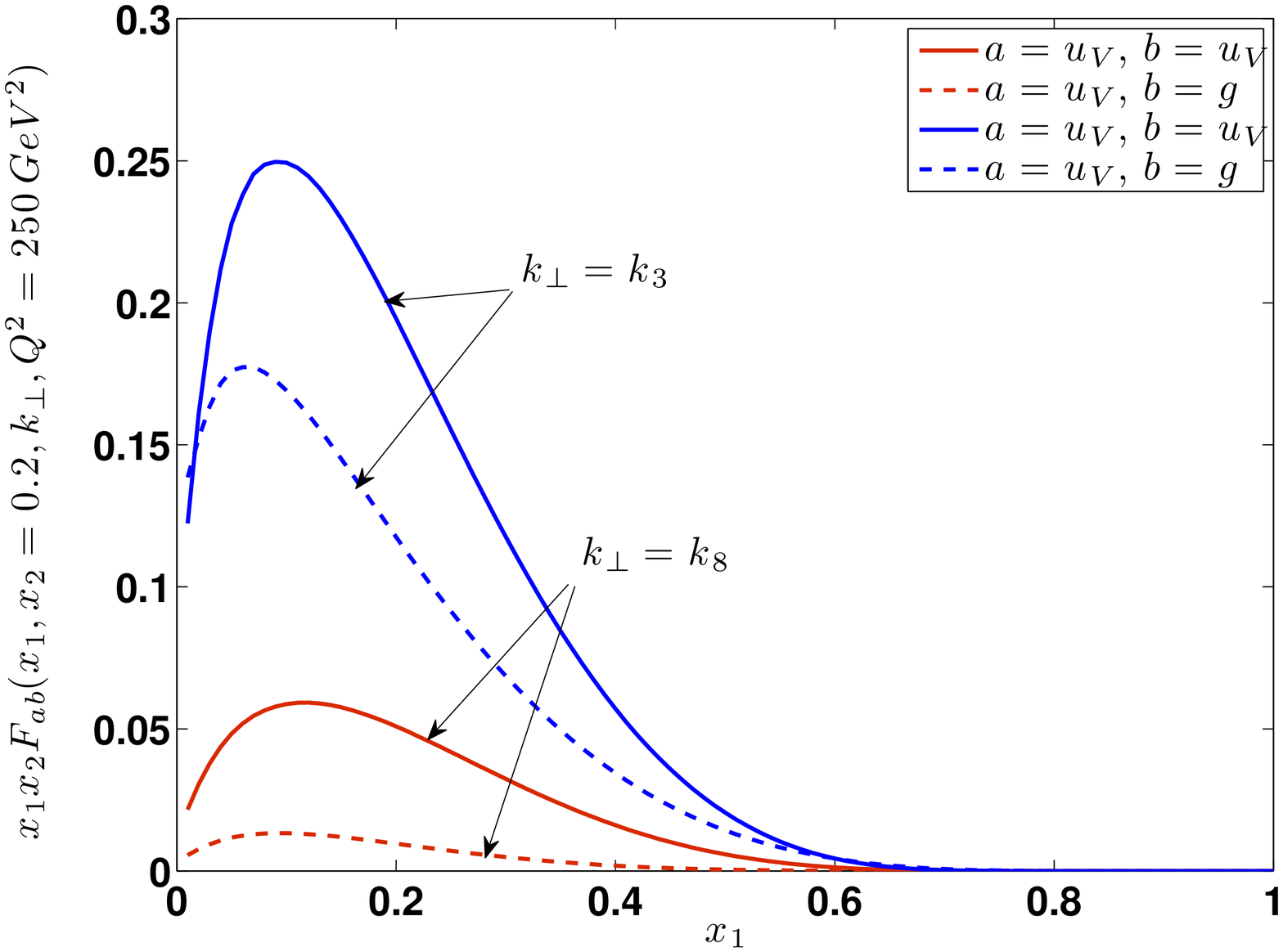}
\centering\includegraphics[width=8.2cm,clip=true,angle=0]
{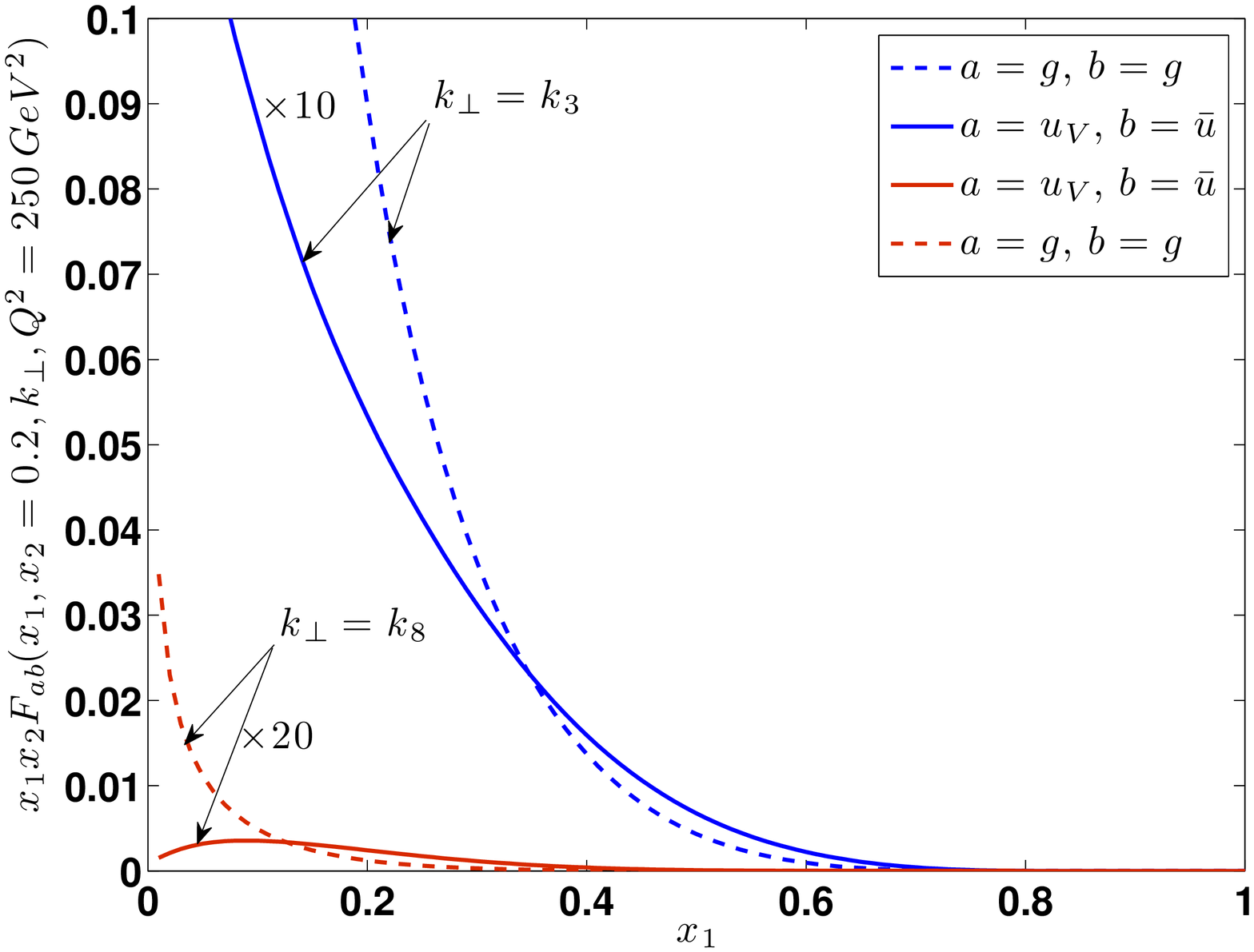}
\caption{\small  {Left panel:} The $x_1x_2F_{u_V u_V}$  
dPDFs obtained from perturbative evolution from the $Q_0^2$ scale to 
$Q^2 = 250$ GeV$^2$ (continuous lines) as function of $x_1$ and at fixed 
$x_2 = 0.2$ (and for two values of $k_\perp$), are compared with  
$x_1x_2 F_{u_V g}$ dPDFs (dashed lines) at the same high scale.{ Right panel:} 
The $x_1x_2 F_{u_V \bar u}$ dPDFs (amplified 10 and 20 
times respectively) at the scale $Q^2 = 250$ GeV$^2$ are compared with 
the $x_1x_2 F_{gg}$ correlations at the same scale, and kinematical 
conditions. 
($k_3 \simeq 0.32$ GeV and $k_8 \simeq 1.68 $ GeV). {The only difference of 
the present Figure with 
Fig. \ref{fig:xxFabmu0Q2k_3_8_x202} is represented by the starting scale 
$Q_0^2 = 1.0$ GeV $\!\!^2$ $> \mu_0^2$}. As a consistency check one can 
verify that the results obtained in the two Figures for $x_1x_2 F_{u_V u_V}$ 
are exactly the same. See text.}
\label{fig:xxFVV-Vg_Vub-gg_Q0Q2k_3_8_x202}
\end{figure}

\noindent where  $u_Vu_V(x_1, x_2,k_\perp,Q_0^2)$ 
is obtained evolving at the scale of the MSTW parametrization, 
$Q_0^2 = 1.0$ GeV$^2$, the LF result (\ref{eq:uVuVpol}), (\ref{eq:uVuVunpol})  
at fixed $k_\perp$. Besides, $u_V(x,Q_0^2)$ is the PDF 
obtained at the same 
scale within the LF approach,  and $\bar u(x,Q_0^2)$ is taken
from the LO MSTW parametrization \cite{MSTW2008}.

By means of the perturbative evolution developed in Appendix \ref{App:pQCD} 
one can now evolve the distribution calculated at low-momentum scale to a 
typical experimental scale. We evolve to $Q^2 = 250$ GeV$^2$, a scale relevant 
to study properties of dPDFs, as shown by
experiments \cite{afs,data0,data2,data3,data4,data5} and by a 
quite recent theoretical study within 
the LF approach \cite{plb}. In two series of figures,
(Figs. \ref{fig:xxFabmu0Q2k_3_8_x202} and 
\ref{fig:xxFVV-Vg_Vub-gg_Q0Q2k_3_8_x202}),
we compare the results obtained evolving directly from the lowest scale 
$\mu_0^2$ where only valence-valence dPDFs are present (cfr. 
Fig. \ref{fig:xxuVuV-kperp}), with those obtained with
the evolution from the scale of the MSTW 
parametrization $Q_0^2 = 1$ GeV$^2$, where also gluon and sea 
dPDFs contribute (cfr. Fig. \ref{fig:xxFuuQ0k_3_8_x20204}). The presence 
of the additional Singlet components is quite relevant, in particular for 
those components containing sea and gluon degrees of freedom, as it appears 
clearly from the comparison of the set of Figures. The Singlet 
components parametrized by means of the factorization procedure 
appear to play a relevant role at low $x$, where the dPDFs can be more easily 
studies by means of proton-proton collisions at very high energy.  
On the other hand the evolution obtained from the lowest 
momentum scale 
has the merit of being directly connected with quark dynamics and 
correlations are generated in a transparent way. 
A detailed study of  the interrelations between non-perturbative correlations,
generated by the dynamics of the model, and perturbative ones,
generated by QCD evolution, is performed,
at low-$x$, in the next sections.

\section{\label{sec:v-ratios} Perturbative and non-perturbative 
two-parton correlations at low-$x$}

In this section we present results obtained within our
LF scheme, aimed at establishing
what kind of error one can do if two-parton correlations are neglected
in treating dPDFs, for example in analyzing collider data.
In previous papers of ours \cite{noi2,plb} we have already emphasized that
this error can be rather sizeable when $x_1, x_2$ lie 
in the valence region. We want now to analyze the low $x$ scenario,
reaching $x$ values as low as $10^{-2}$,
using a full, non-singlet and singlet, LO QCD evolution to the
very high $Q^2$ scales typical of $pp$ scattering at the LHC.
As in Ref. \cite{dkk},
for the moment being, only the homogeneus part of the evolution
of dPDFs is performed. As already said, the $Q^2$ evolution of the $k_\perp$
dependence has not been investigated yet and is still a missing item
in this phenomenology. 

\subsection{
\label{sec:prima}
Characterizing the two-parton correlations at low-$x$
}

To study the relevance of two-parton correlations at low-$x$, 
we found very helpful to show ratios of dPDFs to 
products of PDFs;
in the case of gluon distributions, for example at $x_2=0.01$, this ratio reads
\bea
{\rm ratio}_{gg}(x_1,x_2 = 0.01,k_\perp=0,Q^2) =
 {F_{gg}(x_1,x_2=0.01,k_\perp=0,Q^2) \over g(x_1,Q^2) 
\cdot g(x_2=0.01,Q^2)}~,
\label{eq:ratiogg}
\eea
where $Q^2$ is a phenomenologically relevant scale, chosen
in the following to be $Q^2 = 250$ and $10^4$ GeV$^2$.
These scales are
reached by performing QCD evolution 
of the results obtained within our LF scheme for both sPDFs and sdPDFs,
starting from the hadronic scale
$\mu_0^2$, where only valence degrees of freedom are present.
It is clear that this ratio would be just 1 if it were possible
to approximate the dPDF with the product of two sPDFs.
The difference from 1 of the ratio is a measure of the error
which is done by using that approximation, which amounts to
disregard any kind of two-parton correlations.

In general the ratio can be written
\bea
&&{\rm ratio}_{ab}(x_1,x_2 = 0.01,k_\perp=0,Q^2) = \nonumber \\
&= & {F_{ab}(x_1,x_2=0.01,k_\perp=0,Q^2) + a \to b \over a(x_1,Q^2) 
\cdot b(x_2=0.01,Q^2)+a \to b}~,
\label{eq:ratioab}
\eea
including other kind of partons; in the following, we will analyze
the {\em selected} combinations 
\bea
\{a b\} & = & \{u_V u_V \}, \{u_V  g + g u_V\}, \{ u_V  \bar u 
+ \bar u u_V\}, \nonumber \\
&& \{g g \}, \{ \bar u \bar u\} 
\eea

The symmetrization is mandatory from the point of view of the experimental 
measurements, which cannot distinguish the two combinations. 
Obviously $u_V$ is a $Non-Singlet$-index, as well as $g$ is a $Singlet$-index,
while the sea indexes have no fixed flavor-symmetries;
the different distributions evolve following the corresponding
equations, as discussed in  Appendix \ref{App:pQCD}.
\begin{figure}[tbp]
\centering\includegraphics[width=8.2cm,
clip=true,angle=0]{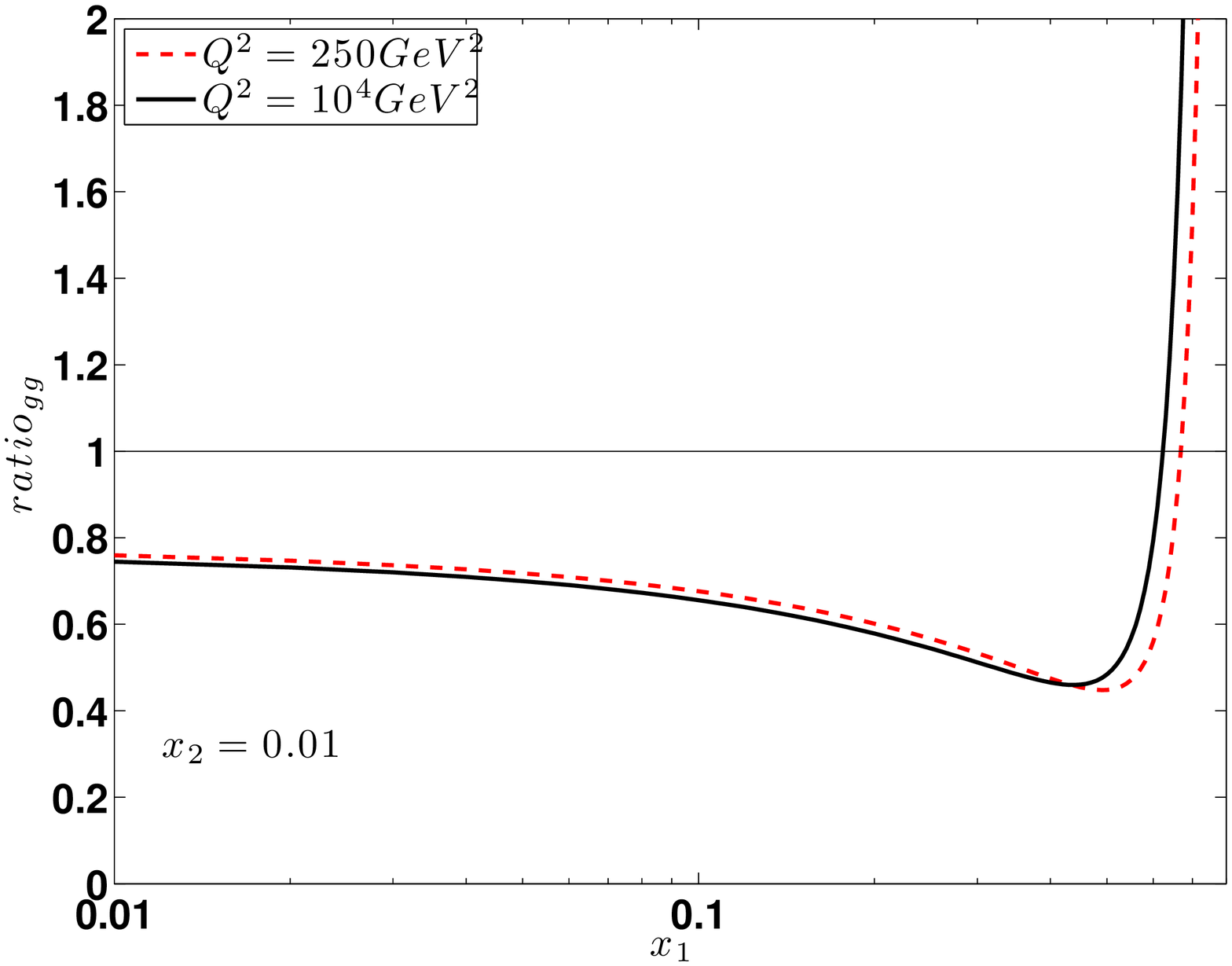}
\centering\includegraphics[width=8.2cm,clip=true,angle=0]
{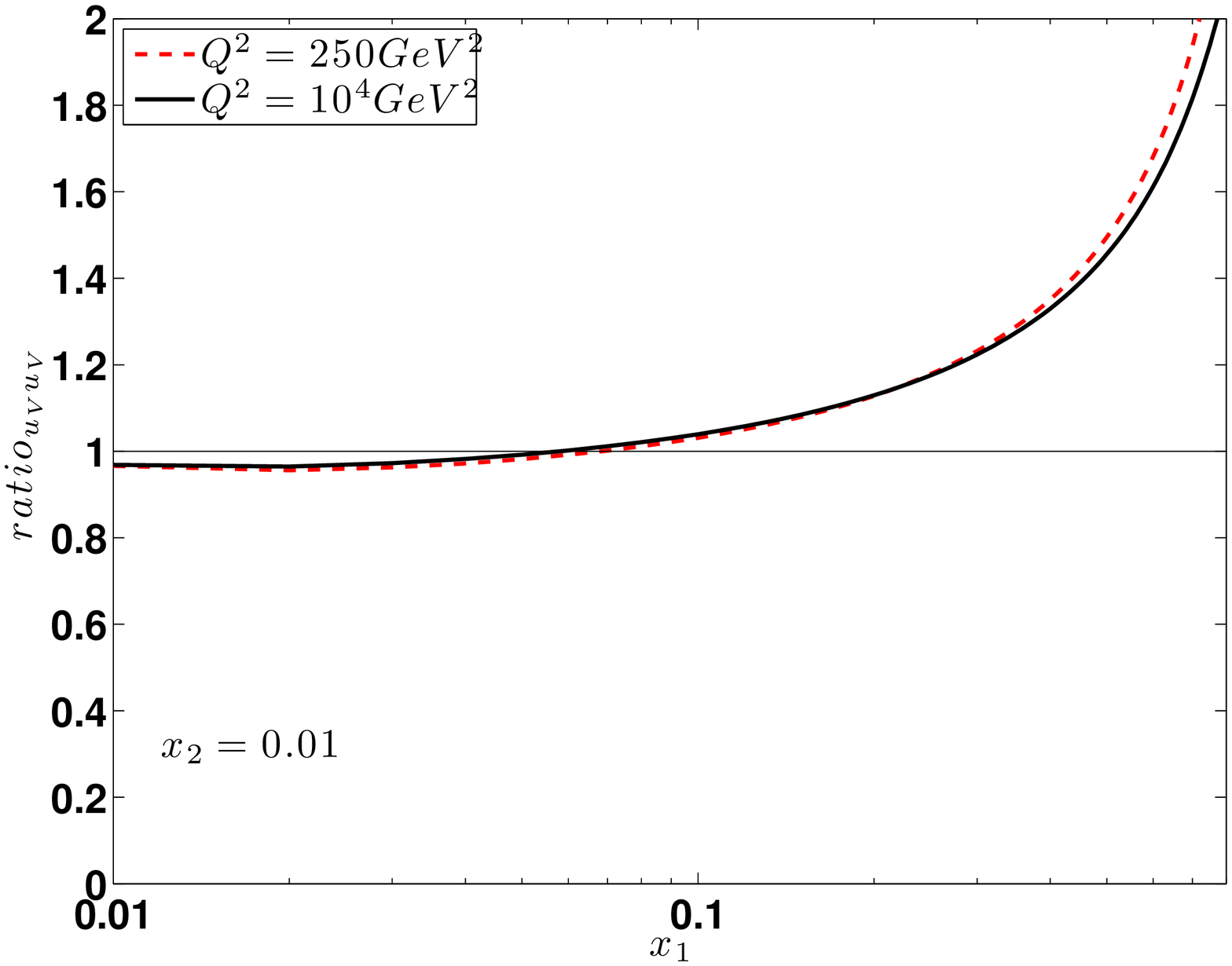}
\caption{\small  {Left panel:} The ratio 
Eq. (\ref{eq:ratiogg}) at different values of $Q^2$ as function of $x_1$ 
at fixed $x_2 = 0.01$. Numerator and denominator are evolved by means of 
dPDF evolution and single parton evolution, respectively. 
The starting point is the low momentum scale $\mu_0^2$. {Right panel:} The same 
ratio 
for the 
valence-valence components within 
the same kinematical and dynamical conditions.}
\label{fig:ratiogg&uVuV_D}
\end{figure}

\begin{figure}[tbp]
\centering\includegraphics[width=8.2cm,clip=true,angle=0]
{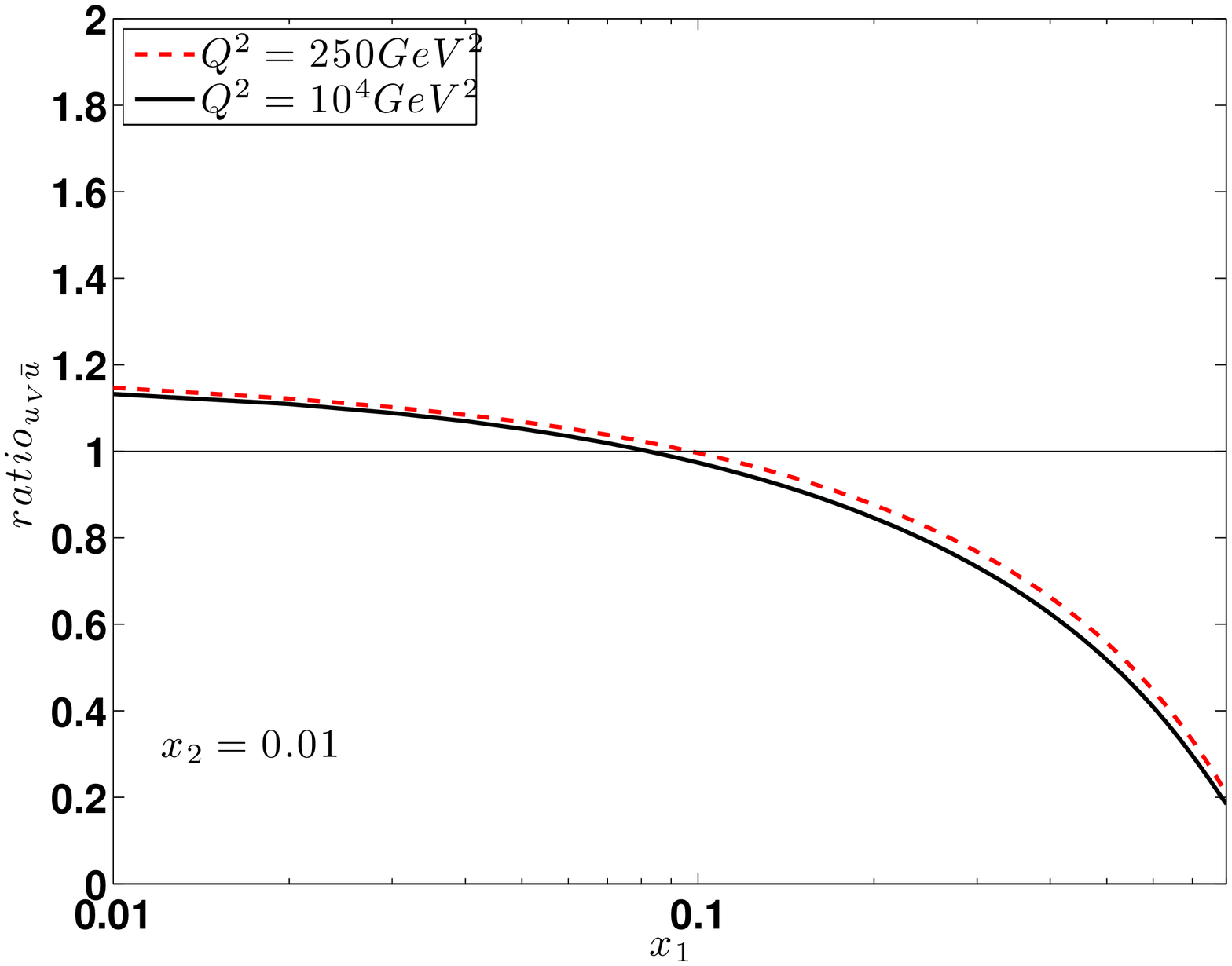}
\centering\includegraphics[width=8.2cm,clip=true,angle=0]
{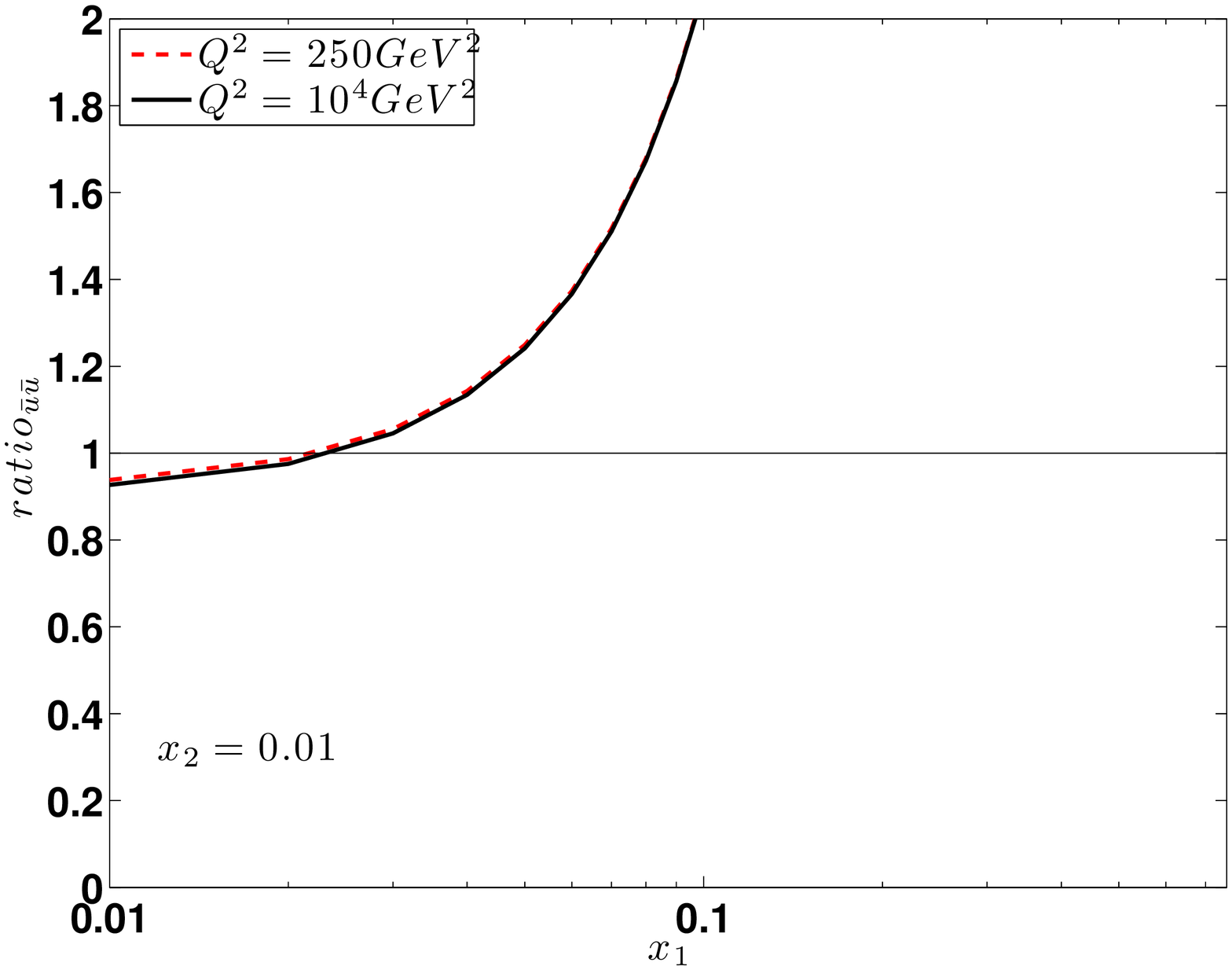}
\caption{\small  As in Fig. \ref{fig:ratiogg&uVuV_D} for the 
ratio involving valence - sea ({left panel}) or sea - sea correlations ({right 
panel}). Notations as in Fig. \ref{fig:ratiogg&uVuV_D}.}
\label{fig:ratioubub&uVub_D}
\end{figure}
\begin{figure}[tbp]
\centering\includegraphics[width=8.2cm,clip=true,angle=0]
{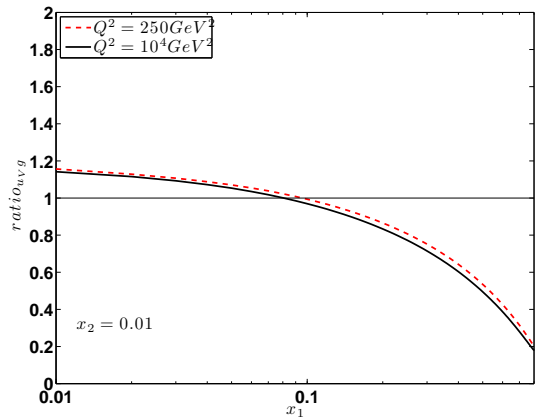}
\caption{\small   As in previous Figures for the ratio involving valence - gluon 
correlations. Notations as in Fig. \ref{fig:ratiogg&uVuV_D}.}
\label{fig:ratiouVg_D}
\vspace{-1.0em}
\end{figure}

Results of the ratio Eq. (\ref{eq:ratioab})
for the flavor combinations $gg$, $u_V u_V$, $u_V \bar u$, $\bar u \bar u$,
$u_V g$
are shown in Figs. \ref{fig:ratiogg&uVuV_D}  -- \ref{fig:ratiouVg_D}.

All the ratios have two common qualitative features:

\begin{itemize}
\item[i)]
results at $Q^2$ = 250 GeV$^2$ do not really differ from those
at $Q^2 = 10^4$ GeV$^2$; the role of correlations does not depend
therefore on the different high momentum scale which is chosen;
 
\item[ii)]
in all flavor combinations, when at least one of the
momentum fractions of the two partons is in the valence
region, correlations are strong and the error which is
done in approximating a dPDF with a product of sPDFs is huge.  

\end{itemize}

When both the momentum fractions of the partons are small, the situation
is more involved. In the valence-valence sector, one finds
negligible correlations and the ratio is basically 1
(cf. Fig. \ref{fig:ratiogg&uVuV_D}, right panel). 
This fact, in the Non-Singlet (NS) sector,
had been already found and discussed in Ref. \cite{noi2}.
In all other cases, where singlet evolution
is playing a role, even at values of $x_1, x_2$ as low as $10^{-2}$,
correlations are found to produce sizable deviations of the ratios from 1.
The maximum effect is found in the gluon-gluon case (cf. Fig. 
\ref{fig:ratiogg&uVuV_D}, left panel), when it reaches 20 $\%$.
One should realize that, if two-parton correlations were present at 
the LHC scale, one could access through
DPS studies novel information on the proton structure.
Our evolved model results show that if one were able to measure
dPDFs at a 20 $\%$ accuracy, a specific dynamical information
would be reachable.
The different behavior of the valence-valence sector from the others,
as well as the fact that the gluon-gluon sector experiences
the biggets effect, are interesting features of our results
and deserve to be understood through a further investigation. 
This is carried on in the next section.

\subsection{\label{sec:PnP} Perturbative versus Non-Perturbative 
Two-Parton Correlations}

In this subsection we will find that the results described in the
previous one can be understood
by disentangling perturbative and non-perturbative effects.

To this aim,
let us consider again the ratio Eq. (\ref{eq:ratioab}) 
\bea
&&{\rm ratio}_{ab} = \nonumber \\
&\phantom{=} & = {F_{ab}(x_1,x_2=0.01,k_\perp=0,Q^2) 
+ a \to b \over a(x_1,Q^2) \cdot b(x_2=0.01,Q^2)+a \to b}\,.
\label{eq:ratioab2}
\eea
In the previous subsection \ref{sec:prima}, results are obtained 
evolving the numerator from $\mu_0^2$ to $Q^2$, considering at the lowest 
scale the dPDFs predicted by our LF-model. The denominator is obtained 
evolving to $Q^2$ the analogous sPDFs of the same LF-model. 

A first  consideration is in order: if the denominator, 
given by the product of single PDFs,
had been evolved by means 
of dPDF-evolution criteria, we would have obtained a simplified approximation 
of the dPDFs at $Q^2$, including {\em perturbative correlations only}.
 
Let us define {{the following quantity}}
\bea
&& \left. F_{ab}(x_1,x_2=0.01,k_\perp=0,Q^2) \right|^{Perturbative} = 
\nonumber \\
&& \left[a(x_1,Q^2) \cdot b(x_2=0.01,Q^2)\right] ^{dPDF \, evolution}\,.
\label{eq:Fab-Perturbative}
\eea

In fact, $\left. F_{ab}(x_1,x_2=0.01,k_\perp=0,Q^2) \right|^{Perturbative}$ 
contains those correlations which come from {{dPDF}} 
perturbative evolution only.

At this point, we could consider three different ratios:

\begin{itemize} 

\item[i)]  the ratio$_{ab}$, Eq. (\ref{eq:ratioab2});

\item[ii)]  the ratio$_{a b}^{Perturbative}$:
\bea &&{\rm ratio}_{ab}^{Perturbative} = \nonumber \\
&= & {\left.F_{ab}(x_1,x_2=0.01,k_\perp=0,Q^2)\right|^{Perturbative} 
+ a \to b \over a(x_1,Q^2) \cdot b(x_2=0.01,Q^2)+a \to b}\,, \nonumber \\
\label{eq:ratioabPerturbative}
\eea
which contains perturbative correlations only; in fact it would be 
strictly 1 if the dPDF-evolution did not include double-parton correlations 
(see the definition Eq. (\ref{eq:Fab-Perturbative})); 

\item[iii)] the ratio$_{a b}^{Non-Perturbative}$
\bea &&{\rm ratio}_{ab}^{Non-Perturbative} = \nonumber \\
&= & {F_{ab}(x_1,x_2=0.01,k_\perp=0,Q^2) + a \to b \over 
\left.F_{ab}(x_1,x_2=0.01,k_\perp=0,Q^2)\right|^{Perturbative}+a \to b}\,,
\nonumber \\
\label{eq:ratioabNonPerturbative}
\eea
which would be strictly 1 if {\em only} perturbative correlations were 
included in the numerator.
\end{itemize}
\begin{figure}[tbp]
\centering\includegraphics[width=8.2cm,clip=true,angle=0]
{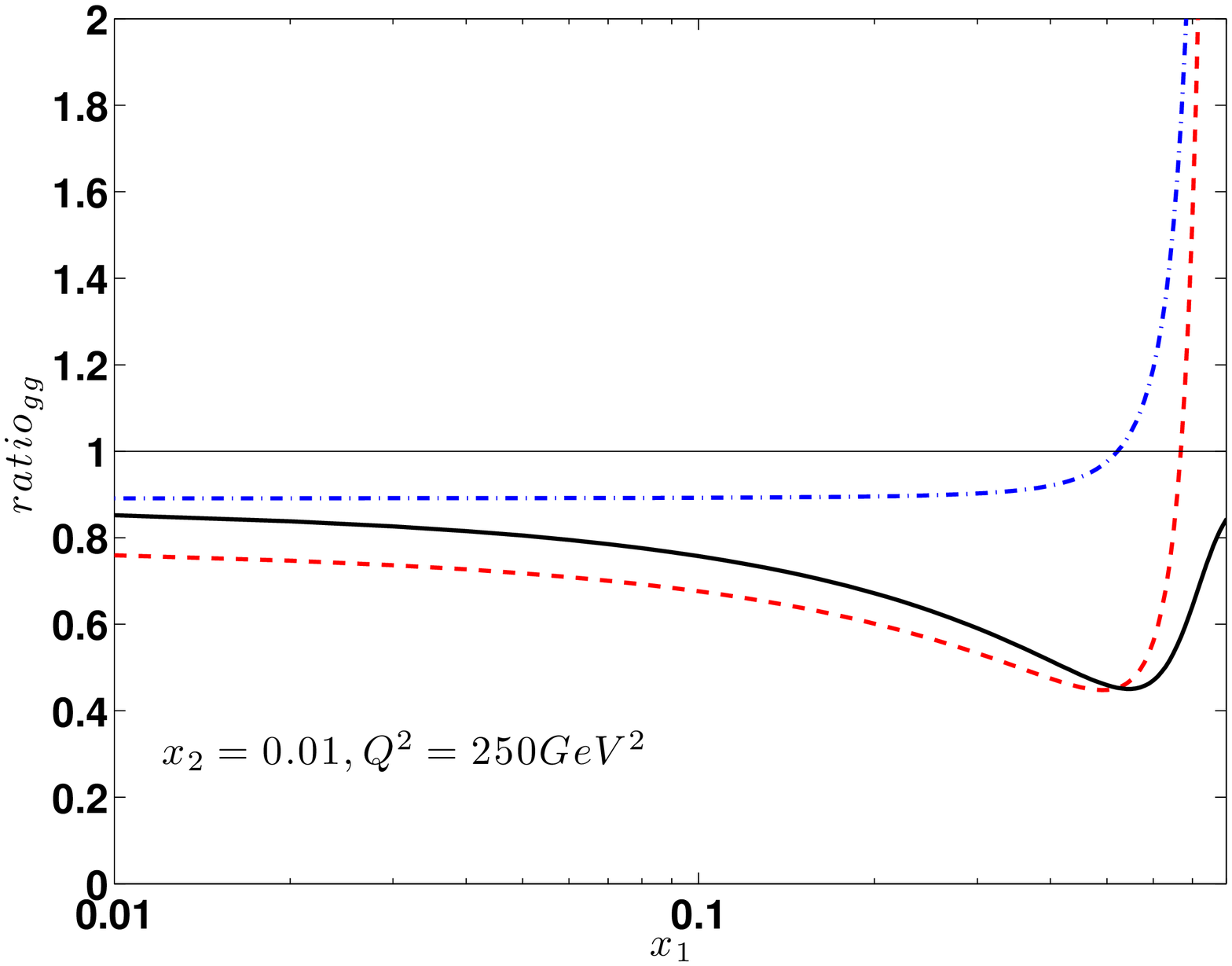}
\centering\includegraphics[width=8.2cm,clip=true,angle=0]
{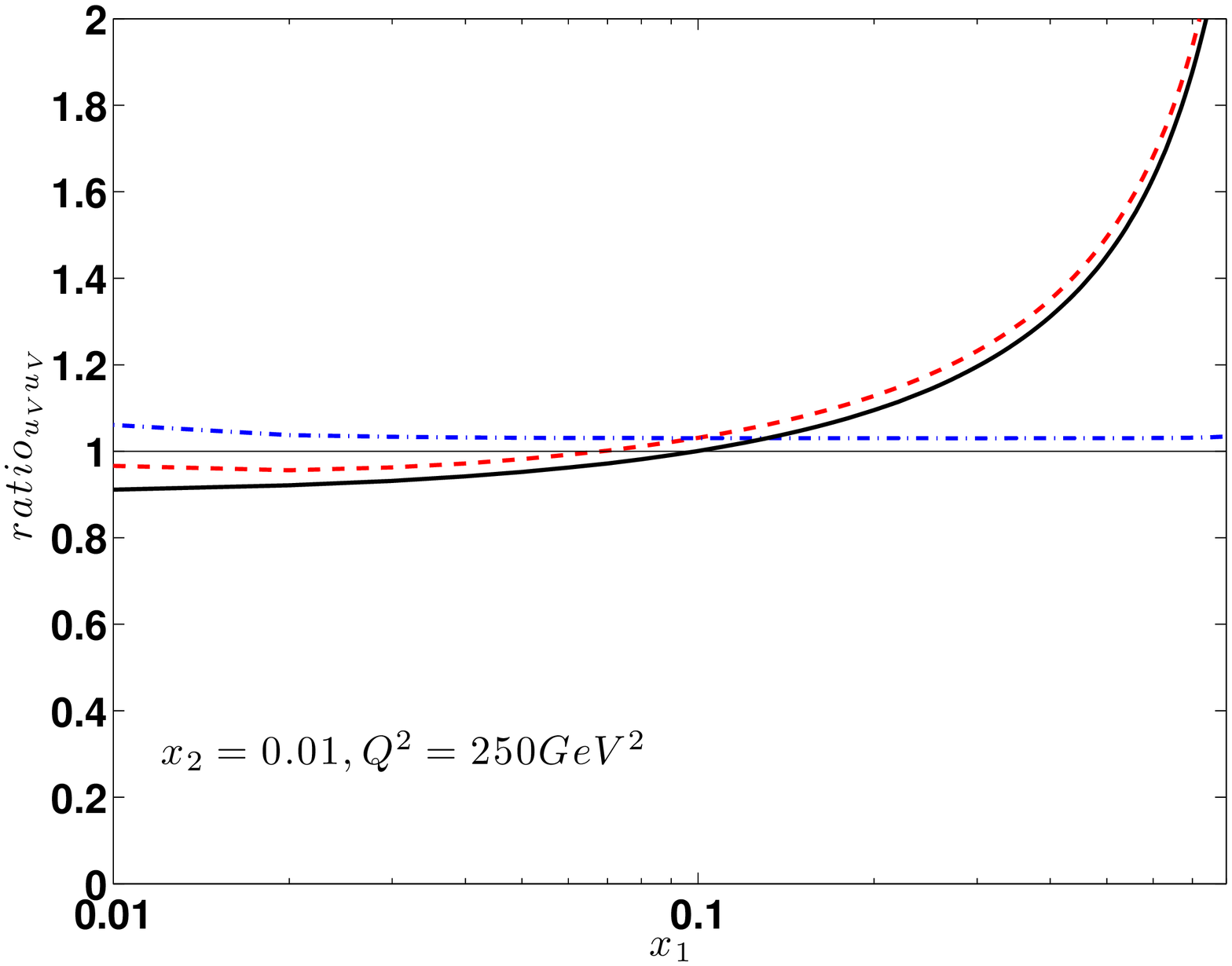}
\caption{\small  {Left panel:} The ratios 
(\ref{eq:ratioab2}) (dashed lines), (\ref{eq:ratioabPerturbative}) 
(dot-dashed lines) and (\ref{eq:ratioabNonPerturbative}) (continuous lines) 
specified for  $(a b) = (g g)$ and at $Q^2 = 250$  GeV$^2$ as function of 
$x_1$ at fixed $x_2 = 0.01$. {Right panel:} The same ratios for the 
valence-valence components 
$(a b) = (u_V u_V)$ within the same kinematical and dynamical conditions.}
\label{fig:ratiogg&uVuV_mCS}
\end{figure}

The three ratios are very useful to disentangle the effects of 
perturbative versus 
non-perturbative double-parton correlations; of course the ratios 
(\ref{eq:ratioab}) or (\ref{eq:ratioab2})  are the most complete, 
including both kind of correlations in a consistent way.

\begin{figure}[tbp]
\centering\includegraphics[width=8.2cm,clip=true,angle=0]
{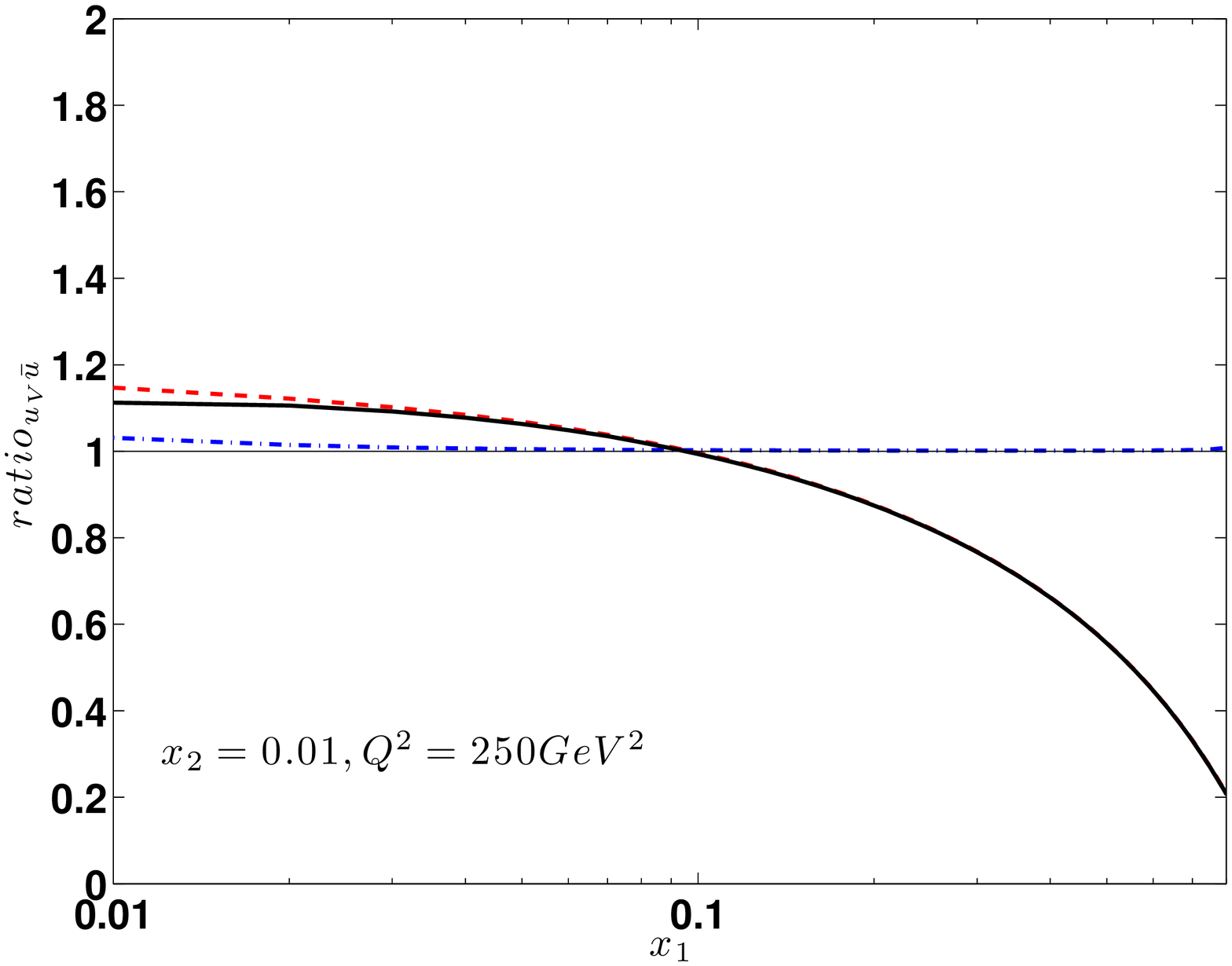}
\centering\includegraphics[width=8.2cm,clip=true,angle=0]
{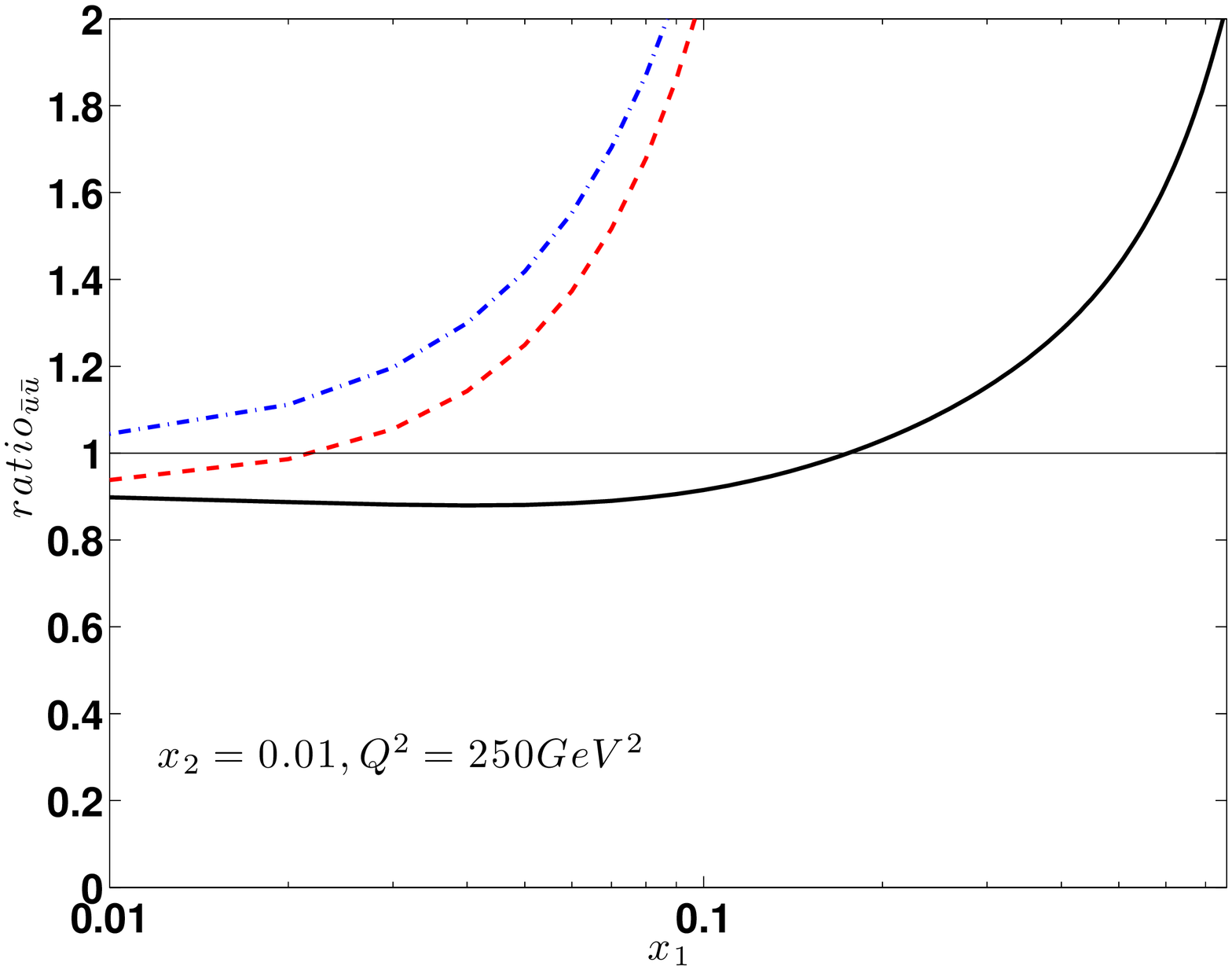}
\caption{\small  As in 
Fig. \ref{fig:ratiogg&uVuV_mCS} 
for the ratio involving valence - sea ({left panel}) or sea - sea correlations 
({right panel}). Notations as in Fig. \ref{fig:ratiogg&uVuV_mCS}.}
\label{fig:ratioubub&uVub_mCS}
\end{figure}

\begin{figure}[tbp]
\centering\includegraphics[width=8.2cm,clip=true,angle=0]
{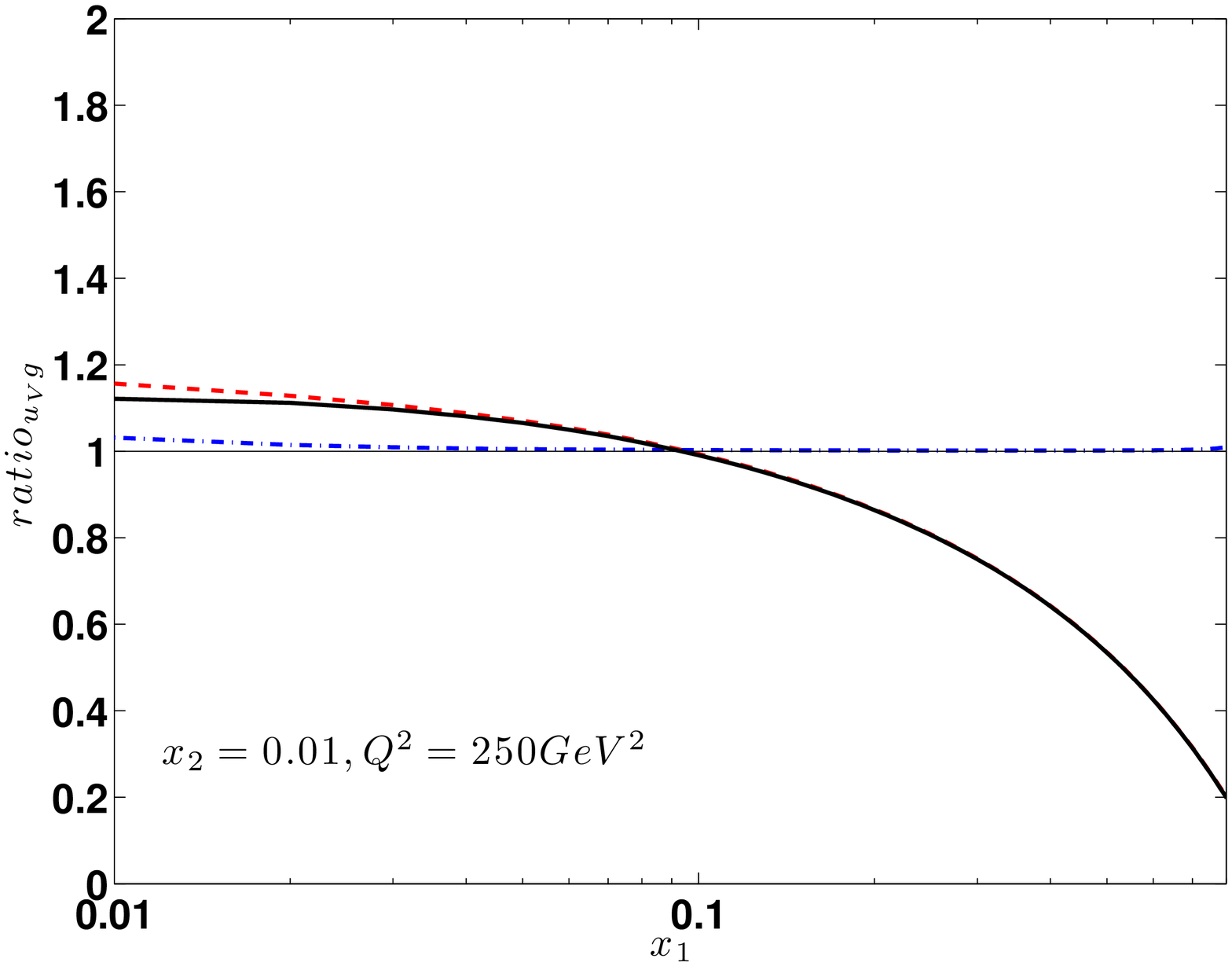}
\caption{\small   As in Figs. 18 and 19, 
for the ratio involving valence - gluon correlations. 
Notations as in Fig. \ref{fig:ratiogg&uVuV_mCS}.}
\label{fig:ratiouVg_mCS}
\vspace{-1.0em}
\end{figure}

In Figs. \ref{fig:ratiogg&uVuV_mCS}, \ref{fig:ratioubub&uVub_mCS} 
and \ref{fig:ratiouVg_mCS}, 
the results for the three ratios are compared at 
the scale $Q^2 = 250$ GeV$^2$, at $x_2 = 0.01$, as functions of $x_1$.

The ratio$_{gg}$, shown in Fig. \ref{fig:ratiogg&uVuV_mCS} (left panel),
is particularly emblematic. The full ratio$_{gg}$ of Eq. (\ref{eq:ratioab2}) 
(dashed line), clearly influenced by both perturbative (dot-dashed line) 
and non-perturbative (continuous line) effects, is compared with those
where perturbative and 
non-perturbative correlations are disentangled,
contributing to the behavior of 
$gluon-gluon$ dPDFs at low values of $x_1$ and $x_2$.
The same comments hold for dPDFs corresponding to the other partons.
An interesting feature of these results, clearly read
in Figs. \ref{fig:ratiogg&uVuV_mCS}, \ref{fig:ratioubub&uVub_mCS} 
and \ref{fig:ratiouVg_mCS}, is that
in few cases the perturbative and non-perturbative components tend to cancel 
(in the case of  $Valence-Valence$ illustrated in 
Fig. \ref{fig:ratiogg&uVuV_mCS} (right panel), 
or $\bar u-\bar u$, as it borns out from Fig. \ref{fig:ratioubub&uVub_mCS} 
(right panel)). In the case of the gluon-gluon sector,
the effect tends instead to sum coherently: this explains
the persistence of correlations in this sector, even at high $Q^2$
and low $x$, observed in the previous subsection.

\begin{figure}[tbp]
\centering\includegraphics[width=8.2cm,clip=true,angle=0]
{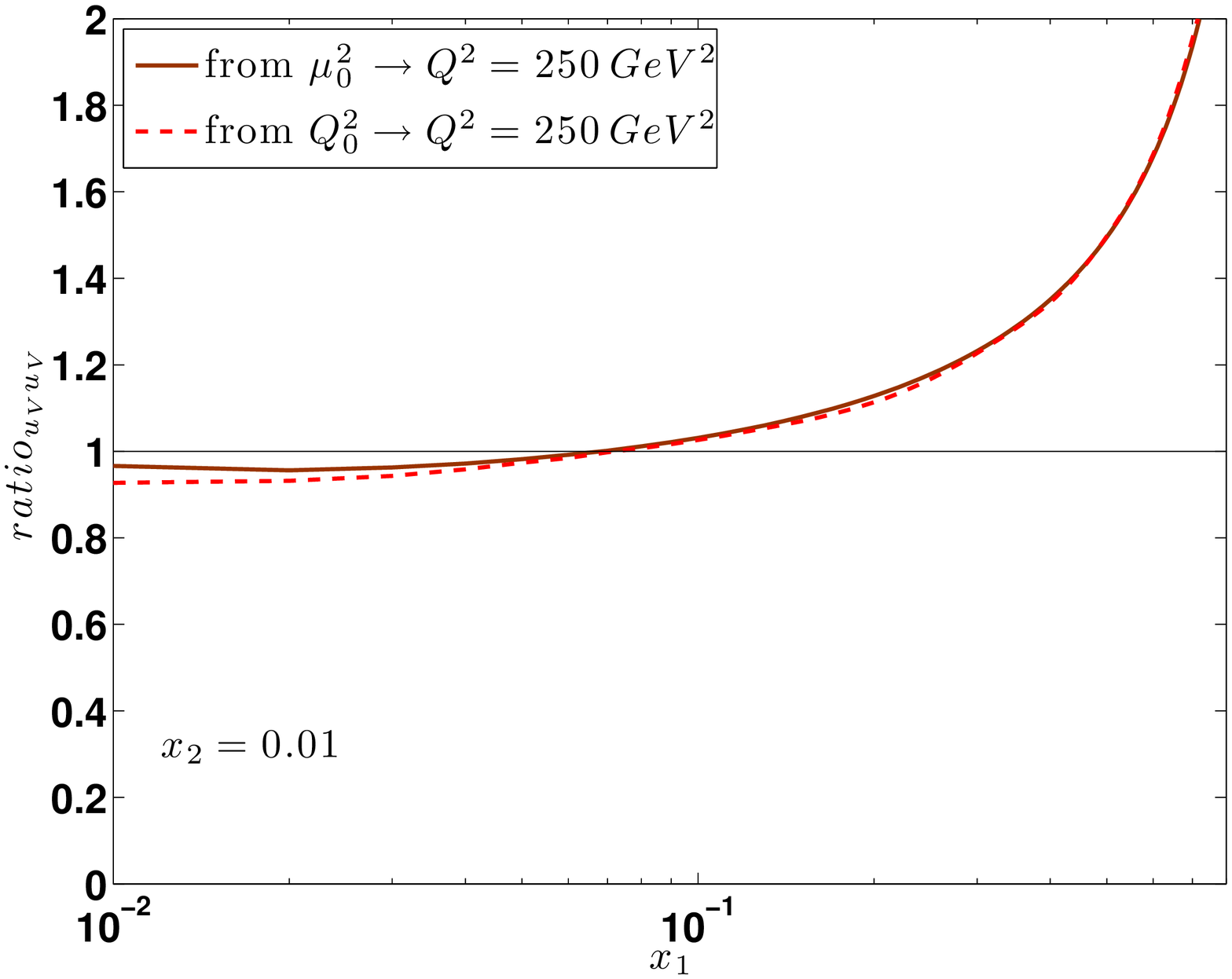}
\centering\includegraphics[width=8.2cm,clip=true,angle=0]
{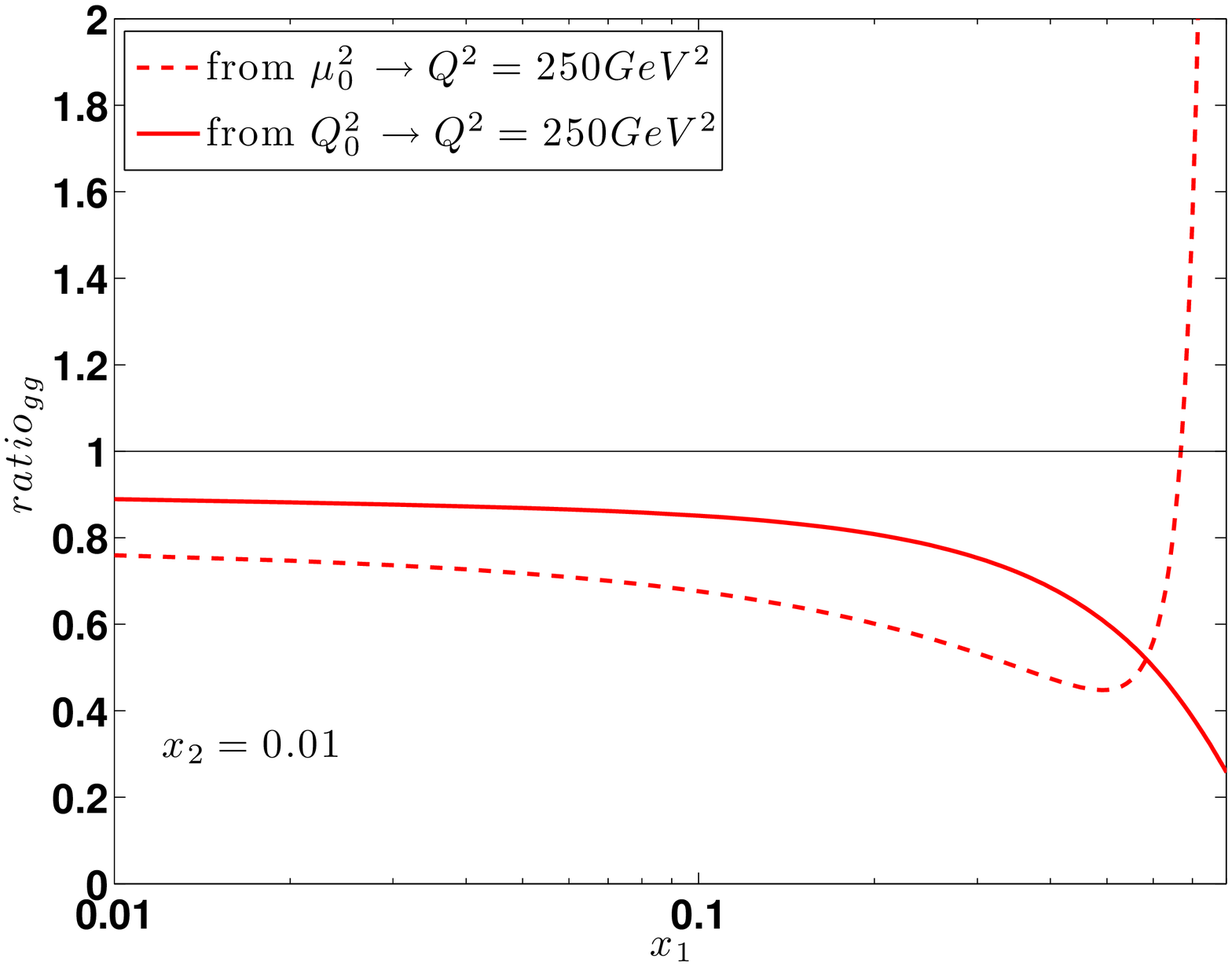}
\caption{\small  {Left panel:} The ratio $ratio_{u_V u_V}$ at 
$Q 2 = 250$ Gev$^2$
as function of $x_1$ at fixed $x_2 = 0.01$. Numerator and denom-
inator are evolved by means of dPDF evolution and single
parton evolution, respectively. The starting points differ for
the two curves: the low momentum scale $\mu_0^2$ (continuous) and the larger
$Q^2_0$ (dashed) . The differences are artifacts due to numerical
uncertainties. Right panel: the same of the left panel but for the 
$ratio_{gg}$.}
\label{fig:Q0}
\end{figure}

In closing this Section, we conclude that correlations
in dPDFs, for some flavor combinations, are present
also at low $x_1$ and $x_2$, even
at the large energy scale of LHC experiments.
This arises because perturbative and non-perturbative effects 
sum coherently. These conclusions
are not artifacts of the specific LF model used.
They hold qualitatively 
also in ratios obtained starting the evolution
from $Q_0^2=1$ GeV$^2$ $>> \mu_o^2$,
using as non-perturbative input the semi-factorized
model of Section IV.
{ In order to illustrate this important point, two more plots are 
included (Fig. (\ref{fig:Q0})). In the first one, the valence-valence ratio is 
shown, in the other the gluon-gluon one. These examples are illustrative, 
indeed, of 
two specific aspects: i) the valence-valence ratios should not depend on the 
starting point because they converge at the same values at the common hadronic 
scale $\mu_0^2$. The small differences which appear in the figures are 
therefore a clear estimate of the errors introduced by our numerical evolution 
and one can appreciate the precision of our approach; ii) the second figure, 
showing the gluon-gluon ratio, is  included because the glue is the dominant 
component at low-$x$ and it contributes in a negligible way to the 
valence region. The correlations induced at low-$x$ still contain a specific 
sign of the correlations introduced in the valence sector and this is due to the 
presence of the valence component in the quark-singlet sector in the evolution 
procedure. The strength of the correlation seems to become smaller but they are 
still 
sizable.}


\section{Conclusions}

Double Parton Scattering (DPS) represents a background in 
several channels for the search of new Physics at the LHC. Its correct description
depends on our ability of modelling double parton distribution functions
(dPDFs).
The knowledge of these quantities would represent also a
novel tool for the study of the three-dimensional nucleon structure,
complementary to possibilities offered by electromagnetic interactions,
in the framework of Generalized Parton Distribution functions.
In this paper we have analyzed dPDFs, using Poincar\'e covariant
predictions obtained, at a low energy scale, 
within a Light-Front model proposed by us in a
recent paper, evolved using
QCD evolution to experimentally relevant scales.
We checked to what extent factorized expressions of dPDFs,
in terms of products or convolutions of one-body densities,
can be used, neglecting, at least in part, two-parton correlations.
Our tests were performed using our model predictions starting
from a scale where only quark degrees of freedom are relevant,
or from  higher scales, modeling sea quark and gluon
contributions.
Our model study demonstrates that factorization procedures
strongly fail in reproducing the calculated dPDFs in the
valence region, where measurements of DPS could really allow
to access two-parton correlations.
Besides, a gaussian behavior for the transverse distance
in coordinate or momentum space seems rather arbitrary.
Anyway, to have contact with measurable processes at existing
facilities, everything has to be pushed to very low values of the 
longitudinal momenta of the interacting partons.
This study has been carried on systematically and represents
the most interesting part of our investigations.
Correlations between pairs of partons of different kind have
been considered, finding that, in some cases, their effect tends
to be washed out at low-$x$, as it happens for the valence,
flavor non-singlet distributions, while they can affect
other distributions in a sizable way, as in the gluon sector, when
they can be as large as 20 $\%$.
We have shown that this different behavior can be understood
in terms of a delicate interference of non-perturbative correlations, 
generated by the dynamics of the model, and perturbative ones,
generated by the model independent evolution procedure.
Our analysis shows that at LHC two-parton correlations can
be relevant in DPS, opening a possibility to observe them for
the first time. Our model dPDFs have now to be used
to predict cross sections in specific channels where
DPS is known to give an important contribution, such as,
for example, the production of two $W$ bosons with the same sign.
Our research is now addressing this final goal.
 
\section{Acknowledgements}

This work was supported in part  
by the Mineco under contract FPA2013-47443-C2-1-P, 
by GVA-PROMETEOII/2014/066 and SEV-2014-0398.
S.S. thanks the Department of Theoretical Physics of 
the University of Valencia for warm hospitality and support. 
M.T. and V.V. thank the INFN, sezione di Perugia and the Department 
of Physics and Geology of the University of Perugia for 
warm hospitality and support. Several discussions with F.A. Ceccopieri 
are gratefully acknowledged.


\appendix


\section{\label{App:pQCD} Appendix:  Perturbative evolution of dPDFs in Mellin 
space}

Following Diehl and Kasemets \cite{DiehlKasemets2013} one has to admit  that 
{\it ``a consistent formulation of factorization for double parton scattering 
does not yet exist, so that it remains unclear how dPDFs should best be 
defined (and how they evolve)"}.  
However some phenomenological aspects of QCD-evolution are known since long 
time (e.g Refs. \cite{Kirschner,ShelestSnigirevZinovjev1982}) and have been 
recently 
retaken \cite{Gauntetal,cec1,cec2}
developing  numerical codes able to solve the evolution equations. In addition 
also theoretical progresses 
have been reported (for example 
the demonstration that the exchange of Glauber gluons cancels for the 
considered observable, a step forward {{in the proof of QCD
factorization for DPS}}
\cite{DiehlGauntOstermeierPloesslSchafer2015}). 

In the following we develop a systematic numerical approach to the 
evolution of dPDFs,  in Mellin space instead of coordinate space, 
restricting ourselves to the, so called, homogenous equation, a restriction 
we share with numerical solutions in coordinate space as applied in several 
contributions by Diehl and other coauthors (see Ref. \cite{Diehl_SantaFe2014} 
and reference therein). 

If we assume equal renormalization scales $Q_1$ and $Q_2$ for the two partons 
(i.e. $Q_1=Q_2=Q$), the LO evolution equation for the unpolarized 
double parton distributions $F_{j_1j_2}(x_1,x_2;  Q^2)$ then reads 
(see Ref. \cite{snigsnigzin})
\bea
&& \left.{d F_{j_1j_2}(x_1,x_2;  Q^2) \over d \log Q^2}\right|_{\rm LO}  = {\left. \alpha_s(Q^2)\right|_{\rm LO} \over 4 \pi} \times \nonumber \\
&&\times \left[ \sum_{j_1'} \int_{x_1}^{1-x_2} {d y_1 \over y_1} F_{j_1' j_2}(y_1,x_2; Q^2) P_{j_1' \to j_1}\left({x_1 \over y_1}\right) +\right.  \nonumber \\
& & + \phantom{[[} \left. \sum_{j_2'} \int_{x_2}^{1-x_1} {d y_2 \over y_2} F_{j_1 j_2'}(x_1,y_2; Q^2) P_{j_2' \to j_2}\left({x_2 \over y_2}\right) +\right.\nonumber \\
&& + \left. \sum_{j'} F_{j'}(x_1+x_2; Q^2) {1 \over x_1+x_2} P_{j' \to j_1 j_2}\left(x_1 \over x_1+x_2\right) \right] \nonumber \\ \label{eq:EV_Fijx}
\eea


The convolution integrals  appearing in Eq. (\ref{eq:EV_Fijx}) have the same structure of the integrals appearing in the evolution of the single parton distributions, namely the {\sl renormalization group equation} (RGE). In order to solve evolution equations, one can perform a $Mellin$-transformation of Eqs. (\ref{eq:EV_Fijx}),
in particular for the first  two terms 
\bea
&& \left.{d M_{j_1j_2}^{n_1 n_2}(Q^2) \over d \log Q^2}\right|_{\rm LO}  = {\left. \alpha_s(Q^2)\right|_{\rm LO} \over 4 \pi} \times \nonumber \\
&& \times \left[ \sum_{i} P^{(0)}_{i j_1}(n_1) M_{i j_2}^{n_1 n_2}(Q^2)  + \right. \nonumber \\
&& \left. \phantom{[[}+ \sum_{k} P^{(0)}_{k j_2}(n_2) M_{j_1 k}^{n_1 n_2}(Q^2) \right] + \nonumber \\
&& \left.\phantom{[]} + {\rm inhomogeneous \,term} \right.\label{eq:EV_Fijn}\,,
\eea
where
\bea
M_{j_1 j_2}^{n_1 n_2}(Q^2) & = & \int_0^1 dx_1 \int_0^1 dx_2 \,\, \theta(1-x_1-x_2) \,  \cdot \nonumber \\
&&  \phantom{} \cdot x_1^{n_1-1} x_2^{n_2-1} \,F_{j_1 j_2}(x_1,x_2; Q^2) \label{eq:Momijn1n2}\\ 
P^{(0)}_{i j}(n) & = & \int_0^1 dx\,x^{n-1} \,  P^{(0)}_{i j}(x)
\eea
and the $\theta(1-x_1-x_2)$ appearing in the definition of the moments Eq.
(\ref{eq:Momijn1n2}) is a direct consequence of the limit of integration in 
Eq. (\ref{eq:EV_Fijx}) and the momentum conservation.
$P_{ij}$ are the evolution kernels or splitting functions. They are calculated perturbatively as a series expansion in $a_s(Q^2)=\alpha_s(Q^2)/(4 \pi)$: 
\be
 P_{ij}\left({x \over y},a_s(Q^2)\right) = \sum_{m=0}^\infty
a_s^{m+1}(Q^2) \, P_{ij}^{(m)}\left({x \over y}\right) \,,
\ee
and $m=0$ indicates the Leading-Order contribution.

(Expressions for $P^{(0)}_{ij}$ in the context of dPDFs can be found (e.g.) in Appendix A of Ref. \cite{DiehlKasemets2013}).

\subsection{dPDF (flavor) decomposition and evolution}
\label{sec:decomposition}

In order to solve Eqs. (\ref{eq:EV_Fijn}) one has to combine the flavor indices in a way consistent with evolution, in particular one has to identify combinations evolving as Singlet and Non-Singlet. The combinations depend on the order of the evolution. At LO and NLO  a  useful transformation is the following
\bea
\Sigma & = & \sum_i q_i^+, \;\;\;\;\;\;\;\;\;\; \,V_i = q_i^- , \nonumber \\
T_3 & = & u^+-d^+, \;\;\;\;\;\;\; T_8 = u^+ + d^+ - 2 s^+\,, \label{eq:decompositionNLO} \\
{\rm with} &&  \nonumber \\
q_i^\pm & = & q_i \pm \bar q_i\,; \nonumber
\eea 
and similar combination if one includes heavier quarks (e.g. Ref. \cite{EllisStirlingWebberBook} section 4.3.3). 
For the up and down quarks, $V_i$ corresponds to the valence  contributions $V_u \equiv u_V$, $V_d \equiv d_V$. 
After performing the evolution, the individual quark and antiquark distributions can be recovered using
\bea
\bar u & = & {1 \over 4} \left({2 \over 3} \Sigma + {1 \over 3} T_8 + T_3\right)-{1\over 2}u_V; \nonumber \\
\bar d & = & {1 \over 4} \left({2 \over 3} \Sigma + {1 \over 3} T_8 - T_3\right)-{1\over 2}d_V. \nonumber \\
s + \bar s & = & {1 \over 3} \left( \Sigma - T_8 \right); \label{eq:NLOinversion}
\eea
Specifically, in the case of dPDFs $F_{ij}$, the same argument holds for indices $i,j$ combined in such a way to produce $T_3$, $T_8$ and $V_i$ structures. Consequently, in addition to $F_{u_ V u_V}$, $F_{d_V d_V}$ and $F_{u_V d_V}$, $F_{d_V u_V}$, also combinations like 
\bea
F_{T_3 T_3}, \,\,F_{T_3 T_8},\,\, F_{u_V T_3},\,\,F_{d_V T_3},\,\, F_{u_V T_8},\,\,F_{d_V T_8},
\eea
will evolve following the simple Non-Singlet rules.

Just to give an example, we will discuss, in the next Section, the evolution of the dPDF 
\bea
F_{V_u T_3} & \equiv & F_{u_V (u+\bar u - d - \bar d)} = \nonumber \\
& = & F_{u_V u} + F_{u_V \bar u} - F_{u_V d} - F_{u_V \bar d} = \nonumber \\
& = & F_{u_V u_V} + 2 F_{u_V \bar u }- F_{u_V d_V} - 2 F_{u_V \bar d}\,.\label{eq:VuT3}
\eea


Neglecting the inhomogeneous term, the solution of Eq. (\ref{eq:EV_Fijn}), 
for the Mellin moments of combination Eq. (\ref{eq:VuT3}) is:
\bea
 M_{V_u T_3}^{n_1 n_2}(Q^2) = \left({a_s \over a_{s0}}\right)^{-{P^{(0)}_{qq}(n_1) + P^{(0)}_{qq}(n_2) \over \beta_0}} \cdot  M_{V_u T_3}^{n_1 n_2}(Q_0^2) \, 
\label{eq:Mn1n2_NS-NS_ev}\nonumber \\
\eea
(compare also the definition Eq. (\ref{eq:Momijn1n2})).

The $Mellin$-inversion completes the solution in $x$-space:
\bea
&& F_{V_u T_3}(x_1,x_2,Q^2)  = \nonumber \\
&& = {1 \over 2 \pi i} \oint_{\cal C} dn_1 \,{1 \over 2 \pi i} \oint_{\cal C} dn_2 \,x_1^{(1-n_1)}\, x_2^{(1-n_2)}\, M_{V_u T_3}^{n_1 n_2}(Q^2)\,. \nonumber \\
\label{eq:solutionVuT3n1n2}
\eea

The procedure described for the example $F_{V_u T_3}$, is valid for each $F_{i j}$ combination ($(i,j) = V_i\,, T_3\,, T_8$).

On the other hand, the $double$-distributions containing $gluons$ and $\Sigma$ evolve mixing the two and each 
index must be evolved in the appropriate way. For example:
\bea
 && 
\left(
\begin{array}{c}
M_{V_u \Sigma}^{n_1 n_2}(Q^2)    \\
 \\
 M_{V_u g}^{n_1 n_2}(Q^2)   \\
\end{array}
\right) = \left({a_s \over a_{s0}}\right)^{-{P^{(0)}_{qq}(n_1) \over \beta_0}}  \times \nonumber \\
&& \nonumber \\
&& \times \left(
\begin{array}{cc}
W^0_{qq}(n_2)  & W^0_{qg}(n_2)    \\
& \\
W^0_{gq}(n_2)  & W^0_{gg}(n_2)    \\
\end{array}
\right) \cdot 
\left(
\begin{array}{c}
M_{V_u \Sigma}^{n_1 n_2}(Q_0^2)    \\
 \\
 M_{V_u g}^{n_1 n_2}(Q_0^2)   \\
\end{array}
\right),\label{eq:Mn1n2_NS-S_ev}
\eea
a result valid also replacing the first index $V_u$ with the other Non-Singlet components, namely  $T_3$ or $T_8$.

Last examples the $Singlet-Singlet$ components:
\bea
M_{\Sigma \Sigma}^{n_1 n_2}(Q^2)  & = & W^0_{qq}(n_1)  W^0_{qq}(n_2) 
M_{\Sigma \Sigma}^{n_1 n_2}(Q_0^2) + \nonumber \\
& + & W^0_{qg}(n_1)   W^0_{qq}(n_2) 
M_{g \Sigma}^{n_1 n_2}(Q_0^2) + \nonumber \\
& + & W^0_{qq}(n_1)   W^0_{qg}(n_2) 
M_{\Sigma g}^{n_1 n_2}(Q_0^2) + \nonumber \\
& + & W^0_{qg}(n_1)   W^0_{qg}(n_2) 
M_{g g}^{n_1 n_2}(Q_0^2) \,;
\label{eq:Mn1n2_S-S_ev}
\eea
\bea
M_{g g}^{n_1 n_2}(Q^2)  & = & W^0_{gg}(n_1)  W^0_{gg}(n_2) 
M_{g g}^{n_1 n_2}(Q_0^2) + \nonumber \\
& + & W^0_{gq}(n_1)   W^0_{gg}(n_2) 
M_{\Sigma g}^{n_1 n_2}(Q_0^2) + \nonumber \\
& + & W^0_{gg}(n_1)   W^0_{gq}(n_2) 
M_{g \Sigma}^{n_1 n_2}(Q_0^2) + \nonumber \\
& + & W^0_{gq}(n_1)   W^0_{gq}(n_2) 
M_{\Sigma \Sigma}^{n_1 n_2}(Q_0^2) \,;
\label{eq:Mn1n2_g-g_ev}
\eea
\bea
M_{\Sigma g}^{n_1 n_2}(Q^2)  & = & W^0_{qq}(n_1)  W^0_{gg}(n_2) 
M_{\Sigma g}^{n_1 n_2}(Q_0^2) + \nonumber \\
& + & W^0_{qg}(n_1)   W^0_{gg}(n_2) 
M_{g g}^{n_1 n_2}(Q_0^2) + \nonumber \\
& + & W^0_{qq}(n_1)   W^0_{gq}(n_2) 
M_{\Sigma \Sigma}^{n_1 n_2}(Q_0^2) + \nonumber \\
& + & W^0_{qg}(n_1)   W^0_{qg}(n_2) 
M_{g \Sigma}^{n_1 n_2}(Q_0^2) \,;
\label{eq:Mn1n2_S-g_ev}
\eea
\bea
M_{g \Sigma}^{n_1 n_2}(Q^2)  & = & W^0_{gq}(n_1)  W^0_{qg}(n_2) 
M_{\Sigma g}^{n_1 n_2}(Q_0^2) + \nonumber \\
& + & W^0_{gg}(n_1)   W^0_{qg}(n_2) 
M_{g g}^{n_1 n_2}(Q_0^2) + \nonumber \\
& + & W^0_{gq}(n_1)   W^0_{qq}(n_2) 
M_{\Sigma \Sigma}^{n_1 n_2}(Q_0^2) + \nonumber \\
& + & W^0_{gg}(n_1)   W^0_{qq}(n_2) 
M_{g \Sigma}^{n_1 n_2}(Q_0^2) \,.
\label{eq:Mn1n2_g-S_ev}
\eea
The $Mellin$-inversion Eq. (\ref{eq:solutionVuT3n1n2}), completes again the procedure.

\subsection{Examples of Flavor decomposition}

\begin{itemize}

\item[A.] Flavor decomposition at the generic scale $Q^2$

\bea
F_{u^-u^-} & = & F_{(u - \bar u)(u- \bar u)} \equiv F_{u_Vu_V} \,;\\
F_{u^+u^+} & = & F_{(u+\bar u)(u+\bar u)} = F_{u_Vu_V}+2 \left[F_{u_V \bar u} + F_{\bar u u_V} \right] + 4 F_{\bar u \bar u} = \nonumber \\
& = & {1 \over 16} \left[ 4 F_{\Sigma \Sigma} + F_{T_8 T_8} + 9 F_{T_3 T_3} + \right. \nonumber \\
& + & 2 (F_{\Sigma T_8} + F_{T_8 \Sigma}) +3 (F_{T_8 T_3} + F_{T_3 T_8})  + \nonumber  \\
& + & \left. 6 (F_{\Sigma T_3} + F_{T_3 \Sigma}) \right]\,; \\
\eea
and the inverse

\bea
F_{\Sigma \Sigma} & = & F_{u^+u^+} + F_{d^+ d^+} + F_{s^+s^+} + \nonumber \\
& + & (F_{u^+d^+} + F_{d^+ u^+}) + (F_{u^+s^+} + F_{s^+ u^+}) + \nonumber \\
& + & (F_{d^+s^+} + F_{s^+ d^+})\,; \\
F_{T_3 T_3} & = & F_{u^+u^+} + F_{d^+ d^+} + \nonumber \\
& - & (F_{u^+d^+} + F_{d^+ u^+}) \,;\\
F_{\Sigma T_8}+F_{T_8 \Sigma} & = & 2 F_{u^+u^+} + 2 F_{d^+ d^+} - 4 F_{s^+s^+} +\\
& + & 2 (F_{u^+d^+} + F_{d^+ u^+}) + \nonumber \\
& - & (F_{u^+s^+} + F_{s^+ u^+}) + \nonumber \\
& - & (F_{d^+s^+} + F_{s^+ d^+})\,;\\
F_{\Sigma T_3}+F_{T_3 \Sigma} & = & 2 F_{u^+u^+} - 2 F_{d^+ d^+} +\\
& - & (F_{u^+s^+} + F_{s^+ u^+}) + \nonumber \\
& - & (F_{d^+s^+} + F_{s^+ d^+})\,;\\
F_{T_3 T_8}+F_{T_8 T_3} & = & 2 F_{u^+u^+} + 2 F_{d^+ d^+} +\\
& - & 2 (F_{u^+s^+} + F_{s^+ u^+}) + \nonumber \\
& + & (F_{d^+s^+} + F_{s^+ d^+})\,;\\
\eea
These relations are generally valid, not only at the specific $Q_0^2$.

\item[B.]{Reduction at the $\mu_0^2$ scale}

At the lowest scale one has:

\bea
F_{u^-u^-} & = & F_{(u - \bar u)(u- \bar u)} \equiv F_{u_Vu_V} = \nonumber \\
& = & u_Vu_V(x_1,x_2,\mu_0^2)\,;\\
F_{u^+u^+} & = & F_{(u+\bar u)(u+\bar u)} = F_{u_Vu_V}+ \nonumber \\
& + & 2 \left[F_{u_V \bar u} + F_{\bar u u_V} \right] + 4 F_{\bar u \bar u} = \nonumber  \\
& = & u_Vu_V(x_1,x_2,\mu_0^2)\,;\\
F_{s^+s^+} & = & 0 \,;\\
F_{gg} & = & 0\,.
\eea
and the inverse

\bea
F_{\Sigma \Sigma} & = & 3 F_{u_V u_V} \,; \\
F_{T_3 T_3} & = & - F_{u_V u_V}\,;  \\
F_{T_8 T_8} & = & 3 F_{u_V u_V}\,;\\
F_{\Sigma T_3}+F_{T_3 \Sigma} & = & 2 F_{u_Vu_V} \,;\\
F_{\Sigma T_8}+F_{T_8 \Sigma} & = & 6 F_{u_Vu_V}\,;\\ 
F_{T_3 T_8}+F_{T_8 T_3} & = & 2 F_{u_Vu_V}\,. \\
\eea
These relations are valid when the contributions at $\mu_0^2$ reduce to valence contributions only.

\end{itemize}

\end{document}